\documentclass[twocolumn,prx,amsmath,amssymb,citeautoscript,nobalancelastpage,aps,10pt,longbibliography]{revtex4-2}

\usepackage[toc,page]{appendix}
\usepackage{enumitem}
\usepackage[dvipsnames]{xcolor}
\usepackage{graphicx}
\usepackage{braket}
\usepackage[UKenglish]{babel}
\usepackage[UKenglish]{isodate}

\usepackage{mathtools}
\usepackage{esint}
\usepackage{comment}
\usepackage{relsize}
\usepackage{dsfont} 
\usepackage{commath}
\usepackage{titlesec}
\usepackage{soul}

\usepackage{hyperref}
\hypersetup{
    unicode=false,          
    pdftoolbar=false,       
    pdfmenubar=true,        
    pdffitwindow=false,     
    pdfstartview={FitH},    
    pdftitle={},            
    pdfauthor={Authors},    
    pdfsubject={},          
    pdfcreator={},          
    pdfproducer={},         
    pdfkeywords={quantum many-body scars}{semiclassical dynamics}{Rydberg atoms},
    pdfnewwindow=true,      
    colorlinks=true,        
    linkcolor=black,        
    citecolor=blue,         
    filecolor=magenta,      
    urlcolor=blue           
}

\usepackage{cleveref}

\Crefformat{equation}{Eq.~(#2#1#3)}
\Crefformat{equation}{Eq.~(#2#1#3)}
\Crefrangeformat{equation}{Eqs.~(#3#1#4--#5#2#6)}
\Crefname{figure}{Fig.}{Fig.}

\def\brakett#1#2{\mathinner{\langle{#1}|{#2}\rangle}}
\def\deriv#1{\mathinner{\mathrm{d}{#1}}}

\def \subname {symmetric } 
\def \qsname {top-band }

\def \M {\ensuremath{\frac{N}{2}}}

\DeclareMathOperator*{\Span}{span}
\DeclareMathOperator*{\Tr}{Tr}

\begin{document}

\title{Correspondence principle for many-body scars in ultracold Rydberg atoms}

\author{C. J. Turner}
\affiliation{School of Physics and Astronomy, University of Leeds, Leeds LS2 9JT, United Kingdom}

\author{J.-Y. Desaules}
\affiliation{School of Physics and Astronomy, University of Leeds, Leeds LS2 9JT, United Kingdom}

\author{K. Bull}
\affiliation{School of Physics and Astronomy, University of Leeds, Leeds LS2 9JT, United Kingdom}

\author{Z. Papi\'c}
\affiliation{School of Physics and Astronomy, University of Leeds, Leeds LS2 9JT, United Kingdom}

\date{\today}

\begin{abstract}
The theory of quantum scarring -- a remarkable violation of quantum unique ergodicity -- rests on two complementary pillars: the existence of unstable classical periodic orbits and the so-called \emph{quasimodes}, i.e., the non-ergodic states that strongly overlap with a small number of the system's eigenstates. Recently, interest in quantum scars has been revived in a many-body setting of Rydberg atom chains. While previous theoretical works have identified periodic orbits for such systems using time-dependent variational principle (TDVP), the link between periodic orbits and quasimodes has been missing.  Here we provide a conceptually simple analytic construction of quasimodes for the non-integrable Rydberg atom model, and prove that they arise from a ``requantisation" of previously established periodic orbits when quantum fluctuations are restored to all orders.
Our results shed light on the TDVP classical system simultaneously playing the role of both the mean-field approximation \emph{and} the system's classical limit, thus allowing us to firm up the analogy between the eigenstate scarring in the Rydberg atom chains and the single-particle quantum systems.

\end{abstract}

\maketitle{}

\section{Introduction}
\label{sec:introduction}

Quantum scars provide a surprising connection between single-particle quantum billiards and their classical counterpart~\cite{Heller:1984zz,HellerLesHouches}. The understanding of quantum scars is built upon the existence of unstable classical periodic orbits and atypical eigenstates. The former was first observed by studying the spreading of wave packets initialised along such orbits~\cite{Heller:1984zz,berry1989quantum} (see also Ref.~\onlinecite{HellerBook}).
In the limit $\hbar \rightarrow 0$,
the classical orbits are recovered, as stipulated by the correspondence principle~\cite{Bohr}. Furthermore, some atypical ``scarred" eigenstates were found to concentrate around such orbits, in contrast to typical eigenstates at a similar energy, which tend to be well-spread across the whole configuration space.
In the fully quantum case, the proof for the existence of scarred eigenstates relied on \emph{approximate} eigenstates or ``quasimodes", which are constructed to be strongly localised around a classical periodic orbit. For example, in the celebrated Bunimovich stadium problem, quasimodes have the form $\psi(x,y)= \phi(x)\sin(n\pi y)$, i.e., they represent a standing wave in one direction with a suitably chosen envelope function in the other direction~\cite{HellerLesHouches,Zelditch:2004fv}. By carefully controlling the density of states, it was possible to show that there exist eigenstates with an anomalously high overlap with a small number of quasimodes, hence ``inheriting" their atypical properties~\cite{OConnor:1988cn,Zelditch:2004fv,Hassell:2010ef}.

Recently, similar non-thermal states have been identified in interacting  quantum systems. These states, named quantum many-body scars by analogy, have been linked to long-lived oscillations measured in an experiment on a chain of Rydberg atoms~\cite{Bernien:2017bp,Turner:2017wu,Ho:2019gv,Choi:2018wn, lin2018exact, Khemani2018, Bull2020,Mark2020exact}. Since then, an increasing variety of models have been associated with  various aspects of many-body scarring~\cite{Calabrese16, Konik1, Konik2, Vafek, Moudgalya2018, IadecolaZnidaric, NeupertScars, Michailidis:2019wj, Schecter2019, Haldar2019, mukherjee2019collapse, sugiura2019manybody,
Moudgalya2019, Iadecola2019_3,  Hudomal2019, Pretko2019, Bull2019, Buca2019, Tindall2019, OnsagerScars, MoudgalyaKrylov, Zhao2020, Lee2020, Mark2020unified, Moudgalya2020,Villasenor2020, Moudgalya2020eta, Mark2020,Van2020,Mizuta2020,Hart2020,Kao2020}.
Moreover, non-stationary quantum dynamics has been identified in other classes of models, such as integrable spin chains~\cite{Buca2019_2} and dissipative systems~\cite{Buca2019}.
In this work we focus on the ``PXP" model -- an idealised effective model of the Rydberg atom experiment~\cite{Lesanovsky2012, FendleySachdev} -- which can be formally expressed as a non-integrable spin-1/2 lattice model without an ``obvious" semiclassical limit.
Inspired by the theory of single-particle scars, an immediate question arises: what are the periodic orbits and quasimodes in the PXP model?

Unlike single-particle billiards, the equivalent of the classical trajectory in the many-body case is less transparent.
One proposal makes use of the Time-Dependent Variational Principle (TDVP)~\cite{Kramer:1981tp}, wherein many-body dynamics is projected onto a restricted space of matrix product states~\cite{TDVP_QL, Altman17, Hallam2018ta}. 
Its application to the PXP model successfully captures the revivals following quenches from specific product states~\cite{Ho:2019gv,Michailidis:2019wj}.
It is thus natural to suggest that this approach defines an effective ``semiclassical'' description of the quantum dynamics, and holds the same relation with the exact PXP model as expected from the correspondence principle.

On the other hand, a family of quasimodes for the PXP model have also been independently constructed by the so-called Forward Scattering Approximation (FSA)~\cite{Turner:2017wu,Turner:2018in}.
This method makes it possible to extract properties of scarred eigenstates in systems much larger than those accessible by exact diagonalisation (ED). 
However, this way of building quasimodes presents some deficiencies, both on technical and conceptual levels.
Indeed, while the computational cost of generating the FSA states is lower than the exponential growth of the Hilbert space, it is still polynomial in the system size.
Furthermore, the link between the FSA and the TDVP is obscure, thus making it difficult to relate the quasimodes with the classical limit of the model, in a manner that had been achieved for single-particle scars.

In this paper we introduce a new way to form quasimodes which resolves the aforementioned problems.
Our construction relies on building a subspace symmetrised over permutations within each of the sublattices comprising even and odd sites in a chain. This approach can be thought of as a ``mean-field" approximation, and makes it possible to obtain closed-form expressions for the projections of states and operators in this symmetric subspace. This enables us to numerically approximate highly-excited eigenstates of chains of length $N\lesssim 800$, far beyond other methods. Further, we show that the path-integral quantisation of the TDVP topological space with a proper resolution of the identity \emph{exactly} reproduces the symmetric subspace in the thermodynamic limit. While adding quantum fluctuations to the TDVP description of the PXP model has been approximated to first order~\cite{Werman:2020vd}, our method includes quantum fluctuations to \emph{all} orders.
This correspondence demonstrates that the symmetric subspace, which defines a mean-field approximation, also serves as a justification for the TDVP dynamical system playing the role of the \emph{bona fide} classical limit for the PXP model.
We also show that the quasimodes corresponding to the scarred eigenstates are localised around the TDVP periodic orbit, in a way that is reminiscent of single-particle quantum scars.
Finally, performing time evolution in the symmetric subspace strongly suggests that the classical periodic orbit found in TDVP is stable to the addition of quantum fluctuations in the thermodynamic limit, at least to the ``mean-field" level as described above.

\section{Rydberg blockade and quantum scars}
\label{sec:background}

The PXP model is an effective model of a chain of atoms in the Rydberg blockade~\cite{Lesanovsky2012}. Denoting by $\ket{\circ}$ and $\ket{\bullet}$ the ground and excited states of each atom, the Hamiltonian is
 \vspace{-0.1cm}
\begin{equation}\label{eq:pxp}
H = \sum_{n=1}^N P_{n-1} X_n P_{n+1}\text{,}
\end{equation}
where $X_n=\ket{\circ}\bra{\bullet}+\ket{\bullet}\bra{\circ}$ is the Pauli $X$ matrix and $P_n = \ket{\circ}\bra{\circ}$ is a local projector. Here $N$ denotes the number of sites in the chain and, unless specified otherwise, we assume periodic boundary conditions (PBCs), $N+1\equiv 1$. The projectors make the model in Eq.~(\ref{eq:pxp}) constrained: an atom can be excited only if both adjacent atoms are in the ground state, e.g., the transition $\ket{\ldots{\circ}{\circ}{\bullet}\ldots} \rightarrow \ket{\ldots {\circ}{\bullet}{\bullet}\ldots}$ is forbidden. This causes a fracturing of the Hilbert space, with a strong impact on its dynamical properties~\cite{Khemani2019, Sala2019,Khemani2019_2,Roy2020,Yang2020,Pancotti2020,Lychkovskiy2020}. 
Below we focus on the largest connected sector of the Hilbert space which contains no state with adjacent excitations.

The PXP model in \Cref{eq:pxp} is non-integrable and thermalising~\cite{Turner:2017wu}, but it exhibits periodic revivals when quenched from certain initial states such as the N\'eel state, $\ket{\mathbb{Z}_2}=\ket{{\bullet}{\circ}{\bullet}{\circ}\ldots}$.
The observed revivals in the density of domain walls~\cite{Bernien:2017bp} are remarkable given that the $\ket{\mathbb{Z}_2}$ state effectively forms an ``infinite-temperature" ensemble for this system, for which the ``strong" Eigenstate Thermalisation Hypothesis (ETH) predicts fast equilibration~\cite{DeutschETH,SrednickiETH,dAlessio2016, Gogolin2016}.
Unlike many-body localised~\cite{basko2006metal,serbyn2013local,HuseNandkishoreOganesyan} and integrable systems~\cite{sutherland2004beautiful} which strongly break ergodicity, the PXP model breaks it only weakly~\cite{Shiraishi:2017ek,Shiraishi2019}. The breaking of ergodicity in the PXP model occurs most prominently due to the existence of a band of $O(N)$ scarred eigenstates, which have equal energy separation, anomalously high overlap with $\ket{\mathbb{Z}_2}$ state, and lowest entanglement among all eigenstates~\cite{Turner:2018in}. These states are representatives of scarred towers of eigenstates (as can be seen in Fig.~\ref{fig:quasi}(a) below), which contain additional atypical eigenstates that exhibit clustering around the same energies and whose  finite-size scaling of inverse participation ratio is different from that of the thermalising eigenstates~\cite{Turner:2017wu}. The scarring properties in all of these special eigenstates are explained by the existence of quasimodes: simpler states with an underlying structure -- the analogues of  ``standing waves" in  billiard systems -- which represent concentration points of the special eigenstates.

\section{Construction of quasimodes}
\label{sec:quasi}

\begin{figure}[t]
\centering
\includegraphics[width=\linewidth]{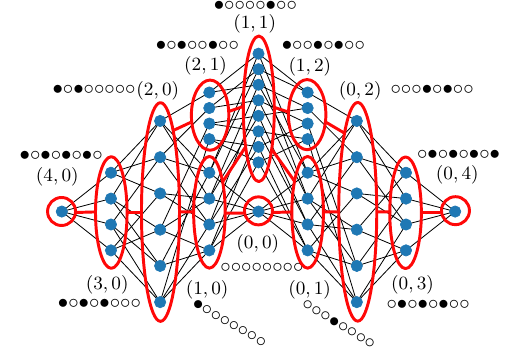}
\caption{
An example of the constrained Hilbert space of the PXP model in Eq.~(\ref{eq:pxp}) for $N=8$. Each blue dot represents an allowed product state of atoms compatible with the Rydberg constraint. The states are grouped into equivalence classes of the symmetric subspace $\mathcal{K}$, denoted by ellipses. Each class is labelled by a representative state, e.g., the class $(0,3)$ contains the representative state ${\circ}{\bullet}{\circ}{\bullet}{\circ}{\bullet}{\circ}{\circ}$ and all other obtained by permuting sites in each of the sublattices.
In order to make explicit the connections between classes, we chose representatives that only differ by a single spin flips between neighbouring classes.}
\label{fig:K}
\end{figure}
We construct quasimodes for the PXP model by first defining a symmetric subspace $\mathcal{K}$ from a set of equivalence classes $(n_1, n_2)$, where integers $n_1$, $n_2$ label the number of excitations on the two sublattices, encompassing the odd and even sites, respectively.
Elements in these classes are equivalent under the action of the product of two symmetric groups $S_{N/2}$ which ``shuffle" the sites in each sublattice.
For example, states $\ket{{\bullet}{\circ}{\bullet}{\circ}{\circ}{\bullet}{\circ}{\circ}{\circ}{\circ}}$ and $\ket{{\circ}{\circ}{\bullet}{\circ}{\circ}{\circ}{\bullet}{\circ}{\circ}{\bullet}}$ belong to the same class as they both have two excitations in the first sublattice and one in the second.
An illustration of the construction of $\mathcal{K}$ for a PXP model of size $N{=}8$ is presented in Fig.~\ref{fig:K}.

The construction of the subspace $\mathcal{K}$ is inspired by the fact that most of the information is retained near the two reviving states (the N\'eel states), while away from them the details about local correlations are erased.
This is a natural generalisation of the free paramagnet, which is recovered by dropping the projectors in Eq.~(\ref{eq:pxp}): to characterise states of the free paramagnet, we only need to know the number of excitations  relative to some reference state, not their positions in the lattice, which is the foundation of a mean-field description.
In the present case of the PXP model, we generalise this mean-field approximation by assuming that the important information is the number of excitations \emph{on each sublattice}, regardless of their positions.
This approach is reminiscent of the fully symmetric subspaces in the context of quantum de Finetti theorem~\cite{Brandao2016} and in studies of fully-connected models~\cite{Sciolla2011, Mori2017}.
However, the key difference here is that the permutation symmetry is broken to the sublattice level, and many of the permutation shuffles violate the Rydberg constraint and therefore have to be excluded, which makes the analysis non-trivial.

We define an orthonormal basis for $\mathcal{K}$ from symmetric combinations of members of each class $(n_1,n_2)$,
\begin{equation}
  \ket{(n_1,n_2)} = \frac{1}{\#(n_1,n_2)} \sum_{x\in(n_1,n_2)}\ket{x}\text{,}
  \label{eq:symm_basis}
\end{equation}
where $\#(n_1,n_2)$ is the size of an equivalence class and the sum runs over all product states $\ket{x}$ that are members of this class.

The class sizes can be calculated analytically because the configurations can be generated by a recursive procedure or transfer-matrix problem which `glues' sites (in the basis) onto the boundary of a smaller version of the same counting problem.
Because the constraint on gluing an additional ${\bullet}$ site is only sensitive to the state of the site immediately on the boundary, we can erase the information about the configurations further away and represent them by a count of all those configurations which are equivalent on the boundary.
This produces a method for calculating $\#(n_1,n_2)$ using dynamic programming in only polynomial time.
It can be shown (see \Cref{sec:mat_elems}) that the class sizes admit the closed-form expression
\begin{align}
  \#(n_1,n_2) &= \frac{\frac{N}{2}\!\left(\frac{N}{2}{-}n_1{-}n_2\right)}{(\frac{N}{2}{-}n_1)(\frac{N}{2}{-}n_2)}
  \binom{\frac{N}{2}{-}n_1}{n_2}\!\binom{\frac{N}{2}{-}n_2}{n_1}\text{,}
  \label{eq:class_cf}
\end{align}
in the case of PBC, with an analogous expression for open boundary condition. For simplicity, we assumed $N$ is even.

Using Hermiticity and the sublattice exchange symmetry, all matrix elements of $H$ in the symmetric subspace can be found by examining only the $(H)_{(n_1,n_2)}^{(n_1{-}1,n_2)} = \bra{(n_1{-}1,n_2)}H\ket{(n_1,n_2)}$ matrix elements.
These can be found using the idea that for each configuration in $(n_1,n_2)$ there are precisely $n_1$ ways to remove an excitation from sublattice $1$.
Accounting for normalisation and using the closed form \Cref{eq:class_cf} yields a remarkably simple effective Hamiltonian within $\mathcal{K}$:
\begin{align}\label{eq:Ham}
  \Big(H\Big)_{(n_1,n_2)}^{(n_1{-}1,n_2)}
  &= \sqrt{\frac{n_1(\frac{N}{2}{-}n_1{-}n_2)(\frac{N}{2}{-}n_1{-}n_2{+}1)}{\frac{N}{2}{-}n_1}}\text{.}
\end{align}

\section{Properties of quasimodes}\label{sec:quasimode_properties}

\begin{figure}[t]
  \includegraphics[width=\linewidth]{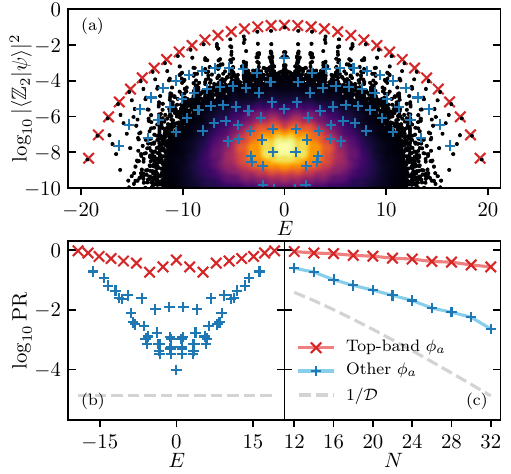}
  \caption{
  Quasimode properties.
    (a) Scatter plot showing energies of all eigenstates vs. their overlap with $|\mathbb{Z}_2\rangle$ state.  Red crosses denote \qsname quasimodes, blue pluses are the remaining quasimodes. Colour indicates the density of data points.
    (b), (c): Localisation of quasimodes in the energy eigenbasis.
    Plots show PR (see text)  as a function of (b) energy $E$ (for $N=32$) and (c) system size $N$.
    In (c) we compare the \qsname quasimode closest to $E=0$ with the mean over the other quasimodes with $0 < E < \sqrt{N}$, all in the zero momentum sector.
    The scaling of PR is different for all  \qsname modes compared to generic states (grey dashed line).
  }
  \label{fig:quasi}
\end{figure}

The quasimodes, or approximations to eigenstates, are formed by diagonalising the PXP Hamiltonian projected to $\mathcal{K}$, \Cref{eq:Ham}.
Despite the apparent simplicity of the matrix elements, we are unaware of an analytical method to diagonalise \Cref{eq:Ham} and we perform this step numerically.
In \Cref{fig:quasi}(a) we compare the quasimodes against the exact eigenstates for system size $N{=}32$.

We measure the localisation of the quasimodes among energy eigenstates using participation ratio,
$\mathrm{PR} = \sum_{E} \left|\brakett{E}{\phi_a}\right|^4$,
which represents the inverse of the number of eigenstates with which a state has `typical' overlap. For random states this is $1/\mathcal{D}$, with $\mathcal{D}$ the Hilbert space dimension.
In \Cref{fig:quasi}(b) we plot the PR of all the quasimodes.
The $N{+}1$ \qsname quasimodes have very large $\mathrm{PR}$ compared to a typical random state, with appreciable overlap with only a small number of eigenstates.
While the remaining quasimodes are also more localised than typical states, their PR is clearly smaller than the top-band ones.
This difference between the two groups is further illustrated in the scaling of $\mathrm{PR}$ with $N$.
In \Cref{fig:quasi}(c) we show this scaling, focusing on the \qsname quasimode closest to the middle of the spectrum ($E=0$) in the zero-momentum sector, alongside other quasimodes with energies $0 {<} E {<} \sqrt{N}$. This avoids the edges of the spectrum where the density of states is much smaller.
From this we conclude that the \qsname quasimodes are localised over a small number of eigenstates compared to the density of states.

Furthermore, the simple structure of the symmetric subspace allows us to characterise the errors in our eigenstate approximations.
We can calculate the energy variance matrix in the symmetric subspace,
$\sigma_H^2 = \hat{\mathcal{K}}(H^2) - (\hat{\mathcal{K}}(H))^2$,
where $\hat{\mathcal{K}}$ is a super-operator which projects linear operators on the full Hilbert space down to the symmetric subspace $\mathcal{K}$.
It can be shown that the energy variance matrix is diagonal in the symmetric basis and has the following closed form (see \Cref{sec:subs_var}):
\begin{align}
  \sigma_H^2 = \frac{\hat{n}_1\hat{n}_2(\hat{n}_2-1)}{\left(\frac{N}{2}-\hat{n}_1-1\right)\left(\frac{N}{2}-\hat{n}_1\right)} +\left(1\leftrightarrow 2\right)\text{,}
  \label{eq:mt_variance_cf}
\end{align}
where $\hat{n}_1,\hat{n}_2$ are the operators that count excitations on their respective sublattices, e.g.,  $\hat{n}_i\ket{(n_1,n_2)}=n_i\ket{(n_1,n_2)}$ for $i=1,2$. 

\begin{figure}[t]
  \includegraphics[width=\linewidth]{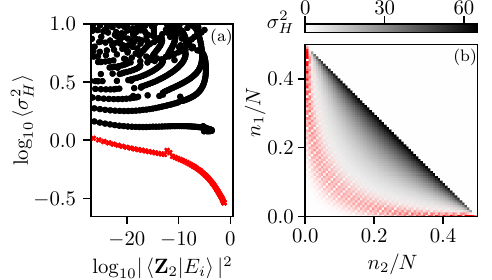}
  \caption{
    (a) Energy variance of the quasimodes for $N=140$. Top-band quasimodes (red crosses) have low energy variance density compared to typical energy scales, indicating that the dynamics restricted to $\mathcal{K}$  remains close to exact evolution at relatively long times.
    (b) Energy variance in the symmetric subspace, \Cref{eq:symm_basis}, as a function of scaled $n_1$ and $n_2$.
    Superimposed (in red) is the probability distribution on $n_1$ and $n_2$ for a \qsname quasimode in the middle of the spectrum (with $E\approx 27.28$). 
  }
  \label{fig:variance}
\end{figure}

As expected, among the quasimodes, we find that the \qsname ones are those with the smallest energy variance, especially those that have high overlap with $|\mathbb{Z}_2\rangle$ state.
This can be seen in \Cref{fig:variance}(a), which slows the energy variance for individual quasimodes in system size $N{=}140$.
The variance is linear in $N$, representing constant density of energy fluctuations, but for the \qsname quasimodes it is much smaller than the typical energy scales.
The average energy fluctuations in a quasimode is asymptotically given by $1/9$, while the same quantity restricted to \qsname quasimodes is found numerically to be $\approx 0.0049$.
This suggests that the time evolution approximated within this subspace will be reasonably accurate for local quantities such as local observable expectations for relatively long times.
The typical \qsname quasimode support can be visualised on an ($n_1$,$n_2$) triangle -- see \Cref{fig:variance}(b), which reveals that in these cases the quasimodes tend to ``avoid" the region with $n_1$ and $n_2$ both large where the energy variance is high. This is analogous to the classical trajectory in the TDVP avoiding ``high-leakage" regions of the phase space~\cite{Ho:2019gv}.

\section{Quantum dynamics}
\label{sec:dyn}

\begin{figure}
  \includegraphics[width=\linewidth]{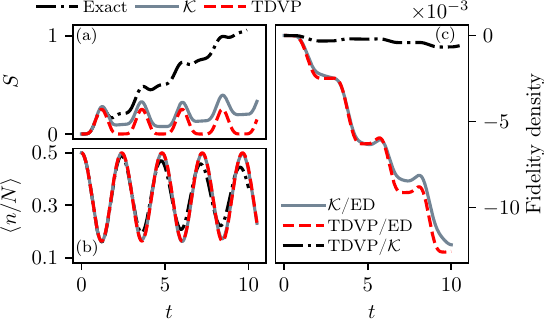}
  \caption{
    Dynamics for $N=32$. 
    (a) Entanglement entropy following a quench for the full quantum dynamics, the dynamics projected into $\mathcal{K}$, and the TDVP dynamics on $\mathcal{M}$.
    The entropy growth shows that for $\mathcal{K}$, the initial state does not revive exactly leaving some residual entropy compared to the more weakly entangled dynamics on $\mathcal{M}$.
    (b) Local observables following the quench are well reproduced.
    Both TDVP and $\mathcal{K}$ subspace approximations fail to  capture a slow decay in the oscillations for $t\gtrsim 5$.
    (c) Pairwise comparisons between the exact dynamics, projected dynamics on $\mathcal{K}$ and TDVP dynamics on $\mathcal{M}$.
    This illustrates that the dynamics on $\mathcal{K}$ is always more accurate than TDVP on $\mathcal{M}$.
  }
  \label{fig:dynamics}
\end{figure}

Our construction of the quasimodes can also be used to approximate the dynamics of the system. This is illustrated in \Cref{fig:dynamics}, which displays a comparison with ED as well as variational (TDVP) results, for a quench from the $\mathbb{Z}_2$ Rydberg crystal.
TDVP approach is based on a restricted class $\mathcal{M}$ of matrix product states $|\tilde\Psi(\theta,\phi)\rangle$ with bond dimension equal to 2, which can be parametrised using matrices~\cite{Bernien:2017bp,Ho:2019gv}
\begin{align}
  \mathcal{A}^{\circ}_i &= \left(\begin{array}{cc} \cos\tfrac{\theta_i}{2} & 0 \\ 1 & 0 \end{array}\right)\text{,} &
  \mathcal{A}^{\bullet}_i &= \left(\begin{array}{cc} 0 & -ie^{i\phi_i}\sin\tfrac{\theta_i}{2} \\ 0 & 0 \end{array}\right),
  \label{eq:gauged_mps_param_main}
\end{align}
where $i$ labels the lattice sites within a unit cell -- i.e. $i=1,2$ for the $|\mathbb{Z}_2\rangle$ quench.
This is either an infinite matrix product in the thermodynamic limit or a traced matrix product for finite and periodic system of size $N$.
The states are asymptotically normalised, $\brakett{\tilde\Psi(\theta,\phi)}{\tilde\Psi(\theta,\phi)} \rightarrow 1$, as $N \rightarrow \infty$~\cite{Ho:2019gv}. 
The time evolution of the TDVP angles in \Cref{fig:dynamics} was obtained using the equations of motions for the infinite system~\cite{Michailidis:2019wj}. 

The differences between the dynamics within $\mathcal{K}$ and the TDVP dynamics on $\mathcal{M}$  can be seen clearly in \Cref{fig:dynamics}(a), which shows the von Neumann entropy for an equal bipartition of the system.
This quantity, although non-local, can also be efficiently calculated using the present method -- see \Cref{sec:supp_entropy}.
We observe that entropy captures the differences between approximation schemes well before the relaxation time set by the energy variance.
In contrast, local quantities such as local expectation values are very consistent -- see \Cref{fig:dynamics}(b) which shows  the density of Rydberg excitations in the system, $\langle n / N\rangle$. 
The accuracy of these different methods can be assessed using co-moving fidelity densities between pairs of evolutions, such as $\frac{1}{N}{\rm log}\left(|\brakett{\psi_\mathrm{ED}(t)}{\psi_\mathcal{K}(t)}|^2\right)$.
Both the TDVP and symmetric subspace approaches generate a very low error density.
The dynamics in the \subname subspace has been simulated in large system sizes (for $N\leq 720$, see \Cref{sec:supp_Ninfty}), and the results are consistent with the revivals of the N\'eel states becoming perfect in the thermodynamic limit.
However, there is a slight deviation from the TDVP orbit~\cite{Ho:2019gv}, in that there is no perfect state transfer between $\ket{{\circ}{ \bullet}{ \circ }{\bullet} \ldots}$ and the N\'eel state with the two sublattices flipped, $\ket{{\bullet}{ \circ}{ \bullet}{ \circ} \ldots}$.

The co-moving fidelities in \Cref{fig:dynamics}(c) reveal a curious  similarity between the evolutions within TDVP and symmetric subspace $\mathcal{K}$. The variational class of states considered in the TDVP approach can be viewed as a Gutzwiller projection of spin coherent states with a unit cell of two sites, where the projector eliminates nearest neighbour excitations~\cite{Bernien:2017bp,Ho:2019gv,Michailidis:2019wj}. The states in this subspace $\mathcal{M}$ are parametrised by two angles which represent probabilities for a site of each sublattice to hold an excitation.
Recall that the basis for $\mathcal{K}$ consisted of states with definite occupations numbers for each sublattice.
In this sense, there is a direct relationship between the two subspaces, reminiscent of the relationship between canonical and grand canonical ensembles in statistical mechanics.
More formally, it can be proven that the linear span of $\mathcal{M}$ is equal to the symmetric subspace $\mathcal{K}$, for every fixed system size $N$ -- see \Cref{sec:TDVP_1,sec:supp_span}.
This implies that the dynamics within $\mathcal{K}$ has a leakage rate, or instantaneous error, which is bounded above by that of the TDVP evolution.
Next we will investigate this connection in more detail and show how $\mathcal{K}$ and $\mathcal{M}$ are related by quantisation.

\section{Quantisation of periodic orbits and the correspondence principle}\label{sec:corr}

Our final step is to establish a correspondence between $\mathcal{K}$
and the TDVP system on $\mathcal{M}$.
The latter is typically conceived as a variational approximation -- how do we reinterpret it as a classical limit?
Conversely, how can we best quantise this classical system, given that quantisation is often a non-unique recipe?
These questions have been conveniently addressed in a path integral framework~\cite{Gazeau:2009vb,Green:2016uw}.
Thus, we first carefully formulate the path integral over our constrained space before arriving at the correspondence principle that relates $\mathcal{M}$ and $\mathcal{K}$.

\subsection{The variational principle and dequantisation} \label{sec:path}
A path integral over a classical system is usually used to define a quantum propagator which associates amplitudes to pairs of path endpoints.
However, in the present case, we follow a reverse procedure: we use the path-integral formulation to ``dequantise" a system by writing its propagator as a path integral~\cite{Gazeau:2009vb}.
This provides a classical state space and an action functional defining a classical system. 
The restriction to the TDVP state space is equivalent to such a dequantisation, if it can be given a resolution of the identity~\cite{Green:2016uw}.
This process provides a correspondence between a classical system such as $\mathcal{M}$ and a quantum system in its linear span (for $\mathcal{M}$, the latter is the symmetric subspace $\mathcal{K}$). 
If the same variational principle is taken for restricting to a vector subspace, such as our symmetric subspace $\mathcal{K}$, then it is equivalent to a projection. 
It is natural that both the quantised and dequantised system satisfy this same same kind of variational optimality.

Crucially, the prescription above relies on being able to associate a measure $\mu$ to some family of paths.
This can be achieved by equipping the classical state space with an integration measure, forming a structure we will refer to as a \emph{frame}~\cite{Christensen2016}.
This structure is characterised by the frame operator,
\begin{equation}\label{eq:frametransformation}
  S_\mu = \int_\mathcal{M} \mathrm{d}\mu_{\theta,\phi} \ket{\tilde\Psi(\theta,\phi)}\bra{\tilde\Psi(\theta,\phi)}\text{,}
\end{equation}
using the parametrisation of states in Eq.~(\ref{eq:gauged_mps_param_main}).
For $(\mathcal{M},\mu)$ to constitute a genuine frame, $S_\mu$ must be bounded and positive definite.
It is this operator which must reproduce the identity to yield a well-defined quantisation.

The strategy we follow is to start from a ``proxy" measure that produces a frame over $\mathcal{M}$ which is reasonably well-behaved, although it may not resolve the identity -- see Appendix~\ref{sec:supp_delta} for details on this measure and the corresponding frame operator $S_\mu$. 
This can then be used to construct a transformed state space, $\overline{\mathcal{M}}$, 
by replacing the states attached to each point according to the linear action,
\begin{equation}
  \ket{\tilde\Psi} \mapsto S^{-1/2}_\mu \ket{\tilde\Psi}\text{.}
\end{equation}
We find the differences between the two spaces decrease as $N$ increases, until in the thermodynamic limit they are equivalent almost everywhere.
This means that we can construct a corrected measure $\nu$ such that in the thermodynamic limit the frame $(\mathcal{M},\nu)$ forms a resolution of the identity.
\begin{figure}[t]
  \includegraphics[width=\linewidth]{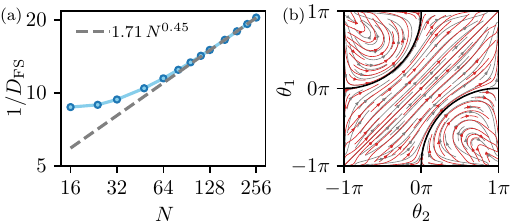}
  \caption{
    (a) Fubini-Study distance $D_\text{FS}$ between the original $\mathcal{M}$ and transformed $\overline{\mathcal{M}}$ state spaces, averaged over the phase-space with measure $\mu$.
    (b) Illustration of the integral curves of (red) the frame-transformed dynamics and (grey) the original dynamical system, for $N=128$.
    The flows are highly similar except towards the corners.
    The black line shows the evolution following a N\'eel state quench in the transformed system.
  }
  \label{fig:pathintegral_ft}
\end{figure}
We demonstrate this numerically in \Cref{fig:pathintegral_ft}(a) by computing the Fubini-Study metric or Bures distance $D_\text{FS}$~\cite{Bengtsson:2006iz}  -- which measures the angle between two states as projective rays -- averaged over the phase space.
We find $D_\text{FS} \sim 1/\sqrt{N}$, supporting the asymptotic equivalence between $\mathcal{M}$ and $\overline{\mathcal{M}}$.
This means that $S_\mu$ asymptotically acts essentially as a scalar on each of the points of $\mathcal{M}$, and can therefore be absorbed into the measure.
This change in measure is unable to affect the behaviour of the classical dynamical system, which can be formulated entirely independent of the measure.

For finite sizes, we find that the transformed space $\overline{\mathcal{M}}$ is a moral equivalent to $\mathcal{M}$, in the sense that they are highly similar around the ``good'' trajectories, i.e., the ones in the vicinity of the $|\mathbb{Z}_2\rangle$ quench trajectory,  while potentially disagreeing about the ``bad'' trajectories. The latter trajectories are the ones that pass through the corners of the phase space, and they are deemed unphysical as the corners represent the zero vector in the quantum picture, which yields a large  geometric error. The disregard of the bad trajectories is also justified because $\mathcal{M}$ is intended to be effective at describing the $|\mathbb{Z}_2\rangle$  state quench dynamics, while in other parts of the phase space the leakage rate or geometric error may be large~\cite{Ho:2019gv,Michailidis:2019wj}.
This is illustrated in \Cref{fig:pathintegral_ft}~(b) for $N{=}128$, where the grey lines are for the original dynamical system and the red lines are the frame-transformed dynamics. The vector fields are for finite-sized systems and are calculated from the action principal represented in the symmetric basis for $\mathcal{K}$.
This method avoids numerical instability found in the canonical treatment~\cite{Kramer:1981tp} around singularities in the Gram matrix.
We observe in \Cref{fig:pathintegral_ft}~(b) that the flows agree around the good trajectories and diverge from one another on approaching the corners.
The black line is an integral curve for the frame-transformed system starting from $|\mathbb{Z}_2\rangle$  state, showing that the trajectory which underlies the quantum scar dynamics is retained.

To summarise, we have shown there is a measure which quantises the original space $\mathcal{M}$ to $\mathcal{K}$ in the thermodynamic limit.
This was achieved by constructing a sequence of transformed spaces for finite sizes which has a common $N{\rightarrow}\infty$ limit as the original space.
For each finite $N$, the transformed space quantises naturally to the Hilbert space $\mathcal{K}$ examined in the previous sections, in which many quantities of interest may be calculated efficiently.
Even for finite system sizes, the transformed space has very similar dynamics to the original space, except around regions which do not contribute to the quench dynamics.


\subsection{Physical interpretation of quasimodes}\label{sec:corrprinc}

Armed with a quantisation--dequantisation correspondence between the classical TDVP system and the symmetric subspace $\mathcal{K}$, we now turn to the consequences for our interpretation of individual quasimodes, and by extension the exact eigenstates.
We will show that the \qsname quasimodes can be viewed as standing waves localised around $|\mathbb{Z}_2\rangle$  state trajectory, making contact with the the phenomena which inspired the term quantum scar in billiard systems.
In these systems, it was found that some eigenfunctions anomalously localise around unstable periodic orbits~\cite{Heller:1984zz}, leading to the picture that the classical trajectory is ``imprinted" upon the quantum states, giving rise to wavefunction scarring. 

First, we will show how the quasimodes can be viewed as wavefunctions.
The frames constructed in the previous section, which resolve the identity, allow us to analyse a quantum state $\ket{\psi} \in \mathcal{K}$ through the inner-product map:  
\begin{align}
  \psi(\theta,\phi) &= \bra{\tilde\Psi(\theta,\phi)} S_\mu^{-1/2} \ket{\psi}, \label{eq:rep}\\
  \ket{\psi} &= \int_{\mathcal{M}} \deriv\mu_{\theta,\phi} \psi(\theta,\phi)\,S_\mu^{-1/2}\ket{\tilde\Psi(\theta,\phi)}\text{.}
\end{align}
This mapping allows us to form a wavefunction representation, $\psi(\theta,\phi)$, and, conversely, to recover the quantum state from that representation.  This serves as a phase-space analogue to a wavefunction, which in the usual Schr\"odinger picture would be defined only on a Lagrangian submanifold such as configuration space.

\begin{figure}[t]
  \includegraphics[width=\linewidth]{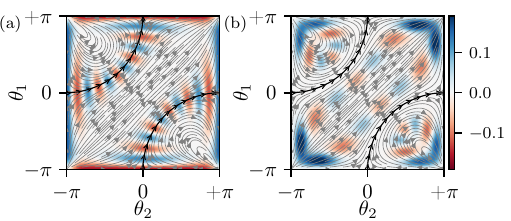}
  \caption{
    Viewing a selected (a) \qsname and (b) non \qsname quasimode as wavefunctions over $\theta_1$ and $\theta_2$ for $N{=}128$. Colour scale represents the real part of the wavefunction.
    Top-band quasimodes concentrate around the classical periodic orbits demonstrating quantum scarring.
    The phase of the wavefuntion winds along the trajectory, and this winding number is related to the approximately equally spaced energies of the quasimodes.
    The other quasimodes avoid this periodic trajectory, due to orthogonality, and concentrate around the trajectories with the corners of the square for endpoints.
  }
  \label{fig:correspondence}
\end{figure}

We restrict attention to a Lagrangian section with $\phi_a=0$, as this typically loses no information and the two-dimensional space is better suited for visualisation~\cite{Michailidis:2019wj}.
The real part of the wavefunction representation in Eq.~(\ref{eq:rep}) is shown in \Cref{fig:correspondence}, for an illustrative example of a  \qsname quasimode (a) and one of the remaining quasimodes (b), for $N{=}128$.
The \qsname quasimodes are found to concentrate around the $|\mathbb{Z}_2\rangle$  classical trajectory in a manner strikingly reminiscent of wavefunction scarring in quantum billiards.
The classical trajectory shown by the black line in \Cref{fig:correspondence} is for the thermodynamic limit. In  \Cref{fig:correspondence}(a) we observe that
the wavefunction spreads out towards the boundary and becomes completely delocalised at the edges, $\theta_i = \pm \pi$, where the points are indistinguishable in the transverse direction.  Following the classical trajectory, the phase of the wavefunction winds, and integrality of this winding number leads to their approximately equal spacing in energy. We recognise this behaviour from quantisation of regular trajectories~\cite{GutzwillerBook}: in the old quantum theory, this underpins Sommerfeld-Wilson quantisation and leads to the de~Broglie standing-wave condition for the Bohr model. These features are inherited by the exact scarred eigenstates given their strong overlap with the quasimodes.
In contrast, the other quasimodes are typically found around the orbits connecting corners of the manifold -- see \Cref{fig:correspondence}(b).
As the geometric error is large around the corners, this may account for the increased energy variance and energy delocalisation found previously in such quasimodes.

We note one dissimilarity between our result in \Cref{fig:correspondence} and wavefunction scarring in the fact that the TDVP phase-space section is entirely regular. It cannot exhibit chaos because a symmetry argument leaves the two-dimensional $\phi_a{=}0$ section dynamically invariant~\cite{Ho:2019gv}.
The other quasimodes are distinguished from the top-band ones not by chaotic nature but because they do not appear to survive the ``injection'' into the full Hilbert space, i.e., while the top-band quasimodes accurately approximate the true eigenstates of the full PXP model, this is not the case for the other quasimodes.
In Ref.~\onlinecite{Michailidis:2019wj}, related ans\"atze were used to study the dynamics of different initial states and in the presence of a chemical potential which makes $\phi_a$ variables dynamical.
In these situations, mixed phase space containing both regular and chaotic regions was found to generally appear~\cite{Michailidis:2019wj}. We expect that if the bond dimension of our ansatz is increased, then the TDVP phase space we consider here can be embedded into this larger phase space wherein chaos might develop.
For the periodic trajectory, which underlies the top-band quasimodes, the integrated error is low and we expect this trajectory to remain regular, while the top-band quasimodes would continue to leave a scar on the eigenstates. However, the other quasimodes, which concentrate around  trajectories with greater integrated variational error, are more sensitive to the addition of extra degrees of freedom and are expected to develop chaos.

\section{Conclusions and discussion}
\label{sec:discussion}
	
We have introduced a symmetric subspace $\mathcal{K}$ which allows for a simple construction of quasimodes in the Rydberg atom model.
We showed that these quasimodes are well-localised in the energy eigenbasis and have greatly suppressed energy fluctuations, thereby making them excellent approximations to the exact eigenstates.
As a consequence the quench dynamics is also well captured in this subspace..
We have argued that our subspace $\mathcal{K}$ can be interpreted as a ``requantisation'' of the variational classical limit~\cite{Ho:2019gv}, when quantum fluctuations are restored to all orders.  This reveals the TDVP classical system is formed of two approximations, one of which is a mean-field type approximation and another which is the classical limit.
Using finite size scaling, we showed that the wavefunction revivals survive in the thermodynamic limit in the mean-field regime, but the state transfer between the two N\'eel state is no longer perfect. This implies that the TDVP periodic trajectories are robust to the addition of quantum fluctuations.

The simplicity of this approach allowed us to obtain closed form results for many important quantities and our method can be easily extended to account for various perturbations to the PXP model that have been considered in the literature~\cite{Turner:2018in,Khemani2018,Choi:2018wn}.
Moreover, we expect the analogue of subspace $\mathcal{K}$ to be particularly useful in studying two-dimensional scarred models~\cite{Lee2020,Lin2020,Michailidis2020}, models of lattice gauge theories~\cite{Surace2020} and the one-dimensional limit of the fractional quantum Hall effect~\cite{Moudgalya2019}, as well as in describing dynamics in other types of constrained models, e.g., quantum dimer models~\cite{Moessner2011}, whose ground states have a similar ``symmetrised" structure at Rokshar-Kivelson points~\cite{Rokhsar1988}.

Our work also relates to several fundamental questions.
Underlying our method is a symmetrisation process which identifies our symmetric subspace by shuffling the lattice sites within the even and odd sublattices, whilst respecting the Rydberg blockade constraint.
It is within this symmetric subspace that the scarred quasimodes are found.
Previously, the structure of the scarred subspace has been argued to form an approximate SU(2) representation~\cite{Turner:2017wu,Choi:2018wn}. The combination of these two structures is suggestive of the Schur-Weyl duality, which classifies SU(2) representations by representations of the symmetric group. At the same time, all the detailed elements of this duality are modified due to the presence of constraints. For example, the constrained Hilbert space cannot be given a permutation group action similar to the tensor space, which has a natural symmetric group action exchanging the tensor factors. We believe that the relationship seen here is an interesting and non-trivial extension of the Schur-Weyl duality that merits further investigation.

Our results around the quantisation of the variational classical system allow us to firm up the analogy between the wavefunction scarring in the PXP model and the single-particle quantum systems. The one outstanding challenge for making contact between the quasimodes and rigorous results on exact eigenstates is the exponential many-body density of states. The approximate permutation group action, identified in our work, could provide a way to control the density of states in order to rigorously prove the existence of quantum non-unique ergodicity in the Rydberg atom chain.

\section{Acknowledgements}

We thank Alexios Michailidis for useful discussions.
We acknowledge support by EPSRC grants EP/R020612/1, EP/R513258/1 and EP/M50807X/1.  Statement of compliance with EPSRC policy framework on research data: This publication is theoretical work that does not require supporting research data.  
Z.P. acknowledges support by the Leverhulme Trust Research Leadership Award RL-2019-015. This work benefited from participation at KITP Follow-Up programme, supported by the National Science Foundation under Grant No. NSF PHY-1748958. 

\appendix

\section{Analytical derivation of class sizes and matrix elements} \label{sec:mat_elems}

\subsection{Class sizes}
 
In the \subname subspace, each basis state is a symmetric superposition of product states in the full (constrained) Hilbert space.
All these states have the same number of excitations on each sublattice, and so they belong to the equivalence class $(n_1,n_2)_N$.
The basis states of the subspace are then defined as
\begin{equation}
\ket{(n_1,n_2)_N} = \frac{1}{\sqrt{\#(n_1,n_2)_N}} \sum_{\phi \in(n_1,n_2)}\ket{\phi}\text{,}
\end{equation}
where $\# (n_1,n_2)_N$ is the size of the corresponding class.
In order to compute the matrix elements of $H$, we now derive an analytical expression for the class size using a recurrence relation for periodic boundary conditions.

We start with the simple case where $n_1=0$, meaning that there are no excitations on the first sublattice. 
Then the Rydberg constraints are irrelevant and the class size corresponds to the number of ways to place $n_2$ identical excitations in $N/2$ sites.
This leads to
\begin{equation} \label{eq:CS_0}
  \#(0,n_2)_N=\frac{(N/2)!}{n_2! (N/2-n_2)!}=\binom{N/2}{n_2}\text{,}
\end{equation}
which is simply a binomial coefficient.

For $n_1\neq 0$, we want to obtain a relation between $\#(n_1,n_2)_N$ and $\#(n_1-1,n_2)_{N-2}$. 
This can be done by counting the number of product states in the class starting with the pattern $\ket{{\bullet}{\circ}....}$ in two different ways. 
This quantity is denoted by $D_{{\bullet}{\circ}} (N,n_1,n_2)$, and can be computed using the operator $S^i_{\bullet \circ}$ , which is a projector on all states having $\bullet \circ$ on sites $i$ and $i+1$. 
So for any product state $\ket{\psi}$ in the full Hilbert space,
\begin{equation}
S^i_{{\bullet}{\circ}}\ket{\psi}=\begin{cases}
\ket{\psi} & \text{if } \ket{\psi}=\ket{\underbrace{ \ldots}_\text{$i-1$}{\bullet}{\circ}\hspace{-0.27cm}\underbrace{ \ldots}_\text{$N-i-2$}\hspace{-0.27cm}}\\
0              & \text{otherwise}.
\end{cases}
\end{equation}
This allows us to compute
\begin{equation}\label{eq:D10_1}
\begin{aligned}
&D_{{\bullet}{\circ}}(N,n_1,n_2)=\hspace{-0.3cm}\sum_{\substack{\ket{\psi} \in \\ (n_1,n_2)_N}}\hspace{-0.3cm}\braket{\psi|S^{1}_{\bullet \circ}|\psi}=\\
&\frac{2}{N}\sum_{j=0}^{N/2-1}\hspace{-0.3cm}\sum_{\substack{\ket{\psi} \in \\ (n_1,n_2)_N}}\hspace{-0.3cm}\braket{\psi|S^{2j+1}_{{\bullet}{\circ}}|\psi}
=\frac{2n_1}{N}\#(n_1,n_2)_N.
\end{aligned}
\end{equation}
Here we used the fact that the sum over the class elements is invariant under translation of 2 sites. 
The two sums can then be swapped and the sum over $j$ simply counts the number of excitations on the first sublattice, and thus will give $n_1$ for every state in this class.

The second way of counting $D_{\bullet \circ}(N,n_1,n_2)$ relies on separating the first two sites from the rest of the chain.
Any state $\ket{\psi}\in (n_1,n_2)_N$ with the pattern $\ket{{\bullet}{\circ}\ldots}$ can be rewritten as $\ket{{\bullet}{\circ}}\otimes \ket{\phi}$, where $\ket{\phi}\in (n_1-1,n_2)_{N-2}$ with its last site not excited.
The fraction of states in $(n_1-1,n_2)_{N-2}$ satisfying this condition is given by $1-\frac{n_2}{N/2-1}=\frac{N/2-1-n_2}{N/2-1}$.
Accounting for the size of this class leads to 
\begin{equation} \label{eq:D10_2}
D_{\bullet \circ}(N,n_1,n_2)=\frac{N/2-1-n_2}{N/2-1}\#(n_1-1,n_2)_{N-2}.
\end{equation}
Putting together equations \Cref{eq:D10_1} and \Cref{eq:D10_2} gives the recurrence relation
\begin{equation}
\#(n_1,n_2)_N=\frac{N}{2n_1}\frac{N/2{-}1{-}n_2}{N/2{-}1}\#(n_1{-}1,n_2)_{N{-}2}.
\end{equation}
Using \Cref{eq:CS_0} as the initial condition, it is straightforward to use this recurrence to obtain the general class size
\begin{align} \label{eq:CS}
\#(n_1,n_2)_N{=} \hspace{-0.03cm}\frac{\frac{N}{2}\!\left(\frac{N}{2}{-}n_1{-}n_2\right)}{(\frac{N}{2}{-}n_1)(\frac{N}{2}{-}n_2)}
\binom{\frac{N}{2}{-}n_1}{n_2}\!\binom{\frac{N}{2}{-}n_2}{n_1}\text{.}
\end{align}

\subsection{Matrix elements}

As the  PXP Hamiltonian removes or adds a single excitation, $\ket{n_1,n_2}$ only has non-zero matrix elements with $\ket{n_1\pm 1,n_2}$ and $\ket{n_1,n_2\pm 1}$.
Using the Hermiticity of $H$ and the exchange symmetry $n_1 \rightleftharpoons n_2$, only one of these elements has to be computed.
For simplicity, we choose
\begin{equation}
\begin{aligned}
&\Big(H\Big)_{(n_1,n_2)}^{(n_1{-}1,n_2)}=\braket{(n_1{-}1,n_2)_N|H|(n_1,n_2)_N} \\
&=\hspace{-0.3cm}\sum\limits_{\substack{\ket{\psi} \in \\ (n_1,n_2)_N}}\hspace{-0.0cm}\sum\limits_{\substack{\ket{\phi} \in \\ (n_1{-}1,n_2)_N}}\hspace{-0.4cm}\frac{\braket{\phi|H|\psi}}{\sqrt{\#(n_1,n_2)_N \#(n_1{-}1,n_2)_N}}.
\end{aligned}
\end{equation}
For each element of $(n_1,n_2)_N$, there are $n_1$ excitations that can be removed to give a product state in $(n_1-1,n_2)_N$.
Because of the orthonormality of the product states, each of these moves simply gives a contribution of 1.
The double sum ends up giving $n_1 \cdot \# (n_1,n_2)_N$ and the matrix element is
\begin{equation} \label{eq:mat_elem_general}
\Big(H\Big)_{(n_1,n_2)}^{(n_1{-}1,n_2)}=n_1\sqrt{\frac{\#(n_1,n_2)_N}{\#(n_1-1,n_2)_N}}.
\end{equation}
This can be rewritten using \Cref{eq:CS} as
\begin{multline}
\hspace{-0.35cm}\Big(H\Big)_{(n_1,n_2)}^{(n_1{-}1,n_2)} \hspace{-0.15cm}=\sqrt{\frac{n_1(\frac{N}{2}{-}n_1{-}n_2)(\frac{N}{2}+1{-}n_1{-}n_2)}{\frac{N}{2}{-}n_1}}\text{.}
\end{multline}
The case where $n_1=\frac{N}{2}$ needs to be treated separately.
Because of the kinetic constraint, a sublattice can only be fully occupied if the other one is empty, hence $n_2=0$.
\Cref{eq:mat_elem_general} can then be expanded using \Cref{eq:CS_0} instead of \Cref{eq:CS} to give 
\begin{equation}
\Big(H\Big)_{(n_1,0)}^{(n_1{-}1,0)}=\sqrt{n_1\left(\frac{N}{2}+1{-}n_1\right)},
\end{equation}
which is well behaved for all values of $n_1$.

\section{Subspace energy variance}\label{sec:subs_var}

The subspace energy variance is defined as
\begin{equation}
\sigma_H^2 = \hat{\mathcal{K}}(H^2) - \left(\hat{\mathcal{K}}(H)\right)^2\text{,}
\end{equation}
where $\hat{\mathcal{K}}$ is a super-operator which projects linear operators, defined on the full Hilbert space, down to the symmetric subspace $\mathcal{K}$.
The matrix elements of $(\hat{\mathcal{K}}(H))^2$ can easily be computed from the results of Sec.~\ref{sec:mat_elems}.
To obtain the elements of $\hat{\mathcal{K}}(H^2)$, it is helpful to decompose $H$ as
\begin{equation} \label{eq:H_pm}
H=H_1^{+}+H_2^{+}+H_1^{-}+H_2^{-},
\end{equation}
where $H_{i}^{\pm}$ adds or removes an excitation on the $i$-th sublattice.
This turns $H^2$ into a product of 16 terms, and we will show that only 4 of them are relevant.
To lighten the notation we define the result for each term as
\begin{equation}
A_{1,2}^{+,-}(n_1,n_2)\coloneqq \Big[ \hat{\mathcal{K}}\left( H_1^{+}H_2^{-} \right) \Big] _{(n_1,n_2)}^{(n_1+1,n_2-1)} .
\end{equation}
In a similar fashion, the terms of $\left(\hat{\mathcal{K}}(H)\right)^2$ are labelled as
\begin{equation}
B_{1,2}^{+,-}(n_1,n_2) \coloneqq  \Big[ \hat{\mathcal{K}}\left( H_1^{+} \right) \hat{\mathcal{K}}\left( H_2^{-} \right) \Big] _{(n_1,n_2)}^{(n_1+1,n_2-1)}.
\end{equation}
As these elements are all computed for a system of fixed size $N$, the subscript referring to it is dropped.

An important simplification comes from the fact that
\begin{equation} \label{eq:H_plus}
H_1^{+}\ket{(n_1,n_2)}=(n_1+1)\sqrt{\frac{\# (n_1+1,n_2)}{\# (n_1,n_2)}}\ket{(n_1+1,n_2)},
\end{equation}
which implies that $H_i^{+}$ does not leak out of the \subname subspace.
This is not the case for $H_1^{-}$, and if it were it would mean that the subspace energy variance is always zero.
The difference between these two operations comes from the Rydberg constraint.
Indeed, consider a state $\ket{\psi} \in (n_1+1,n_2)$ and remove an excitation.
As stated in Sec.~\ref{sec:mat_elems}, this gives $n_1+1$ distinct elements in $(n_1,n_2)$, one per excitation that can be removed.
Let us denote the set of these states by $G_{\psi} \subset (n_1,n_2)$. Formally this means that $G_{\psi}\coloneqq \{\ket{\phi}\in (n_1,n_2) | \bra{\phi}H_1^{-}\ket{\psi}\neq 0 \}$.
When applying $H_1^{+}$ to $\ket{(n_1,n_2)}$, the $n_1+1$ states in $G_{\psi}$ will contribute to $\ket{\psi}$, and the ones in $(n_1,n_2) \setminus G_{\psi}$ will not.
So $H_1^{+}\ket{(n_1,n_2)}$ gives $n_1+1$ contributions to $\ket{\psi}$.
As this is valid for any state $\ket{\psi}$ in $(n_1+1,n_2)$, it means that applying $H_1^{+}$ to $\ket{(n_1,n_2)}$ will give a symmetric superposition of states in $(n_1+1,n_2)$. 
Taking into account the normalisation factors gives equation \Cref{eq:H_plus}. 

For $H_1^{-}$ the same procedure is not possible.
Indeed, the number of excitations that can be added to a state $\ket{\psi} \in  (n_1,n_2)$ is not uniquely defined.
It is actually very easy to find an example of two states in the same class to which we can add a different number of excitations.
Consider the states $\ket{{\bullet}{\circ}{\bullet}{\circ}{\circ}{\bullet}{\circ}{\bullet}{\circ}{\circ}{\circ}{\circ}}$ and $\ket{{\bullet}{\circ}{\circ}{\bullet}{\circ}{\circ}{\bullet}{\circ}{\circ}{\bullet}{\circ}{\circ}}$ that are both in $(2,2)_{12}$.
We can add an excitation on any sublattice to the former but not the latter.

The main consequence of \Cref{eq:H_plus} is that the right actions of $H_i^{+}$ and $\hat{\mathcal{K}}(H_i^{+})$ are the same.
This also holds for the left action of $( H_i^{+})^{\dagger}=H_i^{-}$. 
So $H_1^{+}H_2^{-}$, $H_1^{+}H_1^{-}$ and their $1 \rightleftharpoons 2$ counterparts are the only elements that can differ between $\hat{\mathcal{K}}(H^2)$ and  $(\hat{\mathcal{K}}(H))^2$.

\subsection[Diagonal elements]{Diagonal elements : $H_1^{+}H_1^{-}$}

As mentioned previously, applying $H_i^{-}$ to $\ket{(n_1,n_2)}$ does not give a symmetric superposition.
In order to describe correctly the resulting state we need to introduce more detailed classes.

Let $(n_1,n_2,m_1,m_2)$ be the class of states with $n_i$  excitations on the $i$-th sublattice and $m_i$ excitable sites on it.
By ``excitable", we mean a site that is not excited, but with both neighbours also unexcited.
From this definition, it follows that $(n_1,n_2,m_1,m_2) \subset (n_1,n_2)$ and 
\begin{equation} \label{eq:diag_1}
\sum_{m_1,m_2}\#(n_1,n_2,m_1,m_2)=\#(n_1,n_2).
\end{equation}
Applying $H_1^{+}$ to any state in $(n_1,n_2,m_1,m_2)$ gives $m_1$ distinct states, one for each excitable site.
This allows one to compute the matrix element
\begin{equation}\label{eq:mat_elem_m}
\hspace{-0.1cm}\Big(H\Big)_{(n_1,n_2)}^{(n_1{+}1,n_2)}{=}\frac{\sum_{m_1,m_2}m_1 \#(n_1,n_2,m_1,m_2)}{\sqrt{\#(n_1,n_2)\#(n_1{+}1,n_2)}}.
\end{equation}
Because of the Hermiticity of $H$, \Cref{eq:mat_elem_general} can be used to derive the identity
\begin{equation}\label{eq:diag_2}
\sum_{m_1,m_2}\hspace{-0.2cm}m_1 \#(n_1,n_2,m_1,m_2)=(n_1{+}1)\#(n_1{+}1,n_2).
\end{equation}

Using these new classes and identities, it is now possible to compute the matrix element for $\hat{\mathcal{K}}(H_1^{+}H_1^{-})$ acting on $\ket{(n_1,n_2)}$.
First, the number of states in the class needs to be expressed as $\#(n_1,n_2)=\sum_{m_1,m_2} \#(n_1,n_2,m_1,m_2)$. 
Acting with $H_1^{-}$ simply gives a factor of $n_1$ for all states, leaving us with $n_1\sum_{m_1,m_2}\#(n_1,n_2,m_1,m_2)$ elements. 
It also creates an additional excitable site, which means that acting with $H_1^{+}$ leads to
\begin{equation} 
\begin{aligned}
&n_1\sum_{m_1,m_2} (m_1+1)\#(n_1,n_2,m_1,m_2) \\ 
&=n_1(n_1+1)\#(n_1+1,n_2)+n_1\# (n_1,n_2)
\end{aligned}
\end{equation}
contributions, where identities \Cref{eq:diag_1} and \Cref{eq:diag_2} were used to resolve the sum. 
With the normalisation this amounts to
\begin{equation} \label{eq:diag_A}
\begin{aligned} 
&A_{1,1}^{+,-}(n_1,n_2)=n_1(n_1{+}1)\frac{\#(n_1{+}1,n_2)}{\# (n_1,n_2)}{+}n_1\\
&=n_1\frac{(\frac{N}{2}-n_1-n_2-1)(\frac{N}{2}-n_1-n_2)}{\frac{N}{2}-n_1-1}+n_1.
\end{aligned}
\end{equation}
The equivalent quantity for $\left( \hat{\mathcal{K}}(H) \right)^2$ is
\begin{equation} \label{eq:diag_B}
\begin{aligned}
&B_{1,1}^{+,-}(n_1,n_2)=n_1^2\frac{\#(n_1,n_2)}{\# (n_1-1,n_2)} \\
&=n_1\frac{(\frac{N}{2}-n_1-n_2)(\frac{N}{2}-n_1-n_2+1)}{\frac{N}{2}-n_1}.
\end{aligned}
\end{equation}

\subsection[Off-diagonal elements]{Off-diagonal elements : $H_1^{+}H_2^{-}$}

This case is trickier because it involves the mixing of $m_1$ and $m_2$.
To simplify the computation, it is better to focus on the unnormalised matrix element $\bar{A}_{1,2}^{+,-}(n_1,n_2+1)$ such that 
\begin{equation} 
\hspace{-0.1cm}\bar{A}_{1,2}^{+,-}\hspace{-0.05cm}(n_1{,}n_2{+}1){=}\mathcal{N}\hspace{-0.1cm}
\braket{(n_1{+}1,n_2)|H_1^{+}H_2^{-}|(n_1,n_2{+}1)}\hspace{-0.1cm}\text{,}\quad
\end{equation}
where $\mathcal{N}=\sqrt{\#(n_1{+}1,n_2)\#(n_1,n_2{+}1)}$.
The rightmost state is $\ket{(n_1,n_2{+}1)}$ because the focus of the computation is on the intermediate state $\ket{(n_1,n_2)}$.
When acting with $H_2^{-}$ on $\ket{n_1,n_2{+}1}$, each state in $(n_1,n_2,m_1,m_2)$ gains a factor of $m_2$ because of the number of states contributing.
The same is true for the left half, acting with $H_1^{-}$ on $\ket{n_1{+}1,n_2}$, gives a factor of $m_1$.
Inserting the identity in the middle as $\frac{\ket{(n_1,n_2)}\bra{(n_1,n_2)}}{\# (n_1,n_2)}$
gives 
\begin{equation} \label{eq:cov_1}
\begin{aligned}
\bar{A}_{1,2}^{+,-}(n_1,n_2{+}1)=\hspace{-0.2cm}\sum_{m_1,m_2}\hspace{-0.2cm}\frac{ m_1 m_2 \#(n_1,n_2,m_1,m_2)}{\# (n_1,n_2)} \\
=\frac{\sum_{m_1,m_2} m_1 m_2 \#(n_1,n_2,m_1,m_2)}{\sum_{m_1,m_2} \#(n_1,n_2,m_1,m_2)}.
\end{aligned}
\end{equation}
$m_1$ and $m_2$ can be viewed as random variables with a probability distribution proportional to $\#(n_1,n_2,m_1,m_2)$, then \Cref{eq:cov_1} is simply the expectation value $E[m_1m_2]$.
The equivalent element for $(\hat{\mathcal{K}}(H))^2$ can itself be understood as a product of the expectation values 
\begin{align} \label{eq:cov_2}
\hspace{-0.2cm}E[m_1]E[m_2]&{=} \frac{(n_1{+}1)\#(n_1{+}1,n_2)}{\#(n_1,n_2)}\frac{(n_2{+}1)\#(n_1,n_2{+}1)}{\#(n_1,n_2)} \nonumber \\
&{=} \frac{(\M{-}n_1{-}n_2-1)^2(\M{-}n_1{-}n_2)^2}{(\M{-}n_1-1)(\M{-}n_2-1)},
\end{align}
where the relation between $E[m_i]$ and the matrix elements of $\hat{\mathcal{K}}(H)$ is derived from \Crefrange{eq:mat_elem_m}{eq:diag_2}.

The subspace energy variance for $H_1^{+}H_2^{-}$ is then given by the difference between equations \Cref{eq:cov_1} and \Cref{eq:cov_2}, which is equal to the covariance of $m_1$ and $m_2$ 
\begin{equation}
  Cov[m_1,m_2] = E[m_1m_2] - E[m_1]E[m_2].
\end{equation}
Up to a multiplicative factor, this quantity is equal to
\begin{align}
&\smashoperator{\sum_{m_1,m_2}}\bigg\{m_1m_2\left(\M{-}n_1{-}1\right)\left(\M{-}n_2{-}1\right)- \label{eq:offdig_prepde}\\
&\left(\M{-}n_1{-}n_2\right)^{\hspace{-0.1cm}2}\!\left(\M{-}n_1{-}n_2{-}1\right)^{\hspace{-0.1cm}2} \bigg\}\,\#(n_1,n_2,m_1,m_2)_N\text{.} \nonumber
\end{align}
We will now make the assumption that $\M{-}n_1{-}1 \neq 0$ and $\M{-}n_2{-}1 \neq 0$ and demonstrate that this expression is identically zero.
The cases $n_i=\frac{N}{2}-1$ will be handled separately later.

For the purposes of this proof we will introduce an ordinary generating function for the elaborated class sizes,
\begin{align}
  [z^{N/2} x_1^{n_1}x_2^{n_2}y_1^{m_1}y_2^{m_2}] G
  &= \#(n_1,n_2,m_1,m_2)_N\text{.}
\end{align}
From \Cref{eq:offdig_prepde} we can derive a differential equation for this generating function,
\begin{equation}
 \begin{multlined}[b]
   \bigg(\partial_{y_1}\partial_{y_2}(w\partial_w{+}
   x_1\partial_{x_1}{-}1)(w\partial_w{+}x_2\partial_{x_2}{-}1)
   \\+ (w\partial_w)^2(w\partial_w-1)^2\bigg) G  \Bigg|_{\substack{y_1=1\\y_2=1}}
   \end{multlined}  = 0
 \label{eq:offdiag_pde}
\end{equation}
where we have eliminated $z$ by a change of variable in favour of $w=z/(x_1x_2)$.
This equation is satisfied if and only if \Cref{eq:offdig_prepde} is equal to 0.

\begin{figure}[htb]
	  \includegraphics[width=0.8\linewidth]{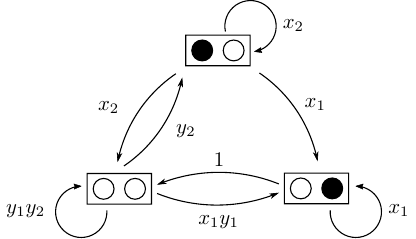}
  \caption{
    Finite state automaton diagram for the elaborated generating function.
    From this diagram we can read off the matrix elements of \Cref{eq:supp_process}.
  }
  \label{fig:offdiag_automaton}
\end{figure}
We will now derive a rational expression for $G$.
Take a look at \Cref{fig:offdiag_automaton}.
Each state of this automaton is the state of the two sites on the (left) boundary of a partial configuration.
Each arrow is an allowed gluing process.
It takes the motif on the target of the arrow and glues it onto the left on the partial configuration producing a configuration with two more sites.
The label on the arrows is a monomial factor which is picked up when following that gluing process.
The $x_1$ and $x_2$ indeterminates mark when an excitation is added to the respective sublattices, whereas $y_1$ and $y_2$ are placed when we realise that we could have placed an excitation.
The sum of all the execution paths of this automaton generates all the allowed configuration and marks them with the appropriate powers of the indeterminates.
A process matrix for this automaton is the following,
\begin{equation}
  M = \left(\begin{array}{ccc}
    y_1 y_2 & y_2 & 1 \\
    x_2     & x_2 & 0 \\
    x_1 y_1 & x_1 & x_1
  \end{array}\right)\text{,}
  \label{eq:supp_process}
\end{equation}
with which we can express the generating function as a Neumann series.
The series clearly is convergent in some neighbourhood of $z=x_1=x_2=0$ and $y_1=y_2=1$ and so we may express it as
\begin{equation}
\begin{aligned}
  &G(x_1,x_2,y_1,y_2;z)
  = \sum_{N/2} z^{N/2} \Tr\{M^{N/2}\} \\
  &= \Tr\{\left(\mathds{1}-z M\right)^{-1}\} 
  = \frac{3 - 2f_1 z - f_2 z^2}{1 - f_1 z - f_2 z^2 - f_3 z^3}\text{,}
\end{aligned}
\end{equation}
where
\begin{align}
  f_1 &= y_1y_2 + x_1 + x_2 \\
  f_2 &= x_1y_1(1{-}y_2) + x_2y_2(1{-}y_1) \\
  f_3 &= (1{-}y_1)(1{-}y_2)\text{.}
\end{align}
This differs by an additive constant from the definition given previously, but this detail is unimportant.
It is a lengthy albeit straightforward calculation to verify that this indeed satisfies the differential equation \Cref{eq:offdiag_pde} as required.

Now we return to the deferred cases of when either $\M-n_1-1=0$ or $\M-n_2-1=0$.
The first term $E[m_1m_2]$ vanishes because in either of these cases, for one of the sublattices, it is not possible to insert an excitation, hence $m_1m_2=0$.
For the second term $E[m_1]E[m_2]$ the same reasoning implies that one of $\#(n_1+1,n_2)$ and $\#(n_1,n_2+1)$ must vanish.
Thus the proposition is shown inclusive of the previously excluded cases.

\subsection{Total subspace energy variance}

The only non-zero matrix elements of the subspace energy variance are the diagonal ones given by \Crefrange{eq:diag_A}{eq:diag_B}.
Taking their difference leads to
\begin{equation}\label{eq:subs_var_diag}
\sigma_H^2 = \frac{n_1n_2(n_2-1)}{\left(\frac{N}{2}-n_1-1\right)\left(\frac{N}{2}-n_1\right)} +\left(1\leftrightarrow 2\right)
\end{equation}
for $\ket{(n_1,n_2)}$.
The total subspace variance can also be obtained by tracing over all basis states
\begin{equation}
\Tr\left[\sigma_H^2\right]=\frac{1}{9}(\tfrac{N}{2}-3)(\tfrac{N}{2}-2)(\tfrac{N}{2}-1).
\end{equation}

\section[Proof that M is a subset of K]{Proof that $\mathcal{M}\subseteq \mathcal{K}$ }
\label{sec:TDVP_1}

We will work in the general setting where there is a collection of $N$ sites, an assignment of them onto $d$ subsets and a projection operation which is diagonal in the computational basis.
The state space $\mathcal{M}$ consists of a coherent state on the highest spin representation formed by fusing all the spins on each subset which is then projected into the constrained Hilbert space.
The vector space $\mathcal{K}$ is the span of the symmetric combinations of the computational basis states with fixed occupation on each of the subsets.
This setting subsumes the cases of the coherent-state TDVP ansatz for the PXP model from Ref.~\onlinecite{Bernien:2017bp} developed further in Ref.~\onlinecite{Ho:2019gv}, and also the generalisations for $\mathbb{Z}_d$ initial states and tree-tensor networks \cite{Michailidis:2019wj}.
The states in $\mathcal{M}$ can be expressed using matrix product states with bond dimension $D=2$.
Similar to Ref.~\onlinecite{Ho:2019gv}, we use the parametrisation 
\begin{equation} \label{eq:TDVP_param}
A^{\bullet}(\theta_j,\phi_j){=}\hspace{-0.1cm}\begin{pmatrix}
0 & -ie^{i\phi_j}{\rm sin}(\frac{\theta_j}{2}) \\ 0 & 0
\end{pmatrix}\hspace{-0.1cm},\,  A^{\circ}(\theta_j,\phi_j){=}\hspace{-0.1cm}\begin{pmatrix}
{\rm cos}(\frac{\theta_j}{2}) & 0 \\ {\rm cos}(\frac{\theta_j}{2}) & 0
\end{pmatrix}\hspace{-0.1cm},
\end{equation}
where $A^{\bullet}$ is for an excited site and $A^{\circ}$ for an unexcited site.
The chain is split into $d$ sublattices, each with two parameters $\theta_j$ and $\phi_j$.
This produces $N/d$ identical unit cells, and the corresponding (unnormalised) quantum state is
\begin{equation} \label{eq:TDVP_prod}
\begin{aligned}
\sum_{\{\sigma \}} {\rm Tr}\big[ A^{\sigma_1}(\theta_1,\phi_1)A^{\sigma_2}(\theta_2,\phi_2)\ldots A^{\sigma_d}(\theta_d,\phi_d) \\ 
\cdot A^{\sigma_{N-d+1}}(\theta_1,\phi_1) \ldots A^{\sigma_N}(\theta_d,\phi_d) \big] \ket{\sigma_1,\ldots \sigma_N},
\end{aligned}
\end{equation}
where the trace enforces periodic boundary conditions.
This is equivalent to fixing sites $j,j+d,j+2d,\ldots$ to ${\rm cos}\left(\frac{\theta_j}{2}\right)\ket{\circ }-ie^{i\phi_j}{\rm sin}\left(\frac{\theta_j}{2}\right)\ket{\bullet}$, with the additional application of a projector that annihilates configurations with neighbouring excitations.

Up to this point, only the \subname subspace with two sublattices ($d=2$) has been considered.
However, it is easy to see how to extend our previous results to $d$ sublattices. The basis states $\ket{(n_1,...,n_d)_N}$ are then symmetric superpositions of product states with $n_i$ excitations on the $i$-th sublattice.
As in Sec.~\ref{sec:mat_elems}, recurrence relations for the class sizes can be found using combinatorics arguments. 
Once these are known, the matrix elements can be computed from them as it was done for $d=2$. 
We will now demonstrate that, for any value of $d$, all states in $\mathcal{M}$ lie in the corresponding $\mathcal{K}$ subspace.

In \Cref{eq:TDVP_prod}, for each configuration of ${\sigma}$'s there is a product of $N$ matrices of size $2\times 2$. 
For the moment, suppose that there is at least one excitation in the configuration.
Then this sequence can be segmented into blocks with the following pattern: the leftmost matrix corresponds to an excitation and all the others are vacancies.
Because of the cyclic property of the trace, we can also assume that the leftmost matrix in \Cref{eq:TDVP_prod} corresponds to an excitation and rewrite the quantity inside the trace as
\begin{equation} \label{eq:blocks}
\prod_{\text{Blocks}} \left[ A^{\bullet}(\theta_j,\phi_j)\prod_{l=1}^m A^{\circ}(\theta_{j+l},\phi_{j+l})\right],
\end{equation}
where the indices of the angles are periodic in $d$, meaning that $\theta_{d+j}=\theta_{j}$.
From \Cref{eq:TDVP_param}, it is easy to compute the product of two vacancy matrices
\begin{equation} \label{eq:mat_down}
A^{\circ}(\theta_k,\phi_k)A^{\circ}(\theta_j,\phi_j)={\rm cos}\left(\frac{\theta_j}{2}\right)
\begin{pmatrix}
{\rm cos}(\frac{\theta_k}{2}) & 0 \\ {\rm cos}(\frac{\theta_k}{2}) & 0
\end{pmatrix}.
\end{equation}
Therefore, in each block the whole trail of vacancy matrices on the right can be reduced to a single vacancy matrix and a product of cosines.
The product between an excitation matrix and a vacancy matrix can also be simplified as
\begin{equation}
A^{\bullet}(\theta_k,\phi_k)A^{\circ}(\theta_j,\phi_j)=\, -ie^{i\phi_k}{\rm sin}\left(\frac{\theta_k}{2}\right){\rm cos}\left(\frac{\theta_j}{2}\right)B,
\end{equation}
where $B\coloneqq \begin{pmatrix}
1 & 0 \\ 0 & 0
\end{pmatrix}$.
Finally, it is easy to see from \Cref{eq:TDVP_param} that the product of two excitation matrices gives 0.
This parametrisation guarantees that each block has at least one vacancy matrix on the right, and that \Cref{eq:blocks} can be written as
\begin{equation}
\prod_{\text{Blocks}} \left[
-B\, ie^{i\phi_j}{\rm sin}\left(\frac{\theta_j}{2}\right)\prod_{l=1}^m {\rm cos}\left(\frac{\theta_{j+l}}{2}\right)\right].
\end{equation}
This is a product of collinear (and hence commuting) matrices, and the effect of the trace on it is trivial.
The last step is to count the number of blocks with a leftmost matrix with index $j$. 
For any configuration, it is simply given by the number of excitations in the corresponding sublattice.
This means that each excitation in the sublattice $j$ gives a factor of $ie^{i\phi_j}\sin(\theta_j/2)$.
The other sites in this sublattice are free and give a factor of  $\cos(\theta_j/2)$ each.
This means that each configuration $n_1,\ldots,n_k$ has a prefactor of 
\begin{equation}
\begin{aligned}
\prod_{j=1}^d c_{\bullet,j}^{n_j} c_{\circ,j}^{N/d-n_{j}}
\ \text{with}\ \begin{array}{l}
c_{\bullet,j} = \, -ie^{i\phi_j}{\rm sin}\left(\frac{\theta_{j}}{2}\right),\\
c_{\circ,j} = {\rm cos}\left(\frac{\theta_{j}}{2}\right).
\end{array}
\end{aligned}
\end{equation}
As this term only depends on the sublattice on which the excitations are placed and not on the actual sites, it can be exactly expressed in the symmetric subspace.
Inserting this result into \Cref{eq:TDVP_prod} and accounting for the number of such configurations in the product basis means that in the \subname subspace the state $\ket{\{\theta_i, \phi_i\}_N}$ can be rewritten as
\begin{equation}\label{eq:TDVP_in_perm}
\hspace{-0.1cm}\sum_{n_1,\ldots,n_d}\hspace{-0.35cm}\sqrt{{\#} \left(n_1,\ldots,n_d\right)}
\left( \prod_{j=1}^d  c_{\bullet j}^{n_j}\;c_{\circ j}^{N/d{-}n_j}\hspace{-0.1cm}\right)\hspace{-0.1cm}\ket{(n_1,\ldots,n_d)_N}.
\end{equation}
Returning to the excluded case where there are no excitations, it is easy to see from \Cref{eq:mat_down} that it will give  
\begin{equation}
\prod_{i=j}^d  \left[  {\rm cos}\left(\frac{\theta_{j}}{2}\right)\right]^{N/d}=
\prod_{j=1}^d  c_{\circ j}^{N/d}.
\end{equation}
As $\sqrt{\# (0,\ldots,0)}=1$, it satisfies \Cref{eq:TDVP_in_perm}.
This concludes the proof that $\mathcal{M}\subseteq \mathcal{K}$, as \Cref{eq:TDVP_in_perm} expresses every point of $\mathcal{M}$ as a linear combination of vectors in $\mathcal{K}$.

\section[Proof that K = span M]{Proof that $\mathcal{K} = \Span \mathcal{M}$}
\label{sec:supp_span}

\Cref{sec:TDVP_1} established that $\mathcal{M} \subseteq \mathcal{K}$. 
We now claim that the relation between these two spaces is stronger, and that $\mathcal{K} = \Span \mathcal{M}$.
For simplicity, in this section we work with the complex parametrisation of $\mathcal{M}$,
\begin{equation}
\ket{\psi_N(z)} = \hspace{-0.1cm}\smashoperator[l]{\sum_{n_1,\ldots,n_d}} \hspace{-0.3cm}\sqrt{\#(n_1,\ldots,n_d)_N}\prod_{a=1}^{d} z_a^{n_a}  \ket{(n_1,\ldots,n_d)_N}\text{.} \\
\label{eq:param_complex}
\end{equation}
This parametrisation, like the coherent-state parametrisation of \Cref{sec:TDVP_1}, may fail to be injective.
This happens for the PXP model when $\theta_i = \pm \theta_{i+1}= \pm \pi/2$.
This is a defect with the space $\mathcal{M}$, where $\mathcal{M}$ fails to be locally Euclidean, rather than with the parametrisations.

We now consider two ordered bases $B$ and $B'$, where $\Span B = \mathcal{K}$ and $\Span B' \subseteq \Span M$.
The first basis $B$ is defined as
\begin{equation}
B_{n_1,\ldots,n_d} = \sqrt{\# (n_1\ldots,n_d)_N} \ket{(n_1,\ldots,n_d)_N}\text{,}
\end{equation}
where the multi-index $n_1,\ldots,n_d$ is taken to include values such as $N/2,\ldots,N/2$ for which the basis vector is a zero-vector.
This makes $B$ an over-complete and linearly dependent basis for $\mathcal{K}$; this is done to provide the multi-index with a natural Cartesian product structure.

In order to define the second basis $B'$ we will first need to choose a set of distinct coordinate values $z_{a,(1)}, \ldots, z_{a,(N/2+1)}$, for each dimension $a=1,\ldots,d$.
We define the basis $B'$ using the complex parametrisation as
\begin{equation}
B'_{i_1,\ldots,i_d} = \ket{\psi(z_{1,(i_1)},\ldots,z_{d,(i_d)})}\text{.}
\end{equation}
We can interpret this basis as coordinate-aligned grid of points, where we have sampled the $\mathcal{M}$-valued polynomial $\ket{\psi(z_1,\ldots,z_d)}$ to provide our basis vectors.

Disregarding possible linear dependence in $B$ and $B'$, our goal of showing that $\Span B' = \Span B$ is equivalent to how the coefficients of a polynomial may be determined by evaluating at a sufficient number of distinct points.
This motivates introducing for each $a=1,\ldots,d$ a Vandermonde matrix
\begin{equation}
Q_a =
\left(\begin{array}{ccccc}
1 & z_{a,(1)} & z_{a,(1)}^2 & \cdots & z_{a,(1)}^{|a|} \\
1 & z_{a,(2)} & z_{a,(2)}^2 & \cdots & z_{a,(2)}^{|a|)} \\
\vdots & \vdots & \vdots &  & \vdots \\
1 & z_{a,(|a|+1)} & z_{a,(|a|+1)}^2 & \cdots & z_{a,(|a|+1)}^{|a|}
\end{array}\right)\text{,}
\end{equation}
where $|a|$ is the number of sites in sublattice $|a|$.
Since the $z_{a,(1)},\ldots,z_{a,(N/2+1)}$ are all distinct each $Q_a$ is invertible.

These matrices can be used to relate the two ordered bases using $Q = \bigotimes_{a=1}^{d} Q_a$ providing two equations,
\begin{align}
B' &= Q B &\text{and}&& B &= Q^{-1} B \text{.}
\end{align}
These can be read to mean that each basis vector in $B'$ is a linear combination of those in $B$, and vice-versa, i.e $\Span B = \Span B'$.
Recalling that $B$ is a complete basis for $\mathcal{K}$ and that every vector in $B'$ is contained in $\mathcal{M}$, we find the containment, $\Span \mathcal{M} \supseteq \Span B' = \Span B = \mathcal{K}$, which completes the proof.

\section{Bipartite entropy in the \subname subspace}
\label{sec:supp_entropy}

Computing the von Neumann bipartite entropy in the \subname subspace is not straightforward due to the non-local definition of the basis states.
Consider a spin chain of length $N$ with periodic boundary conditions and cut it into two subsystems of size $N_L$ and $N_R$, respectively.
Let quantities related to the left part be denoted by the subscript $L$ and the ones related to the right part by the subscript $R$.
The simplest way to compute the bipartite entropy for this cut for a basis state $\ket{(n_1,n_2)_N}$ would be to express it in the full Hilbert space and proceed from there.
However this is suboptimal as it would strongly limit our ability to study large systems. 
Instead, the structure of the subspace can be exploited in order to write the states in the right and left subsystems in their respective \subname subspaces. 
To simplify the computations, we only consider cases where the right and left subsystems are of even size. 
If nothing is specified, we consider the cut into two subsystems of equal size, meaning that the full system has size $N=4M, \ M\in \mathbb{N}$, and the two subsystems have size $N_L=N_R=2M$.

While the full system has periodic boundary conditions, the subsystems must have open boundary conditions.
Consider the cut
 $\ket{{\bullet}{\circ}{\circ}{\bullet}{\circ}{\bullet}{\circ}{\circ}}=\ket{{\bullet}{\circ}{\circ}{\bullet}}\otimes\ket{{\circ}{\bullet}{\circ}{\circ}}$, it is clear that the left part violates the Rydberg constraint if periodic boundary conditions are assumed.
Conversely, the two subsystems $\ket{{\circ}{\circ}{\circ}{\bullet}}$ and $\ket{{\bullet}{\circ}{\circ}{\circ}}$ respect the Rydberg constraints individually but their tensor product does not.
This simple case shows that we must further subdivide the classes in the left and right subsystems based on the value of the extremal sites.

Let us denote by $(n_1,n_2,a,b)_N, \ a,b\in{0,1}$ the class of states in $(n_1,n_2)_N$ with the additional constraint that the leftmost site is defined by $a$ and the rightmost site is by $b$.
The convention used is that the leftmost site is unexcited if $a=0$ and excited if $a=1$, and the same hold for $b$ and the rightmost site.  
Then $\ket{(n_1,n_2,a,b)_N}$ is the normalised symmetric superposition of all product sates in $(n_1,n_2,a,b)_N$.
Using these states, the relevant Schmidt decomposition for $\ket{(n_1,n_2)_N}$ is
\begin{multline} \label{eq:Schmidt_2}
\hspace{-0.35cm}\sum_{\substack{l_1,l_2 \\ r_1,r_2}}\sum_{\substack{a_L,b_L \\ a_R,b_R}}
\hspace{-0.2cm}\alpha
\ket{(l_1,l_2,a_L,b_L)_{N_L}}{\otimes} \ket{(r_1,r_2,a_R,b_R)_{N_R}},
\end{multline}
where the sums on $l_i$ and $r_i$ run from 0 to $\frac{N}{2}$, the sums on all $a$ and $b$ run from 0 to 1, and  $\alpha$ depends on $n_1,n_2$ and on all the summation indices.
This decomposition includes terms that are forbidden by Rydberg constraints such as $\ket{(\frac{N_L}{2},\frac{N_L}{2},a_L,b_L)_{N_L}}$.
However we will see later that the coefficient $\alpha$ corresponding to these will naturally be zero. 

To compute $\alpha$, selection rules can be stated.
The most obvious one is the conservation of the number of excitations on each sublattice, which tells us that $l_i+r_i=n_i$.
More formally, it means that $\alpha$ needs to include a term
\begin{equation}\label{eq:beta}
\beta^{n_1,n_2}_{l_1,l_2,r_1,r_2}=\delta_{l_1+r_1,n_1}\delta_{l_2+r_2,n_2},
\end{equation}
where $\delta_{i,j}$ is the Kronecker delta.
The selection rule on the $a$ and $b$ is also simple: $a_L$ and $b_R$ cannot both be 1 and neither can $b_L$ and $a_R$ as this would correspond to neighbouring excitations. 
This can be expressed as 
\begin{equation}\label{eq:eta}
\eta_{a_L,b_L}^{a_R,b_R}=\left(1-a_L b_R\right)\left(1-b_L a_R\right).
\end{equation}
The last step is to compute the contribution of each class $(n_1,n_2,a,b)_N$. 
Each product state in $(n_1,n_2)_N$ corresponds to a single state in $(l_1,l_2,a_L,b_L)_{N_L}\otimes (r_1,r_2,a_R,b_R)_{N_R}$.
So the only multiplicative factors that enter the equation are the normalisation ones.
For each combination of the $l_i,r_i,a,b$ the contribution to $\alpha$ is simply given by
\begin{equation} \label{eq:S_norm}
\frac{\sqrt{\#(l_1,l_2,a_L,b_L)_{N_L}} \sqrt{\#(r_1,r_2,a_R,b_R)_{N_R}}}{\sqrt{\#(n_1,n_2)_N}}.
\end{equation}
These class sizes can be computed using the same arguments as in section \ref{sec:mat_elems}, keeping in mind that the two subsystems have open boundary conditions.
To separate the contributions of the left and right parts we define
\begin{equation} \label{eq:gamma}
\gamma^N_{n_1,n_2,a,b}=\sqrt{\#(n_1,n_2,a,b)_N}.
\end{equation}
This coefficient ensures that all states in the Schmidt decomposition respect the Rydberg constraint.
Indeed, if a state is incompatible with it, its class size must be 0 and so is the corresponding $\gamma$.

Using \Crefrange{eq:beta}{eq:gamma}, the coefficients $\alpha$ from \Cref{eq:Schmidt_2} can be written as
\begin{equation} \label{eq:Schmidt_coeff}
\alpha=\frac{\beta^{n_1,n_2}_{l_1,l_2,r_1,r_2}\eta_{a_L,b_L}^{a_R,b_R}\gamma^{N_L}_{l_1,l_2,a_L,b_L}\gamma^{N_R}_{r_1,r_2,a_R,b_R}}{\sqrt{\#(n_1,n_2)_N}}.
\end{equation}
From this Schmidt decomposition it is straightforward to compute the entanglement spectrum and the von Neumann entropy.

\section[Revivals in large-N limit]{Revivals in large-$N$ limit}
\label{sec:supp_Ninfty}

The main feature of the PXP model is the revivals of the wavefunction that can be seen for some initial product states. 
This effect is strongest when starting from the N\'{e}el state $\ket{\mathbb{Z}_2}=\ket{{\bullet}{\circ}{\bullet}{\circ}\ldots}$ or the anti-N\'{e}el state $\ket{\bar{\mathbb{Z}}_2}=\ket{{\circ}{\bullet}{\circ}{\bullet}\ldots}$.
When starting in one of these states, after a time $T$ the wavefunction approximately comes back to itself. 
Furthermore, when starting in the state $\ket{\mathbb{Z}_2}$, after a half-period the wavefunction will be $\ket{\bar{\mathbb{Z}}_2}$, and vice-versa.
In the remainder of this section we will refer to this as state transfer.
However, these revivals (and state transfers) are not perfect and decay with time.
On the other hand, TDVP shows a periodic orbit going through both the N\'{e}el and anti-N\'{e}el states for an infinite system.
In this section we study the finite-size dependence of revivals in the \subname subspace.

\begin{figure}
  \includegraphics[width=\linewidth]{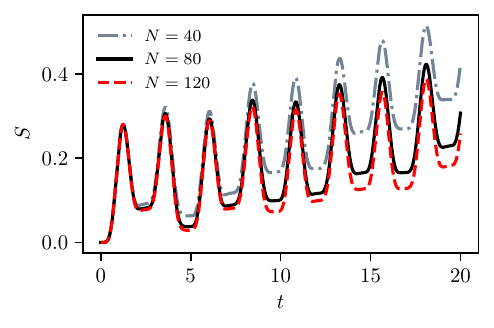}
  \caption{
    Bipartite von Neumann entropy of $e^{-iHt}\ket{\mathbf{Z}_2}$ in the \subname subspace. As $N$ increases, the entropy growth with time becomes slower.
  }
  \label{fig:S_scaling}
\end{figure}

\begin{figure}
  \includegraphics[width=\linewidth]{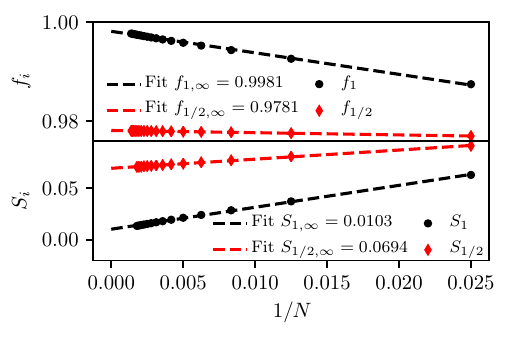}
  \caption{
    Wavefunction fidelity and bipartite entanglement entropy at the first state transfer ($f_{1/2},S_{1/2}$) and revival ($f_1,S_1$). The fits shown are of the form $y=a\cdot 1/N+b$. The data is for sizes up to $N=720$ for $f_{1/2},f_1$ and for sizes up to $N=560$ for $S_{1/2},S_1$.
  }
  \label{fig:f_S_scaling}
\end{figure}

To measure the quality of the revivals, the wavefunction fidelity and the entanglement entropy are used.
Starting from the N\'{e}el state $\ket{(\frac{N}{2},0)_N}{=}\ket{\mathbb{Z}_2}$, the Schr\"odinger time evolution is generated by $e^{-iHt}$.
The fidelity of revival is then 
\begin{equation}
f_{Rev}(t)=\abs{\braket{(N/2,0)_N|e^{-iHt}|(N/2,0)_N}}^2
\end{equation}
and the fidelity of state transfer is 
\begin{equation}
f_{Trans}(t)=\abs{\braket{(0,N/2)_N|e^{-iHt}|(N/2,0)_N}}^2.
\end{equation}
For the entanglement entropy we use the von Neumann bipartite entropy, as computed in Sec.~\ref{sec:supp_entropy}.
We also restrict our study to systems of size $N=4M, \ M\in \mathbb{N}$, and we cut them into two equally sized subsystems. 
From this we define the entropy as
\begin{equation}
S(t)= S\big(e^{-iHt}\ket{(N/2,0)_N}\big).
\end{equation}
As both the N\'{e}el and anti-N\'{e}el states are product states in the full Hilbert space, their entanglement entropy is 0.
From the exact PXP model, we expect the first state transfer to happen around $t=2.35$ and the first revival around $t=4.70$~\cite{Turner:2017wu}.
This can be seen in the entropy plot in \Cref{fig:S_scaling}.
From this plot it is clear that the revivals get enhanced as $N$ increases, but the quality of the first state transfer varies very little with system size.
To obtain  a more rigorous scaling, let us define the times of the first revival and state transfer, respectively, as
\begin{align}
t_1 &= \underset{3<t<6}{\rm arg\, max}\left( f_{ \rm Rev}(t) \right), \\
t_{1/2} &= \underset{1<t<4}{\rm arg\, max}\left( f_{\rm Trans}(t) \right).
\end{align}
This leads to their respective fidelity and entropy
\begin{align}
f_1&=f_{\rm Rev}(t_1) \\
f_{1/2}&=f_{\rm Trans}(t_{1/2}) \\
S_1&=S(t_1) \\
S_{1/2}&=S(t_{1/2})
\end{align}
The scaling of these quantities with $N$ can be seen in \Cref{fig:f_S_scaling}. The results for $f_1$ are consistent with the symmetric subspace having perfect revivals in the thermodynamic limit. However, the state transfer fidelity $f_{1/2}$ clearly does not converge towards 1.

To understand this behaviour we need to look at  different symmetry sectors as well as different bands of states.
With periodic boundary conditions, the relevant symmetry of the PXP model is translation. 
Because of the periodicity of the N\'{e}el state, it is only composed of eigenstates with momentum $k=0$ or $k=\pi$.
The scarred states of the top band belong alternatively to these two sectors, with the condition that the ground state always has $k=0$.
In the \subname subspace, the quasimodes show the same behaviour with respect to momentum sectors.

Because our method allows us to access larger systems, more can be said about the \qsname eigenstates. 
For example, scaling the system size shows that there is a constant fraction of them with an equal energy spacing around $E=0$.
As $N\rightarrow \infty$, the number of \qsname quasimodes with a significant support on $\ket{\mathbb{Z}_2}$ scales as $\mathcal{O}(\sqrt{N})$.
This means that, as we approach the thermodynamic limit, the only \qsname eigenstates that have an non-zero overlap with the N\'eel state are evenly spaced.

Equal spacing of the energies with a difference $\Delta E$ implies perfect revivals with a period $T=2\pi/\Delta E$.
Without loss of generality, let us assume that the energy of the $j$-th eigenstate is equal to $j \cdot \Delta E$.
After half a period, the phases of the even and odd eigenstates are 
\begin{align}
\phi_{2n}&={\rm exp}\left(-i\, \pi 2n E \right)=1, \\
\phi_{2n{+}1}&={\rm exp}\left(-i\, \pi (2n+1) E \right)=\, -1.
\end{align}
Because of the alternation of momentum, only the $k=\pi$ states get a $-1$ factor as they are the odd ones.
This exactly corresponds to the anti-N\'{e}el state as
\begin{equation}
\begin{aligned}
\ket{\bar{\mathbb{Z}}_2}=\hat T\ket{\mathbb{Z}_2}&=\sum_{\ket{E} \in k=0}\hspace{-0.3cm} \alpha^0_E \hat T\ket{E}+\hspace{-0.3cm}\sum_{\ket{E} \in k=\pi} \hspace{-0.3cm}\alpha^{\pi}_E \hat T\ket{E} \\
&=\sum_{\ket{E} \in k=0}\hspace{-0.3cm} \alpha^0_E \ket{E}-\hspace{-0.3cm}\sum_{\ket{E} \in k=\pi} \hspace{-0.3cm}\alpha^{\pi}_E \ket{E},
\end{aligned}
\end{equation}
where $\hat T$ is the translation operator.
This means that if only the top band were relevant there would be perfect revivals and perfect state transfer in $\mathcal{K}$.

\begin{figure}
  \includegraphics[width=\linewidth]{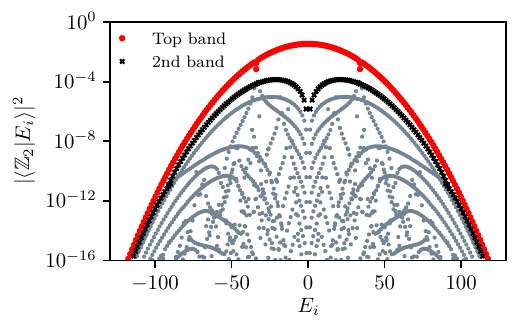}
  \caption{
   Overlap of the quasimodes with the N\'eel state for $N=440$. The top-band is shown in red, and the second band in black.
  }
  \label{fig:band_sep}
\end{figure}

\begin{figure}
  \includegraphics[width=\linewidth]{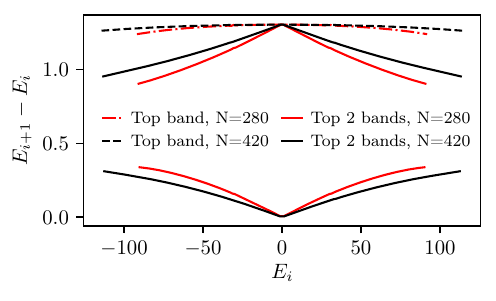}
  \caption{
    Energy difference between consecutive states in the top band and in the top two bands.
    Near $E=0$, the two bands converge in energy, as can be seen from $E_{i+1}-E_i$ alternating between 0 and $\Delta E_{\rm Top \ band}$.
  }
  \label{fig:band_E_diff}
\end{figure}

However, the overlap between $\ket{\mathbb{Z}_2}$ and the quasimodes not belonging to the top band is not negligible. In particular, we can explain the discrepancy between revivals and state transfers by looking at the second band.
By "second band", we refer to the band of states with the highest overlap with $\ket{\mathbb{Z}_2}$ after the \qsname states (see figure \ref{fig:band_sep}).
The quasimodes in this band have a much lower overlap with the N\'eel state than the \qsname ones, but they also display the alternation between the two momentum sectors.  
Near $E=0$, the two top bands converge towards the same energy values (\Cref{fig:band_E_diff}), but with the opposite momentum value.
Hence, in this region there are two eigenstates for each energy value, one with $k=0$ and the other with $k=1$.
Because of that, the previous argument for state transfer no longer holds.
However this has no effect on the energy spacing itself, so perfect revivals are still possible.

To summarise, in the symmetric subspace revivals get better as system size increases. While the finite size scaling is consistent with revival becoming perfect in the thermodynamic limit, it appears that the state transfer is never perfect because of the presence of the second band.  This is to be contrasted with TDVP where both revivals and state transfers are exact.

\section{Path integral dequantisation}
\label{sec:supp_dequantise}

If we seek to model a particular quantum dynamics with a classical analogue, we need a good understanding of the relationship between the two.
In the main text we stated that in certain situations we can view the TDVP classical system as the dequantisation of the original dynamics.
This viewpoint was put forward in Ref.~\onlinecite{Green:2016uw}, where it was applied to non-uniform matrix product states. In the following, we review and justify this procedure using our frame-theoretic language.
We will do this from the point of view of path-integral quantisation and with the proviso that from the state space we can form a good resolution of the identity.

The quantum path integral comes in a number of conventional representations: the position or momentum representations,  corresponding to the Schr\"odinger formulation, and coherent-state representation, corresponding to the phase space formulation.
In each case, the quantisation procedure involves an integral over all admissible paths, including those with non-stationary action, which assigns to each path a weight given by a measure function and a phase provided by the classical action.
Our approach is closely modelled on the way the path integrals are conventionally defined for various types of coherent states, see for example Ref.~\onlinecite{Gazeau:2009vb}.
Because our starting point is a quantum system, we will instead take the backwards view where the path integral is defined by a classical limit.
We are also provided with a collection of states $\ket{x}$ for which we assume that we can pick a measure $\mu$ to form a resolution of the identity,
\begin{equation}
  \mathds{1} = \int \deriv{\mu(x)} \ket{x}\bra{x}\text{.}
  \label{eq:good_resid}
\end{equation}
This allows us to represent the time evolution operator as an integral operator,
\begin{equation}
  K_t(x',x) = \bra{x'} e^{-i H t} \ket{x}\text{,}
\end{equation}
which satisfies the homomorphism condition,
\begin{equation}
  K_{a+b}(x'',x) = \int \deriv{\mu(x')} K_a(x'',x') K_b(x',x)\text{.}
\end{equation}
In this way, the finite time propagator can be broken into a product of infinitesimal generators,
\begin{align}
  K_\epsilon(x_{n{+}1},x_n)
  &= \brakett{x_{n{+}1}}{x_n}\left(1 -i\epsilon\frac{\bra{x_{n{+}1}}H\ket{x_n}}{\brakett{x_{n{+}1}}{x_n}} + o(\epsilon^2)\right) \nonumber \\
  &= \exp\left(-i \epsilon \check{H} + \log T \right) + o(\epsilon^2)\text{,}
\end{align}
where
\begin{align}
  \check{H}(x_{n+1},x_n)
  &= \frac{\bra{x_{n+1}}H\ket{x_n}}{\brakett{x_{n+1}}{x_n}}
\end{align}
and $T(x_{n+1},x_n)=\brakett{x_{n{+}1}}{x_n}$.
To deal with inner products across different times in a symmetric manner, half-steps can be introduced along the path,
\begin{align}
  T
  &=
  \big(\!\bra{x_{n{+}\frac{1}{2}}} + \frac{\epsilon}{2}\bra{\dot x_{n{+}\frac{1}{2}}}\big)
  \big(\ket{x_{n{+}\frac{1}{2}}} - \frac{\epsilon}{2}\ket{\dot x_{n{+}\frac{1}{2}}}\big) + o(\epsilon^2) \nonumber \\
  &=
  \brakett{x_{n{+}\frac{1}{2}}}{x_{n{+}\frac{1}{2}}} + \frac{\epsilon}{2}\big(\!\brakett{x_{n{+}\frac{1}{2}}}{\dot x_{n{+}\frac{1}{2}}} - \text{h.c.}\big) + o(\epsilon^2)\text{.} \nonumber
\end{align}
The first term here gives another contribution to the path integral measure, and the second produces the kinetic term in the resulting Lagrangian,
\begin{equation}
  \mathcal{L}(x,\dot x) = 
  \frac{i}{2} \frac{\brakett{x}{\dot x} - \brakett{\dot x}{x}}{\brakett{x}{x}} - \check{H} \text{.}
\end{equation}
Therefore, in this scenario where we can assume \Cref{eq:good_resid}, the TDVP action describes a valid classical limit and this path-integral provides a way to quantise it to recover the full quantum dynamics.
Many of the familiar manifolds of quantum states, such as spin coherent states and matrix product states come with a natural choice for this measure which is the Haar measure for a transitive group action.
However, for the constrained coherent states of $\mathcal{M}$ there is no obvious choice.
In the following Section, we derive a suitable measure from an analogue to the large-spin correspondence principle.

\section{Fluctuation bounds for the frame operator}
\label{sec:supp_delta}

In this section, we prove convergence between the transformed and untransformed classical systems in the thermodynamic limit, and shows how a measure can be constructed with which the untransformed system quantises naturally.
This involves establishing an asymptotic bound on the fluctuations in frame transformation, which is then used to reduce the frame transformation to a choice of measure.

As mentioned in the main text we start from a measure that does not properly resolve the identity. 
Our starting point is the gauged coherent state parametrisation for $\mathcal{M}$, Eq.~(\ref{eq:gauged_mps_param_main}), and  
we want to find a measure such that $S_\mu$ [see \Cref{eq:frametransformation}] is bounded.
The choice of measure is to an extent arbitrary, since its details will be transformed away later, but the challenge of ensuring that $S_\mu$ is bounded can be understood from the following heuristic argument.
Many of the properties of $S_{\mu}$ can be seen by considering its expectation value for a point $x$ of $\mathcal{M}$.
A contribution to this integral will be significant if it is not too distinguishable from $x$, and therefore the integral can be estimated by the volume of the ``fuzzy'' set indistinguishable from $x$.
For almost all $x$ and as $N$ increases, the fuzzy neighbourhood retracts around $x$ to a point, with a typical fuzzy size $O(1/\sqrt{N})$.
This phenomena also occurs on the manifold of spin-coherent states with increasing spin~\cite{Gazeau:2009vb}, where it is a manifestation of the large-spin correspondence principle.
We can imagine each state as a biased coin from which we observe $N$ flips.
The more flips we observe, the more accurate our estimate of its bias will be.
Consequently, we become better at distinguishing coins with different biases as $N$ increases.
A small part of the space doesn't behave like this -- for example if $\theta_1=\pi$ then $\theta_2$ is entirely indistinguishable regardless of $N$.
These regions can be suppressed by adopting the measure
\begin{equation}
  \deriv\mu = \left(\frac{\frac{N}{2}+1}{4\pi}\right)^2\prod_{i=1,2} \deriv\theta_i \deriv\phi_i \, \sin \theta_i \cos^2\frac{\theta_i}{2}\text{.}
  \label{eq:measure_mu}
\end{equation}
Relative to the product of spherical measures, the additional $\cos^2 \frac{\theta_i}{2}$  is chosen to vanish rapidly enough at the edges of the phase space that the singularity is removed. 

The integral over the azimuthal parameters $\phi_a$ is oscillatory, ensuring that only the diagonal matrix elements are non-zero.
The diagonal matrix elements are now computed,
\begin{align}
  \big(S_\mu\big)_{n_1\!,n_2}
  &{=} \#(n_1,n_2) \smashoperator[l]{\prod_{i=1}^2} \int_0^\pi \!\!\!\! \deriv\theta_i \cos^{N{-}2n{+}3} \frac{\theta_i}{2} \sin^{2n_i{+}1}\frac{\theta_i}{2} \nonumber\\
  &{=} \!\#(n_1,n_2) \prod_{i{=}1}^{2}\! \frac{\Gamma(n_i{+}1)\Gamma(\!\tfrac{N}{2}{-}n{+}2)}{\Gamma(\!\tfrac{N}{2}{-}n_i{+}3)}\text{,}
\end{align}
where $n = n_1+n_2$ and the integration is performed by taking a substitution $t_i = \cos^2(\theta_i/2)$ to place it in the form of an Euler integral.
For periodic boundary conditions, we arrive at the rational function
\begin{align}
  \big(S_\mu\big)_{n_1\!,n_2}
  &{=}
  \frac{\tfrac{N}{2}(\tfrac{N}{2}{+}1)^2(\tfrac{N}{2}{-}n)(\tfrac{N}{2}{-}n{+}1)^2}
       {(\tfrac{N}{2}-n_1)^{(3)}(\tfrac{N}{2}-n_2)^{(3)}}
  \text{,}
\label{eq:smatelem_app}
\end{align}
where $(x)^{(n)}$ is the rising factorial or Pochhammer polynomial~\cite{Abramovich}.
This is best thought of as a function of the densities, $n_a/N$.
Each linear factor in the numerator and denominator is a zero line or pole line, respectively, for the frame operator.
For finite $N$ and with the exception of the N\'eel states, these lines are outside of the domain of allowed densities.
On the N\'eel states, the zero line $(N/2-n)$ and pole line $(N/2-n_a)$ intersect and cancel.
In the vicinity of the N\'eel configurations, there are three `corner' pole lines with fixed $n_a$ and three `diagonal' zero lines with fixed $n$, all within a distance of order $1/N$.
As $N$ increases the pole lines may become closer, but proportionally the zero lines also become closer.
The scaling imparted by the zero lines to the residues of the approaching poles removes the divergence that would otherwise be caused by their approach.
The additional cosine factors in the measure have the effect of adding two diagonal zero lines and one pole line passing-by each N\'eel state, and so without this there would be a residual pole at each N\'eel state, spoiling boundedness.
Along the edge close to $(N-n)=0$, there are no poles to cancel the zero lines so we can find matrix elements which are arbitrarily small as $N$ increases.
This provides a frame operator which is bounded and positive definite, even in the $N \rightarrow \infty$ limit, except along the triangle edge.
This last point is not too concerning because we know that the dynamics avoids this part of $\mathcal{K}$, and is greatly preferable to leaving a simple pole on the most physically important states.

\begin{figure}
  \includegraphics[width=\linewidth]{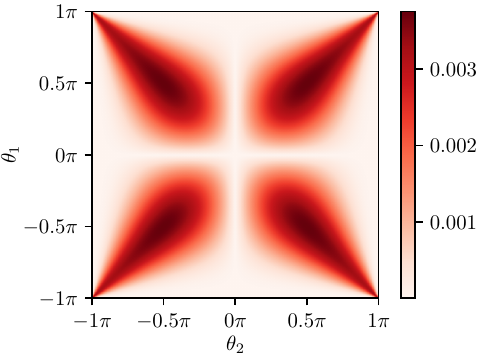}
  \caption{
    Transformed measure $\nu$, \Cref{eq:nu_measure}, in the thermodynamic limit, is non-negative and smooth away from the corner points.
  }
  \label{fig:pathintegral_nu}
\end{figure}

\begin{figure}
  \includegraphics[width=\linewidth]{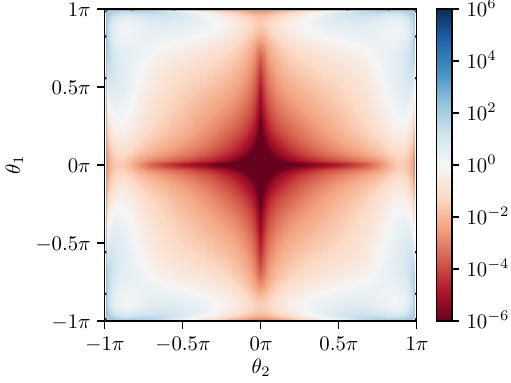}
  \caption{
    Variance in the $S_\mu^{-1/2}$ observable -- see \Cref{eq:var_S_half} -- divided by the expectation value, in the $\phi_a=0$ section of the state space $\mathcal{M}$ for $N=128$.
    The fluctuations are strong only towards the corners in a region which shrinks as $N$ increases.
  }
  \label{fig:pathintegral_fluctuations}
\end{figure}

As stated in the main text, the frame operator found in this way is not suitable for quantisation.
Its bounded and essentially positive-definite nature, however, leaves a measure which is qualitatively close to one in which a good resolution of the identity is found.
Using the resulting frame operator we defined a transformed frame $\overline{\mathcal{M}}$, which produces a resolution of the identity from a congruence,
\begin{align}
  \mathds{1}_\mathcal{K}
  &= \int \deriv{\mu_{\theta,\phi}} S_\mu^{-1/2} \ket{\Psi(\theta,\phi)}\bra{\Psi(\theta,\phi)} S_\mu^{-1/2}\text{.}
\end{align}
This frame transformation potentially disturbs the states and obscures their physical meaning.
We remedy this situation by approximating the effect of the frame transformation with a mere measure transformation,
\begin{align}
  \mathds{1}_\mathcal{K} \approx S_\nu = \int \deriv{\nu_{\theta,\phi}} \ket{\Psi(\theta,\phi)}\bra{\Psi(\theta,\phi)}\text{,}
\end{align}
with the transformed measure,
\begin{equation}
  \deriv{\nu} = \bigg(\frac{\bra{\Psi}S_\mu^{-1/2}\ket{\Psi}}{\brakett{\Psi}{\Psi}}\bigg)^2 \deriv{\mu}\text{,}
  \label{eq:nu_measure}
\end{equation}
which we claim is asymptotically exact.
This measure in the thermodynamic limit is shown in \Cref{fig:pathintegral_nu}, where it is seen to be smooth away from the corners.
This allows the classical system produced by the retraction onto the gauged constrained coherent states to quantise naturally to the quantum dynamics projected into $\mathcal{K}$, as described in the previous section.

The error in the approximation can be characterised by the quantity,
\begin{equation}
  \mathrm{var}_\Psi[S_\mu^{-\tfrac{1}{2}}] = \frac{\bra{\Psi}S_\mu^{-1}\ket{\Psi}}{\brakett{\Psi}{\Psi}} - \frac{\bra{\Psi}S_\mu^{-1/2}\ket{\Psi}^2}{\brakett{\Psi}{\Psi}^2}\text{.}
  \label{eq:var_S_half}
\end{equation}
This is the norm of the difference between the action of $S^{-1/2}$ on a state $\ket{\Psi(\theta,\phi)}$ and the action of multiplying by the state by the expectation value of that operator.
It can be used to bound the difference between the measure-transformed frame operator $S_\nu$ and the identity, and also the pointwise differences between untransformed and frame-transformed states.

In the remainder of this appendix we will show that $\mathrm{var}_\Psi[S_\mu^{-1/2}] = O(1/N)$, which establishes the asymptotic equivalence between the frame transformation and a change of measure.
The essential idea involves viewing each state $\Psi(\theta,\phi)$ as a stationary ergodic Markov process.
These correlations in $\ket{\Psi}$ are controlled by the correlation length $\xi$ of connected correlation functions, which can be determined from the eigenvalues of the transfer matrix.
This length-scale is finite almost everywhere on $\mathcal{M}$ but diverges as the corner points are approached.
According to the G\"artner-Ellis theorem, such a process satisfies a large deviation principle~\cite{Ellis:1984iy}.
Therefore, large deviations such as extensive fluctuations in $n_1$ and $n_2$ are suppressed exponentially and typical fluctuations are of magnitude $O(\sqrt{N})$.
This suggests that the frame-transformation, which is diagonal in the density basis, would also become `diagonal' in the frame $\mathcal{M}$ and act as a scalar.
In \Cref{fig:pathintegral_fluctuations}, we show how the variance behaves over $\mathcal{M}$ for a fixed system size $N{=}128$.
The fluctuations become strong only in the corners where $\xi$ diverges.

We will use \Cref{eq:smatelem_app} to view the operators $S_\mu^{-1}$ and $S_\mu^{-1/2}$ as functions of densities $n_1/N$ and $n_2/N$ which are then extended to the complex domain.
We use $f$ to refer to either of these functions in what follows.
This function is shifted in the argument such that the density expectation value is at the origin and shifted in value with a constant term so that $f$ vanishes there.
In some places, it may be necessary to regularise the frame operator by introducing some $\epsilon > 0$ and replacing $S_\mu \mapsto S_\mu + \epsilon$ before the inverse, in order to displace any singularities into the complex domain. This is then removed at a later point by taking the $\epsilon {\rightarrow} 0$ limit after calculating some physical quantity insensitive to the singularity.
For example, when applied to the frame operator for the transformed frame, this procedure yields a projector into the subspace of $\mathcal{K}$ with the non-physical states removed.

Now, we model $f$ as a Taylor series truncated at first order in each variable, and form a Cauchy bound to the remainder term.
Pick a polydisc $D$ centred around the mean densities with radii $0 < r_1 < \rho_1$ and $0 < r_2 < \rho_2$.
We also choose some $0 < \beta < 1$, defining inner radii $\beta r_1$ and $\beta r_2$.
If possible, the radii should chosen to include all subextensive density fluctuations for each finite $N$, but otherwise can be varied for an optimal bound in the final result.
Since $f$ is complex differentiable, its value interior to the disc can be related to that on the boundary by the multivariate Cauchy integral formula,
\begin{align}
\nonumber   f(z_1,z_2)
  &= \oiint_{\partial D} \frac{\deriv{w_1}\deriv{w_2}}{(2\pi i)^2}\frac{f(w_1,w_2)}{(w_1-z_1)(w_2-z_2)} \\
  &= f^{(1,0)}z_1 {+} f^{(0,1)}z_2 {+} f^{(1,1)}z_1z_2 {+} R(z)\text{,}
\end{align}
where $R(z)$ is the remainder term and the other terms are the truncated Taylor series.
The remainder term is found by expanding the geometric series and is bounded by,
\begin{align}
  | R(z) |
  &{\le} \bigg| \oiint\limits_{\partial  D} \!\!\frac{f(w_1,w_2)\deriv{w_1}\deriv{w_2}}{(w_1 {-} z_1)(w_2 {-} z_2)(2\pi i)^2}\!\bigg(\!\frac{z_1^2}{w_1^2} {+} \frac{z_2^2}{w_2^2} {-} \frac{z_1^2z_2^2}{w_1^2w_2^2}\bigg)\!\bigg| \nonumber\\
  &\le \frac{M_{\partial D}}{r_1r_2(1 - \beta)^2}
  \bigg(\frac{|z_1|^2}{r_1^2} {+} \frac{|z_2|^2}{r_2^2} {+} \frac{|z_1 z_2|^2}{(r_1 r_2)^2}\bigg)\text{,}
  \label{eq:R_bound}
\end{align}
where $M_{\partial D}$ appears as a uniform bound $|f(w_1,w_2)| \le M_{\partial D}$ for all $(w_1,w_2) \in \partial D$.
The constant factors in this bound are optimally independent of $N$ (for sufficiently large $N$) because the smallest radii only admissible for larger $N$ are bounded by at least $M_{\partial D} = O(r_1+r_2)$.

Next, we relate these properties of $S^{-1}$ and $S^{-1/2}$ as functions to their properties as operators.
This is done with the Dunford-Riesz functional calculus~\cite{Kadison:1983to}, which is a map $\mathbf{\Phi} : \mathrm{Hol}(D) \rightarrow \mathcal{L}(K)$ from holomorphic functions on a domain $D$ to continuous linear operators on a Hilbert space $K$.
It assigns to each function the Cauchy integral formula but with the geometric series replaced by resolvents of $n_1$ and $n_2$, effectively promoting the indeterminants to operators.
The result is a power series in the expectation values density fluctuation operators $\delta_i {=} (n_i {-} \langle n_i \rangle)/N$; equivalently, these are the central moments of the density observable distribution.
In the spectral subspace outside the inner radii, the series may not converge, thereby failing to correspond to the original operator, and also the bound established does not apply.
This region, however, corresponds to extensive fluctuations in the densities which are suppressed in probability exponentially with $N$ due to the large deviation principle, wherever the correlation length is finite.
The series divergence cannot overcome this, so we may safely use our Taylor series and bound the remainder as if they were entire, whilst incurring only an error exponentially small in $N$.
The first order terms vanish in expectation, hence the only remaining contributions provide a bound,
\begin{align}
  \big|\langle\mathbf{\Phi}(f)\rangle\big|
  &\le |f^{(1\!,1)}\!\langle\delta_1\delta_2\rangle|
  + C\bigg|\!\frac{\langle\delta_1^2\rangle}{r_1^2} {+} \frac{\langle\delta_2^2\rangle}{r_2^2} {+} \frac{\langle\delta_1^2 \delta_2^2\rangle}{r_1^2 r_2^2}\!\bigg|\text{,}
\end{align}
where $C$ is a constant factor from \Cref{eq:R_bound}.
The moments appearing here are finite-order connected correlations functions which are all $O(1/N)$.
This implies that the differences between the expectation value of e.g. $S_\mu^{-1}$ and its value as a function at the expected densities is only $O(1/N)$, and therefore so is the variance of $S_\mu^{-1/2}$.

In summary, we have shown that the frame operator fluctuations vanish as $O(1/\sqrt{N})$ for almost any point in $\mathcal{M}$ in the thermodynamic limit.
This is in agreement with what was numerically found in \Cref{fig:pathintegral_ft} in the main text, where the Fubini-Study distance between untransformed and transformed state spaces was integrated for different finite sizes $N$.
That calculation would also pick up a contribution due to the region of unphysical states with extensive correlation lengths which would decrease as the measure of this region decreases.
Given the agreement with the $O(1/\sqrt{N})$ theory, it appears that the dominant contribution is instead the normal fluctuations discussed in this section.

\bibliography{references}

\begin{thebibliography}{90}%
\makeatletter
\providecommand \@ifxundefined [1]{%
 \@ifx{#1\undefined}
}%
\providecommand \@ifnum [1]{%
 \ifnum #1\expandafter \@firstoftwo
 \else \expandafter \@secondoftwo
 \fi
}%
\providecommand \@ifx [1]{%
 \ifx #1\expandafter \@firstoftwo
 \else \expandafter \@secondoftwo
 \fi
}%
\providecommand \natexlab [1]{#1}%
\providecommand \enquote  [1]{``#1''}%
\providecommand \bibnamefont  [1]{#1}%
\providecommand \bibfnamefont [1]{#1}%
\providecommand \citenamefont [1]{#1}%
\providecommand \href@noop [0]{\@secondoftwo}%
\providecommand \href [0]{\begingroup \@sanitize@url \@href}%
\providecommand \@href[1]{\@@startlink{#1}\@@href}%
\providecommand \@@href[1]{\endgroup#1\@@endlink}%
\providecommand \@sanitize@url [0]{\catcode `\\12\catcode `\$12\catcode
  `\&12\catcode `\#12\catcode `\^12\catcode `\_12\catcode `\%12\relax}%
\providecommand \@@startlink[1]{}%
\providecommand \@@endlink[0]{}%
\providecommand \url  [0]{\begingroup\@sanitize@url \@url }%
\providecommand \@url [1]{\endgroup\@href {#1}{\urlprefix }}%
\providecommand \urlprefix  [0]{URL }%
\providecommand \Eprint [0]{\href }%
\providecommand \doibase [0]{https://doi.org/}%
\providecommand \selectlanguage [0]{\@gobble}%
\providecommand \bibinfo  [0]{\@secondoftwo}%
\providecommand \bibfield  [0]{\@secondoftwo}%
\providecommand \translation [1]{[#1]}%
\providecommand \BibitemOpen [0]{}%
\providecommand \bibitemStop [0]{}%
\providecommand \bibitemNoStop [0]{.\EOS\space}%
\providecommand \EOS [0]{\spacefactor3000\relax}%
\providecommand \BibitemShut  [1]{\csname bibitem#1\endcsname}%
\let\auto@bib@innerbib\@empty
\bibitem [{\citenamefont {Heller}(1984)}]{Heller:1984zz}%
  \BibitemOpen
  \bibfield  {author} {\bibinfo {author} {\bibfnamefont {E.~J.}\ \bibnamefont
  {Heller}},\ }\bibfield  {title} {\bibinfo {title} {Bound-state eigenfunctions
  of classically chaotic {Hamiltonian} systems: {Scars} of periodic orbits},\
  }\href {https://doi.org/10.1103/PhysRevLett.53.1515} {\bibfield  {journal}
  {\bibinfo  {journal} {Phys. Rev. Lett.}\ }\textbf {\bibinfo {volume} {53}},\
  \bibinfo {pages} {1515} (\bibinfo {year} {1984})}\BibitemShut {NoStop}%
\bibitem [{\citenamefont {Heller}(1991)}]{HellerLesHouches}%
  \BibitemOpen
  \bibfield  {author} {\bibinfo {author} {\bibfnamefont {E.~J.}\ \bibnamefont
  {Heller}},\ }\bibfield  {title} {\bibinfo {title} {Wavepacket dynamics and
  quantum chaology},\ }in\ \href@noop {} {\emph {\bibinfo {booktitle} {Chaos
  and quantum physics}}},\ Vol.~\bibinfo {volume} {52}\ (\bibinfo  {publisher}
  {North-Holland: Amsterdam},\ \bibinfo {year} {1991})\BibitemShut {NoStop}%
\bibitem [{\citenamefont {Berry}(1989)}]{berry1989quantum}%
  \BibitemOpen
  \bibfield  {author} {\bibinfo {author} {\bibfnamefont {M.~V.}\ \bibnamefont
  {Berry}},\ }\bibfield  {title} {\bibinfo {title} {Quantum scars of classical
  closed orbits in phase space},\ }\href
  {https://doi.org/10.1098/rspa.1989.0052} {\bibfield  {journal} {\bibinfo
  {journal} {Proc. R. Soc. Lond. A}\ }\textbf {\bibinfo {volume} {423}},\
  \bibinfo {pages} {219} (\bibinfo {year} {1989})}\BibitemShut {NoStop}%
\bibitem [{\citenamefont {Heller}(2018)}]{HellerBook}%
  \BibitemOpen
  \bibfield  {author} {\bibinfo {author} {\bibfnamefont {E.}~\bibnamefont
  {Heller}},\ }\href {https://books.google.co.uk/books?id=oGuYDwAAQBAJ} {\emph
  {\bibinfo {title} {The Semiclassical Way to Dynamics and Spectroscopy}}}\
  (\bibinfo  {publisher} {Princeton University Press},\ \bibinfo {year}
  {2018})\BibitemShut {NoStop}%
\bibitem [{\citenamefont {Bohr}(1976)}]{Bohr}%
  \BibitemOpen
  \bibfield  {author} {\bibinfo {author} {\bibfnamefont {N.}~\bibnamefont
  {Bohr}},\ }\bibfield  {title} {\bibinfo {title} {The correspondence principle
  (1918--1923)},\ }in\ \href@noop {} {\emph {\bibinfo {booktitle} {Niels Bohr
  Collected Works}}},\ Vol.~\bibinfo {volume} {3}\ (\bibinfo  {publisher}
  {Amsterdam: North-Holland Publishing},\ \bibinfo {year} {1976})\BibitemShut
  {NoStop}%
\bibitem [{\citenamefont {Zelditch}(2004)}]{Zelditch:2004fv}%
  \BibitemOpen
  \bibfield  {author} {\bibinfo {author} {\bibfnamefont {S.}~\bibnamefont
  {Zelditch}},\ }\bibfield  {title} {\bibinfo {title} {Note on quantum unique
  ergodicity},\ }\href {https://doi.org/10.1090/S0002-9939-03-07298-8}
  {\bibfield  {journal} {\bibinfo  {journal} {Proc. Am. Math. Soc.}\ }\textbf
  {\bibinfo {volume} {132}},\ \bibinfo {pages} {1869} (\bibinfo {year}
  {2004})}\BibitemShut {NoStop}%
\bibitem [{\citenamefont {O'Connor}\ and\ \citenamefont
  {Heller}(1988)}]{OConnor:1988cn}%
  \BibitemOpen
  \bibfield  {author} {\bibinfo {author} {\bibfnamefont {P.~W.}\ \bibnamefont
  {O'Connor}}\ and\ \bibinfo {author} {\bibfnamefont {E.~J.}\ \bibnamefont
  {Heller}},\ }\bibfield  {title} {\bibinfo {title} {Quantum localization for a
  strongly classically chaotic system},\ }\href
  {https://doi.org/10.1103/PhysRevLett.61.2288} {\bibfield  {journal} {\bibinfo
   {journal} {Phys. Rev. Lett.}\ }\textbf {\bibinfo {volume} {61}},\ \bibinfo
  {pages} {2288} (\bibinfo {year} {1988})}\BibitemShut {NoStop}%
\bibitem [{\citenamefont {Hassell}(2010)}]{Hassell:2010ef}%
  \BibitemOpen
  \bibfield  {author} {\bibinfo {author} {\bibfnamefont {A.}~\bibnamefont
  {Hassell}},\ }\bibfield  {title} {\bibinfo {title} {Ergodic billiards that
  are not quantum unique ergodic},\ }\href
  {http://doi.org/10.4007/annals.2010.171.605} {\bibfield  {journal} {\bibinfo
  {journal} {Ann. Math.}\ }\textbf {\bibinfo {volume} {171}},\ \bibinfo {pages}
  {605} (\bibinfo {year} {2010})}\BibitemShut {NoStop}%
\bibitem [{\citenamefont {Bernien}\ \emph {et~al.}(2017)\citenamefont
  {Bernien}, \citenamefont {Schwartz}, \citenamefont {Keesling}, \citenamefont
  {Levine}, \citenamefont {Omran}, \citenamefont {Pichler}, \citenamefont
  {Choi}, \citenamefont {Zibrov}, \citenamefont {Endres}, \citenamefont
  {Greiner}, \citenamefont {Vuleti{\'c}},\ and\ \citenamefont
  {Lukin}}]{Bernien:2017bp}%
  \BibitemOpen
  \bibfield  {author} {\bibinfo {author} {\bibfnamefont {H.}~\bibnamefont
  {Bernien}}, \bibinfo {author} {\bibfnamefont {S.}~\bibnamefont {Schwartz}},
  \bibinfo {author} {\bibfnamefont {A.}~\bibnamefont {Keesling}}, \bibinfo
  {author} {\bibfnamefont {H.}~\bibnamefont {Levine}}, \bibinfo {author}
  {\bibfnamefont {A.}~\bibnamefont {Omran}}, \bibinfo {author} {\bibfnamefont
  {H.}~\bibnamefont {Pichler}}, \bibinfo {author} {\bibfnamefont
  {S.}~\bibnamefont {Choi}}, \bibinfo {author} {\bibfnamefont {A.~S.}\
  \bibnamefont {Zibrov}}, \bibinfo {author} {\bibfnamefont {M.}~\bibnamefont
  {Endres}}, \bibinfo {author} {\bibfnamefont {M.}~\bibnamefont {Greiner}},
  \bibinfo {author} {\bibfnamefont {V.}~\bibnamefont {Vuleti{\'c}}},\ and\
  \bibinfo {author} {\bibfnamefont {M.~D.}\ \bibnamefont {Lukin}},\ }\bibfield
  {title} {\bibinfo {title} {Probing many-body dynamics on a 51-atom quantum
  simulator},\ }\href {https://doi.org/10.1038/nature24622} {\bibfield
  {journal} {\bibinfo  {journal} {Nature}\ }\textbf {\bibinfo {volume} {551}},\
  \bibinfo {pages} {579} (\bibinfo {year} {2017})}\BibitemShut {NoStop}%
\bibitem [{\citenamefont {Turner}\ \emph
  {et~al.}(2018{\natexlab{a}})\citenamefont {Turner}, \citenamefont
  {Michailidis}, \citenamefont {Abanin}, \citenamefont {Serbyn},\ and\
  \citenamefont {{Papi{\'c}}}}]{Turner:2017wu}%
  \BibitemOpen
  \bibfield  {author} {\bibinfo {author} {\bibfnamefont {C.~J.}\ \bibnamefont
  {Turner}}, \bibinfo {author} {\bibfnamefont {A.~A.}\ \bibnamefont
  {Michailidis}}, \bibinfo {author} {\bibfnamefont {D.~A.}\ \bibnamefont
  {Abanin}}, \bibinfo {author} {\bibfnamefont {M.}~\bibnamefont {Serbyn}},\
  and\ \bibinfo {author} {\bibfnamefont {Z.}~\bibnamefont {{Papi{\'c}}}},\
  }\bibfield  {title} {\bibinfo {title} {Weak ergodicity breaking from quantum
  many-body scars},\ }\href {https://doi.org/10.1038/s41567-018-0137-5}
  {\bibfield  {journal} {\bibinfo  {journal} {Nat. Phys.}\ }\textbf {\bibinfo
  {volume} {14}},\ \bibinfo {pages} {745} (\bibinfo {year}
  {2018}{\natexlab{a}})}\BibitemShut {NoStop}%
\bibitem [{\citenamefont {Ho}\ \emph {et~al.}(2019)\citenamefont {Ho},
  \citenamefont {Choi}, \citenamefont {Pichler},\ and\ \citenamefont
  {Lukin}}]{Ho:2019gv}%
  \BibitemOpen
  \bibfield  {author} {\bibinfo {author} {\bibfnamefont {W.~W.}\ \bibnamefont
  {Ho}}, \bibinfo {author} {\bibfnamefont {S.}~\bibnamefont {Choi}}, \bibinfo
  {author} {\bibfnamefont {H.}~\bibnamefont {Pichler}},\ and\ \bibinfo {author}
  {\bibfnamefont {M.~D.}\ \bibnamefont {Lukin}},\ }\bibfield  {title} {\bibinfo
  {title} {Periodic orbits, entanglement, and quantum many-body scars in
  constrained models: Matrix product state approach},\ }\href
  {https://doi.org/10.1103/PhysRevLett.122.040603} {\bibfield  {journal}
  {\bibinfo  {journal} {Phys. Rev. Lett.}\ }\textbf {\bibinfo {volume} {122}},\
  \bibinfo {pages} {040603} (\bibinfo {year} {2019})}\BibitemShut {NoStop}%
\bibitem [{\citenamefont {Choi}\ \emph {et~al.}(2019)\citenamefont {Choi},
  \citenamefont {Turner}, \citenamefont {Pichler}, \citenamefont {Ho},
  \citenamefont {Michailidis}, \citenamefont {Papi{\'c}}, \citenamefont
  {Serbyn}, \citenamefont {Lukin},\ and\ \citenamefont {Abanin}}]{Choi:2018wn}%
  \BibitemOpen
  \bibfield  {author} {\bibinfo {author} {\bibfnamefont {S.}~\bibnamefont
  {Choi}}, \bibinfo {author} {\bibfnamefont {C.~J.}\ \bibnamefont {Turner}},
  \bibinfo {author} {\bibfnamefont {H.}~\bibnamefont {Pichler}}, \bibinfo
  {author} {\bibfnamefont {W.~W.}\ \bibnamefont {Ho}}, \bibinfo {author}
  {\bibfnamefont {A.~A.}\ \bibnamefont {Michailidis}}, \bibinfo {author}
  {\bibfnamefont {Z.}~\bibnamefont {Papi{\'c}}}, \bibinfo {author}
  {\bibfnamefont {M.}~\bibnamefont {Serbyn}}, \bibinfo {author} {\bibfnamefont
  {M.~D.}\ \bibnamefont {Lukin}},\ and\ \bibinfo {author} {\bibfnamefont
  {D.~A.}\ \bibnamefont {Abanin}},\ }\bibfield  {title} {\bibinfo {title}
  {Emergent {SU(2)} dynamics and perfect quantum many-body scars},\ }\href
  {https://doi.org/10.1103/PhysRevLett.122.220603} {\bibfield  {journal}
  {\bibinfo  {journal} {Phys. Rev. Lett.}\ }\textbf {\bibinfo {volume} {122}},\
  \bibinfo {pages} {220603} (\bibinfo {year} {2019})}\BibitemShut {NoStop}%
\bibitem [{\citenamefont {Lin}\ and\ \citenamefont
  {Motrunich}(2019)}]{lin2018exact}%
  \BibitemOpen
  \bibfield  {author} {\bibinfo {author} {\bibfnamefont {C.-J.}\ \bibnamefont
  {Lin}}\ and\ \bibinfo {author} {\bibfnamefont {O.~I.}\ \bibnamefont
  {Motrunich}},\ }\bibfield  {title} {\bibinfo {title} {Exact quantum many-body
  scar states in the {Rydberg}-blockaded atom chain},\ }\href
  {https://doi.org/10.1103/PhysRevLett.122.173401} {\bibfield  {journal}
  {\bibinfo  {journal} {Phys. Rev. Lett.}\ }\textbf {\bibinfo {volume} {122}},\
  \bibinfo {pages} {173401} (\bibinfo {year} {2019})}\BibitemShut {NoStop}%
\bibitem [{\citenamefont {Khemani}\ \emph {et~al.}(2019)\citenamefont
  {Khemani}, \citenamefont {Laumann},\ and\ \citenamefont
  {Chandran}}]{Khemani2018}%
  \BibitemOpen
  \bibfield  {author} {\bibinfo {author} {\bibfnamefont {V.}~\bibnamefont
  {Khemani}}, \bibinfo {author} {\bibfnamefont {C.~R.}\ \bibnamefont
  {Laumann}},\ and\ \bibinfo {author} {\bibfnamefont {A.}~\bibnamefont
  {Chandran}},\ }\bibfield  {title} {\bibinfo {title} {Signatures of
  integrability in the dynamics of {Rydberg-blockaded} chains},\ }\href
  {https://doi.org/10.1103/PhysRevB.99.161101} {\bibfield  {journal} {\bibinfo
  {journal} {Phys. Rev. B}\ }\textbf {\bibinfo {volume} {99}},\ \bibinfo
  {pages} {161101(R)} (\bibinfo {year} {2019})}\BibitemShut {NoStop}%
\bibitem [{\citenamefont {Bull}\ \emph {et~al.}(2020)\citenamefont {Bull},
  \citenamefont {Desaules},\ and\ \citenamefont {Papi\'c}}]{Bull2020}%
  \BibitemOpen
  \bibfield  {author} {\bibinfo {author} {\bibfnamefont {K.}~\bibnamefont
  {Bull}}, \bibinfo {author} {\bibfnamefont {J.-Y.}\ \bibnamefont {Desaules}},\
  and\ \bibinfo {author} {\bibfnamefont {Z.}~\bibnamefont {Papi\'c}},\
  }\bibfield  {title} {\bibinfo {title} {Quantum scars as embeddings of weakly
  "broken" {Lie} algebra representations},\ }\href
  {https://doi.org/10.1103/PhysRevB.101.165139} {\bibfield  {journal} {\bibinfo
   {journal} {Phys. Rev. B}\ }\textbf {\bibinfo {volume} {101}},\ \bibinfo
  {pages} {165139} (\bibinfo {year} {2020})}\BibitemShut {NoStop}%
\bibitem [{\citenamefont {Mark}\ \emph {et~al.}(2020)\citenamefont {Mark},
  \citenamefont {Lin},\ and\ \citenamefont {Motrunich}}]{Mark2020exact}%
  \BibitemOpen
  \bibfield  {author} {\bibinfo {author} {\bibfnamefont {D.~K.}\ \bibnamefont
  {Mark}}, \bibinfo {author} {\bibfnamefont {C.-J.}\ \bibnamefont {Lin}},\ and\
  \bibinfo {author} {\bibfnamefont {O.~I.}\ \bibnamefont {Motrunich}},\
  }\bibfield  {title} {\bibinfo {title} {Exact eigenstates in the {L}esanovsky
  model, proximity to integrability and the {PXP} model, and approximate scar
  states},\ }\href {https://link.aps.org/doi/10.1103/PhysRevB.101.094308}
  {\bibfield  {journal} {\bibinfo  {journal} {Phys. Rev. B}\ }\textbf {\bibinfo
  {volume} {101}},\ \bibinfo {pages} {094308} (\bibinfo {year}
  {2020})}\BibitemShut {NoStop}%
\bibitem [{\citenamefont {Kormos}\ \emph {et~al.}(2016)\citenamefont {Kormos},
  \citenamefont {Collura}, \citenamefont {Tak{\'a}cs},\ and\ \citenamefont
  {Calabrese}}]{Calabrese16}%
  \BibitemOpen
  \bibfield  {author} {\bibinfo {author} {\bibfnamefont {M.}~\bibnamefont
  {Kormos}}, \bibinfo {author} {\bibfnamefont {M.}~\bibnamefont {Collura}},
  \bibinfo {author} {\bibfnamefont {G.}~\bibnamefont {Tak{\'a}cs}},\ and\
  \bibinfo {author} {\bibfnamefont {P.}~\bibnamefont {Calabrese}},\ }\bibfield
  {title} {\bibinfo {title} {Real-time confinement following a quantum quench
  to a non-integrable model},\ }\href {https://doi.org/10.1038/nphys3934}
  {\bibfield  {journal} {\bibinfo  {journal} {Nat. Phys.}\ }\textbf {\bibinfo
  {volume} {13}},\ \bibinfo {pages} {246} (\bibinfo {year} {2016})}\BibitemShut
  {NoStop}%
\bibitem [{\citenamefont {James}\ \emph {et~al.}(2019)\citenamefont {James},
  \citenamefont {Konik},\ and\ \citenamefont {Robinson}}]{Konik1}%
  \BibitemOpen
  \bibfield  {author} {\bibinfo {author} {\bibfnamefont {A.~J.~A.}\
  \bibnamefont {James}}, \bibinfo {author} {\bibfnamefont {R.~M.}\ \bibnamefont
  {Konik}},\ and\ \bibinfo {author} {\bibfnamefont {N.~J.}\ \bibnamefont
  {Robinson}},\ }\bibfield  {title} {\bibinfo {title} {Nonthermal states
  arising from confinement in one and two dimensions},\ }\href
  {https://doi.org/10.1103/PhysRevLett.122.130603} {\bibfield  {journal}
  {\bibinfo  {journal} {Phys. Rev. Lett.}\ }\textbf {\bibinfo {volume} {122}},\
  \bibinfo {pages} {130603} (\bibinfo {year} {2019})}\BibitemShut {NoStop}%
\bibitem [{\citenamefont {Robinson}\ \emph {et~al.}(2019)\citenamefont
  {Robinson}, \citenamefont {James},\ and\ \citenamefont {Konik}}]{Konik2}%
  \BibitemOpen
  \bibfield  {author} {\bibinfo {author} {\bibfnamefont {N.~J.}\ \bibnamefont
  {Robinson}}, \bibinfo {author} {\bibfnamefont {A.~J.~A.}\ \bibnamefont
  {James}},\ and\ \bibinfo {author} {\bibfnamefont {R.~M.}\ \bibnamefont
  {Konik}},\ }\bibfield  {title} {\bibinfo {title} {Signatures of rare states
  and thermalization in a theory with confinement},\ }\href
  {https://doi.org/10.1103/PhysRevB.99.195108} {\bibfield  {journal} {\bibinfo
  {journal} {Phys. Rev. B}\ }\textbf {\bibinfo {volume} {99}},\ \bibinfo
  {pages} {195108} (\bibinfo {year} {2019})}\BibitemShut {NoStop}%
\bibitem [{\citenamefont {Vafek}\ \emph {et~al.}(2017)\citenamefont {Vafek},
  \citenamefont {Regnault},\ and\ \citenamefont {Bernevig}}]{Vafek}%
  \BibitemOpen
  \bibfield  {author} {\bibinfo {author} {\bibfnamefont {O.}~\bibnamefont
  {Vafek}}, \bibinfo {author} {\bibfnamefont {N.}~\bibnamefont {Regnault}},\
  and\ \bibinfo {author} {\bibfnamefont {B.~A.}\ \bibnamefont {Bernevig}},\
  }\bibfield  {title} {\bibinfo {title} {Entanglement of exact excited
  eigenstates of the {Hubbard} model in arbitrary dimension},\ }\href
  {https://doi.org/10.21468/SciPostPhys.3.6.043} {\bibfield  {journal}
  {\bibinfo  {journal} {SciPost Phys.}\ }\textbf {\bibinfo {volume} {3}},\
  \bibinfo {pages} {043} (\bibinfo {year} {2017})}\BibitemShut {NoStop}%
\bibitem [{\citenamefont {Moudgalya}\ \emph {et~al.}(2018)\citenamefont
  {Moudgalya}, \citenamefont {Regnault},\ and\ \citenamefont
  {Bernevig}}]{Moudgalya2018}%
  \BibitemOpen
  \bibfield  {author} {\bibinfo {author} {\bibfnamefont {S.}~\bibnamefont
  {Moudgalya}}, \bibinfo {author} {\bibfnamefont {N.}~\bibnamefont
  {Regnault}},\ and\ \bibinfo {author} {\bibfnamefont {B.~A.}\ \bibnamefont
  {Bernevig}},\ }\bibfield  {title} {\bibinfo {title} {Entanglement of exact
  excited states of {Affleck-Kennedy-Lieb-Tasaki} models: Exact results,
  many-body scars, and violation of the strong eigenstate thermalization
  hypothesis},\ }\href {https://doi.org/10.1103/PhysRevB.98.235156} {\bibfield
  {journal} {\bibinfo  {journal} {Phys. Rev. B}\ }\textbf {\bibinfo {volume}
  {98}},\ \bibinfo {pages} {235156} (\bibinfo {year} {2018})}\BibitemShut
  {NoStop}%
\bibitem [{\citenamefont {Iadecola}\ and\ \citenamefont {\ifmmode
  \check{Z}\else \v{Z}\fi{}nidari\ifmmode~\check{c}\else
  \v{c}\fi{}}(2019)}]{IadecolaZnidaric}%
  \BibitemOpen
  \bibfield  {author} {\bibinfo {author} {\bibfnamefont {T.}~\bibnamefont
  {Iadecola}}\ and\ \bibinfo {author} {\bibfnamefont {M.}~\bibnamefont
  {\ifmmode \check{Z}\else \v{Z}\fi{}nidari\ifmmode~\check{c}\else
  \v{c}\fi{}}},\ }\bibfield  {title} {\bibinfo {title} {Exact localized and
  ballistic eigenstates in disordered chaotic spin ladders and the
  {Fermi-Hubbard} model},\ }\href
  {https://doi.org/10.1103/PhysRevLett.123.036403} {\bibfield  {journal}
  {\bibinfo  {journal} {Phys. Rev. Lett.}\ }\textbf {\bibinfo {volume} {123}},\
  \bibinfo {pages} {036403} (\bibinfo {year} {2019})}\BibitemShut {NoStop}%
\bibitem [{\citenamefont {Ok}\ \emph {et~al.}(2019)\citenamefont {Ok},
  \citenamefont {Choo}, \citenamefont {Mudry}, \citenamefont {Castelnovo},
  \citenamefont {Chamon},\ and\ \citenamefont {Neupert}}]{NeupertScars}%
  \BibitemOpen
  \bibfield  {author} {\bibinfo {author} {\bibfnamefont {S.}~\bibnamefont
  {Ok}}, \bibinfo {author} {\bibfnamefont {K.}~\bibnamefont {Choo}}, \bibinfo
  {author} {\bibfnamefont {C.}~\bibnamefont {Mudry}}, \bibinfo {author}
  {\bibfnamefont {C.}~\bibnamefont {Castelnovo}}, \bibinfo {author}
  {\bibfnamefont {C.}~\bibnamefont {Chamon}},\ and\ \bibinfo {author}
  {\bibfnamefont {T.}~\bibnamefont {Neupert}},\ }\bibfield  {title} {\bibinfo
  {title} {Topological many-body scar states in dimensions one, two, and
  three},\ }\href {https://doi.org/10.1103/PhysRevResearch.1.033144} {\bibfield
   {journal} {\bibinfo  {journal} {Phys. Rev. Research}\ }\textbf {\bibinfo
  {volume} {1}},\ \bibinfo {pages} {033144} (\bibinfo {year}
  {2019})}\BibitemShut {NoStop}%
\bibitem [{\citenamefont {Michailidis}\ \emph {et~al.}(2020)\citenamefont
  {Michailidis}, \citenamefont {Turner}, \citenamefont {Papi\'c}, \citenamefont
  {Abanin},\ and\ \citenamefont {Serbyn}}]{Michailidis:2019wj}%
  \BibitemOpen
  \bibfield  {author} {\bibinfo {author} {\bibfnamefont {A.~A.}\ \bibnamefont
  {Michailidis}}, \bibinfo {author} {\bibfnamefont {C.~J.}\ \bibnamefont
  {Turner}}, \bibinfo {author} {\bibfnamefont {Z.}~\bibnamefont {Papi\'c}},
  \bibinfo {author} {\bibfnamefont {D.~A.}\ \bibnamefont {Abanin}},\ and\
  \bibinfo {author} {\bibfnamefont {M.}~\bibnamefont {Serbyn}},\ }\bibfield
  {title} {\bibinfo {title} {Slow quantum thermalization and many-body revivals
  from mixed phase space},\ }\href {https://doi.org/10.1103/PhysRevX.10.011055}
  {\bibfield  {journal} {\bibinfo  {journal} {Phys. Rev. X}\ }\textbf {\bibinfo
  {volume} {10}},\ \bibinfo {pages} {011055} (\bibinfo {year}
  {2020})}\BibitemShut {NoStop}%
\bibitem [{\citenamefont {Schecter}\ and\ \citenamefont
  {Iadecola}(2019)}]{Schecter2019}%
  \BibitemOpen
  \bibfield  {author} {\bibinfo {author} {\bibfnamefont {M.}~\bibnamefont
  {Schecter}}\ and\ \bibinfo {author} {\bibfnamefont {T.}~\bibnamefont
  {Iadecola}},\ }\bibfield  {title} {\bibinfo {title} {Weak ergodicity breaking
  and quantum many-body scars in spin-1 {$XY$} magnets},\ }\href
  {https://doi.org/10.1103/PhysRevLett.123.147201} {\bibfield  {journal}
  {\bibinfo  {journal} {Phys. Rev. Lett.}\ }\textbf {\bibinfo {volume} {123}},\
  \bibinfo {pages} {147201} (\bibinfo {year} {2019})}\BibitemShut {NoStop}%
\bibitem [{\citenamefont {Haldar}\ \emph {et~al.}(2019)\citenamefont {Haldar},
  \citenamefont {Sen}, \citenamefont {Moessner},\ and\ \citenamefont
  {Das}}]{Haldar2019}%
  \BibitemOpen
  \bibfield  {author} {\bibinfo {author} {\bibfnamefont {A.}~\bibnamefont
  {Haldar}}, \bibinfo {author} {\bibfnamefont {D.}~\bibnamefont {Sen}},
  \bibinfo {author} {\bibfnamefont {R.}~\bibnamefont {Moessner}},\ and\
  \bibinfo {author} {\bibfnamefont {A.}~\bibnamefont {Das}},\ }\href@noop {}
  {\bibinfo {title} {Scars in strongly driven {Floquet} matter: resonance vs
  emergent conservation laws}} (\bibinfo {year} {2019}),\ \Eprint
  {https://arxiv.org/abs/1909.04064} {arXiv:1909.04064 [cond-mat.other]}
  \BibitemShut {NoStop}%
\bibitem [{\citenamefont {Mukherjee}\ \emph {et~al.}(2020)\citenamefont
  {Mukherjee}, \citenamefont {Nandy}, \citenamefont {Sen}, \citenamefont
  {Sen},\ and\ \citenamefont {Sengupta}}]{mukherjee2019collapse}%
  \BibitemOpen
  \bibfield  {author} {\bibinfo {author} {\bibfnamefont {B.}~\bibnamefont
  {Mukherjee}}, \bibinfo {author} {\bibfnamefont {S.}~\bibnamefont {Nandy}},
  \bibinfo {author} {\bibfnamefont {A.}~\bibnamefont {Sen}}, \bibinfo {author}
  {\bibfnamefont {D.}~\bibnamefont {Sen}},\ and\ \bibinfo {author}
  {\bibfnamefont {K.}~\bibnamefont {Sengupta}},\ }\bibfield  {title} {\bibinfo
  {title} {Collapse and revival of quantum many-body scars via {Floquet}
  engineering},\ }\href {https://doi.org/10.1103/PhysRevB.101.245107}
  {\bibfield  {journal} {\bibinfo  {journal} {Phys. Rev. B}\ }\textbf {\bibinfo
  {volume} {101}},\ \bibinfo {pages} {245107} (\bibinfo {year}
  {2020})}\BibitemShut {NoStop}%
\bibitem [{\citenamefont {{Sugiura}}\ \emph {et~al.}(2021)\citenamefont
  {{Sugiura}}, \citenamefont {{Kuwahara}},\ and\ \citenamefont
  {{Saito}}}]{sugiura2019manybody}%
  \BibitemOpen
  \bibfield  {author} {\bibinfo {author} {\bibfnamefont {S.}~\bibnamefont
  {{Sugiura}}}, \bibinfo {author} {\bibfnamefont {T.}~\bibnamefont
  {{Kuwahara}}},\ and\ \bibinfo {author} {\bibfnamefont {K.}~\bibnamefont
  {{Saito}}},\ }\bibfield  {title} {\bibinfo {title} {{Many-body scar state
  intrinsic to periodically driven system}},\ }\href
  {https://doi.org/10.1103/PhysRevResearch.3.L012010} {\bibfield  {journal}
  {\bibinfo  {journal} {Phys. Rev. Research}\ }\textbf {\bibinfo {volume}
  {3}},\ \bibinfo {eid} {L012010} (\bibinfo {year} {2021})}\BibitemShut
  {NoStop}%
\bibitem [{\citenamefont {{Moudgalya}}\ \emph
  {et~al.}(2020{\natexlab{a}})\citenamefont {{Moudgalya}}, \citenamefont
  {{Bernevig}},\ and\ \citenamefont {{Regnault}}}]{Moudgalya2019}%
  \BibitemOpen
  \bibfield  {author} {\bibinfo {author} {\bibfnamefont {S.}~\bibnamefont
  {{Moudgalya}}}, \bibinfo {author} {\bibfnamefont {B.~A.}\ \bibnamefont
  {{Bernevig}}},\ and\ \bibinfo {author} {\bibfnamefont {N.}~\bibnamefont
  {{Regnault}}},\ }\bibfield  {title} {\bibinfo {title} {{Quantum many-body
  scars in a Landau level on a thin torus}},\ }\href
  {https://doi.org/10.1103/PhysRevB.102.195150} {\bibfield  {journal} {\bibinfo
   {journal} {\prb}\ }\textbf {\bibinfo {volume} {102}},\ \bibinfo {eid}
  {195150} (\bibinfo {year} {2020}{\natexlab{a}})}\BibitemShut {NoStop}%
\bibitem [{\citenamefont {Iadecola}\ and\ \citenamefont
  {Schecter}(2020)}]{Iadecola2019_3}%
  \BibitemOpen
  \bibfield  {author} {\bibinfo {author} {\bibfnamefont {T.}~\bibnamefont
  {Iadecola}}\ and\ \bibinfo {author} {\bibfnamefont {M.}~\bibnamefont
  {Schecter}},\ }\bibfield  {title} {\bibinfo {title} {Quantum many-body scar
  states with emergent kinetic constraints and finite-entanglement revivals},\
  }\href {https://doi.org/10.1103/PhysRevB.101.024306} {\bibfield  {journal}
  {\bibinfo  {journal} {Phys. Rev. B}\ }\textbf {\bibinfo {volume} {101}},\
  \bibinfo {pages} {024306} (\bibinfo {year} {2020})}\BibitemShut {NoStop}%
\bibitem [{\citenamefont {Hudomal}\ \emph {et~al.}(2020)\citenamefont
  {Hudomal}, \citenamefont {Vasi{\'c}}, \citenamefont {Regnault},\ and\
  \citenamefont {Papi{\'c}}}]{Hudomal2019}%
  \BibitemOpen
  \bibfield  {author} {\bibinfo {author} {\bibfnamefont {A.}~\bibnamefont
  {Hudomal}}, \bibinfo {author} {\bibfnamefont {I.}~\bibnamefont {Vasi{\'c}}},
  \bibinfo {author} {\bibfnamefont {N.}~\bibnamefont {Regnault}},\ and\
  \bibinfo {author} {\bibfnamefont {Z.}~\bibnamefont {Papi{\'c}}},\ }\bibfield
  {title} {\bibinfo {title} {Quantum scars of bosons with correlated hopping},\
  }\href {https://doi.org/10.1038/s42005-020-0364-9} {\bibfield  {journal}
  {\bibinfo  {journal} {Commun. Phys.}\ }\textbf {\bibinfo {volume} {3}},\
  \bibinfo {pages} {1} (\bibinfo {year} {2020})}\BibitemShut {NoStop}%
\bibitem [{\citenamefont {Pai}\ and\ \citenamefont
  {Pretko}(2019)}]{Pretko2019}%
  \BibitemOpen
  \bibfield  {author} {\bibinfo {author} {\bibfnamefont {S.}~\bibnamefont
  {Pai}}\ and\ \bibinfo {author} {\bibfnamefont {M.}~\bibnamefont {Pretko}},\
  }\bibfield  {title} {\bibinfo {title} {Dynamical scar states in driven
  fracton systems},\ }\href {https://doi.org/10.1103/PhysRevLett.123.136401}
  {\bibfield  {journal} {\bibinfo  {journal} {Phys. Rev. Lett.}\ }\textbf
  {\bibinfo {volume} {123}},\ \bibinfo {pages} {136401} (\bibinfo {year}
  {2019})}\BibitemShut {NoStop}%
\bibitem [{\citenamefont {Bull}\ \emph {et~al.}(2019)\citenamefont {Bull},
  \citenamefont {Martin},\ and\ \citenamefont {Papi\ifmmode~\acute{c}\else
  \'{c}\fi{}}}]{Bull2019}%
  \BibitemOpen
  \bibfield  {author} {\bibinfo {author} {\bibfnamefont {K.}~\bibnamefont
  {Bull}}, \bibinfo {author} {\bibfnamefont {I.}~\bibnamefont {Martin}},\ and\
  \bibinfo {author} {\bibfnamefont {Z.}~\bibnamefont
  {Papi\ifmmode~\acute{c}\else \'{c}\fi{}}},\ }\bibfield  {title} {\bibinfo
  {title} {Systematic construction of scarred many-body dynamics in {1D}
  lattice models},\ }\href {https://doi.org/10.1103/PhysRevLett.123.030601}
  {\bibfield  {journal} {\bibinfo  {journal} {Phys. Rev. Lett.}\ }\textbf
  {\bibinfo {volume} {123}},\ \bibinfo {pages} {030601} (\bibinfo {year}
  {2019})}\BibitemShut {NoStop}%
\bibitem [{\citenamefont {Buca}\ \emph {et~al.}(2019)\citenamefont {Buca},
  \citenamefont {Tindall},\ and\ \citenamefont {Jaksch}}]{Buca2019}%
  \BibitemOpen
  \bibfield  {author} {\bibinfo {author} {\bibfnamefont {B.}~\bibnamefont
  {Buca}}, \bibinfo {author} {\bibfnamefont {J.}~\bibnamefont {Tindall}},\ and\
  \bibinfo {author} {\bibfnamefont {D.}~\bibnamefont {Jaksch}},\ }\bibfield
  {title} {\bibinfo {title} {Non-stationary coherent quantum many-body dynamics
  through dissipation},\ }\href {https://doi.org/10.1038/s41467-019-09757-y}
  {\bibfield  {journal} {\bibinfo  {journal} {Nat. Commun.}\ }\textbf {\bibinfo
  {volume} {10}},\ \bibinfo {pages} {1730} (\bibinfo {year}
  {2019})}\BibitemShut {NoStop}%
\bibitem [{\citenamefont {Tindall}\ \emph {et~al.}(2019)\citenamefont
  {Tindall}, \citenamefont {Bu\ifmmode~\check{c}\else \v{c}\fi{}a},
  \citenamefont {Coulthard},\ and\ \citenamefont {Jaksch}}]{Tindall2019}%
  \BibitemOpen
  \bibfield  {author} {\bibinfo {author} {\bibfnamefont {J.}~\bibnamefont
  {Tindall}}, \bibinfo {author} {\bibfnamefont {B.}~\bibnamefont
  {Bu\ifmmode~\check{c}\else \v{c}\fi{}a}}, \bibinfo {author} {\bibfnamefont
  {J.~R.}\ \bibnamefont {Coulthard}},\ and\ \bibinfo {author} {\bibfnamefont
  {D.}~\bibnamefont {Jaksch}},\ }\bibfield  {title} {\bibinfo {title}
  {Heating-induced long-range $\ensuremath{\eta}$ pairing in the {Hubbard}
  model},\ }\href {https://doi.org/10.1103/PhysRevLett.123.030603} {\bibfield
  {journal} {\bibinfo  {journal} {Phys. Rev. Lett.}\ }\textbf {\bibinfo
  {volume} {123}},\ \bibinfo {pages} {030603} (\bibinfo {year}
  {2019})}\BibitemShut {NoStop}%
\bibitem [{\citenamefont {Shibata}\ \emph {et~al.}(2020)\citenamefont
  {Shibata}, \citenamefont {Yoshioka},\ and\ \citenamefont
  {Katsura}}]{OnsagerScars}%
  \BibitemOpen
  \bibfield  {author} {\bibinfo {author} {\bibfnamefont {N.}~\bibnamefont
  {Shibata}}, \bibinfo {author} {\bibfnamefont {N.}~\bibnamefont {Yoshioka}},\
  and\ \bibinfo {author} {\bibfnamefont {H.}~\bibnamefont {Katsura}},\
  }\bibfield  {title} {\bibinfo {title} {Onsager's scars in disordered spin
  chains},\ }\href {https://doi.org/10.1103/PhysRevLett.124.180604} {\bibfield
  {journal} {\bibinfo  {journal} {Phys. Rev. Lett.}\ }\textbf {\bibinfo
  {volume} {124}},\ \bibinfo {pages} {180604} (\bibinfo {year}
  {2020})}\BibitemShut {NoStop}%
\bibitem [{\citenamefont {Moudgalya}\ \emph {et~al.}(2019)\citenamefont
  {Moudgalya}, \citenamefont {Prem}, \citenamefont {Nandkishore}, \citenamefont
  {Regnault},\ and\ \citenamefont {Bernevig}}]{MoudgalyaKrylov}%
  \BibitemOpen
  \bibfield  {author} {\bibinfo {author} {\bibfnamefont {S.}~\bibnamefont
  {Moudgalya}}, \bibinfo {author} {\bibfnamefont {A.}~\bibnamefont {Prem}},
  \bibinfo {author} {\bibfnamefont {R.}~\bibnamefont {Nandkishore}}, \bibinfo
  {author} {\bibfnamefont {N.}~\bibnamefont {Regnault}},\ and\ \bibinfo
  {author} {\bibfnamefont {B.~A.}\ \bibnamefont {Bernevig}},\ }\href@noop {}
  {\bibinfo {title} {Thermalization and its absence within {Krylov} subspaces
  of a constrained {Hamiltonian}}} (\bibinfo {year} {2019}),\ \Eprint
  {https://arxiv.org/abs/1910.14048} {arXiv:1910.14048 [cond-mat.str-el]}
  \BibitemShut {NoStop}%
\bibitem [{\citenamefont {Zhao}\ \emph {et~al.}(2020)\citenamefont {Zhao},
  \citenamefont {Vovrosh}, \citenamefont {Mintert},\ and\ \citenamefont
  {Knolle}}]{Zhao2020}%
  \BibitemOpen
  \bibfield  {author} {\bibinfo {author} {\bibfnamefont {H.}~\bibnamefont
  {Zhao}}, \bibinfo {author} {\bibfnamefont {J.}~\bibnamefont {Vovrosh}},
  \bibinfo {author} {\bibfnamefont {F.}~\bibnamefont {Mintert}},\ and\ \bibinfo
  {author} {\bibfnamefont {J.}~\bibnamefont {Knolle}},\ }\bibfield  {title}
  {\bibinfo {title} {Quantum many-body scars in optical lattices},\ }\href
  {https://doi.org/10.1103/PhysRevLett.124.160604} {\bibfield  {journal}
  {\bibinfo  {journal} {Phys. Rev. Lett.}\ }\textbf {\bibinfo {volume} {124}},\
  \bibinfo {pages} {160604} (\bibinfo {year} {2020})}\BibitemShut {NoStop}%
\bibitem [{\citenamefont {Lee}\ \emph {et~al.}(2020)\citenamefont {Lee},
  \citenamefont {Melendrez}, \citenamefont {Pal},\ and\ \citenamefont
  {Changlani}}]{Lee2020}%
  \BibitemOpen
  \bibfield  {author} {\bibinfo {author} {\bibfnamefont {K.}~\bibnamefont
  {Lee}}, \bibinfo {author} {\bibfnamefont {R.}~\bibnamefont {Melendrez}},
  \bibinfo {author} {\bibfnamefont {A.}~\bibnamefont {Pal}},\ and\ \bibinfo
  {author} {\bibfnamefont {H.~J.}\ \bibnamefont {Changlani}},\ }\bibfield
  {title} {\bibinfo {title} {Exact three-colored quantum scars from geometric
  frustration},\ }\href {https://doi.org/10.1103/PhysRevB.101.241111}
  {\bibfield  {journal} {\bibinfo  {journal} {Phys. Rev. B}\ }\textbf {\bibinfo
  {volume} {101}},\ \bibinfo {pages} {241111(R)} (\bibinfo {year}
  {2020})}\BibitemShut {NoStop}%
\bibitem [{\citenamefont {{Mark}}\ \emph {et~al.}(2020)\citenamefont {{Mark}},
  \citenamefont {{Lin}},\ and\ \citenamefont {{Motrunich}}}]{Mark2020unified}%
  \BibitemOpen
  \bibfield  {author} {\bibinfo {author} {\bibfnamefont {D.~K.}\ \bibnamefont
  {{Mark}}}, \bibinfo {author} {\bibfnamefont {C.-J.}\ \bibnamefont {{Lin}}},\
  and\ \bibinfo {author} {\bibfnamefont {O.~I.}\ \bibnamefont {{Motrunich}}},\
  }\bibfield  {title} {\bibinfo {title} {{Unified structure for exact towers of
  scar states in the Affleck-Kennedy-Lieb-Tasaki and other models}},\ }\href
  {https://doi.org/10.1103/PhysRevB.101.195131} {\bibfield  {journal} {\bibinfo
   {journal} {\prb}\ }\textbf {\bibinfo {volume} {101}},\ \bibinfo {eid}
  {195131} (\bibinfo {year} {2020})}\BibitemShut {NoStop}%
\bibitem [{\citenamefont {{Moudgalya}}\ \emph
  {et~al.}(2020{\natexlab{b}})\citenamefont {{Moudgalya}}, \citenamefont
  {{O'Brien}}, \citenamefont {{Bernevig}}, \citenamefont {{Fendley}},\ and\
  \citenamefont {{Regnault}}}]{Moudgalya2020}%
  \BibitemOpen
  \bibfield  {author} {\bibinfo {author} {\bibfnamefont {S.}~\bibnamefont
  {{Moudgalya}}}, \bibinfo {author} {\bibfnamefont {E.}~\bibnamefont
  {{O'Brien}}}, \bibinfo {author} {\bibfnamefont {B.~A.}\ \bibnamefont
  {{Bernevig}}}, \bibinfo {author} {\bibfnamefont {P.}~\bibnamefont
  {{Fendley}}},\ and\ \bibinfo {author} {\bibfnamefont {N.}~\bibnamefont
  {{Regnault}}},\ }\bibfield  {title} {\bibinfo {title} {{Large classes of
  quantum scarred Hamiltonians from matrix product states}},\ }\href
  {https://doi.org/10.1103/PhysRevB.102.085120} {\bibfield  {journal} {\bibinfo
   {journal} {\prb}\ }\textbf {\bibinfo {volume} {102}},\ \bibinfo {eid}
  {085120} (\bibinfo {year} {2020}{\natexlab{b}})}\BibitemShut {NoStop}%
\bibitem [{\citenamefont {Villase{\~n}or}\ \emph {et~al.}(2020)\citenamefont
  {Villase{\~n}or}, \citenamefont {Pilatowsky-Cameo}, \citenamefont
  {Bastarrachea-Magnani}, \citenamefont {Lerma}, \citenamefont {Santos},\ and\
  \citenamefont {Hirsch}}]{Villasenor2020}%
  \BibitemOpen
  \bibfield  {author} {\bibinfo {author} {\bibfnamefont {D.}~\bibnamefont
  {Villase{\~n}or}}, \bibinfo {author} {\bibfnamefont {S.}~\bibnamefont
  {Pilatowsky-Cameo}}, \bibinfo {author} {\bibfnamefont {M.~A.}\ \bibnamefont
  {Bastarrachea-Magnani}}, \bibinfo {author} {\bibfnamefont {S.}~\bibnamefont
  {Lerma}}, \bibinfo {author} {\bibfnamefont {L.~F.}\ \bibnamefont {Santos}},\
  and\ \bibinfo {author} {\bibfnamefont {J.~G.}\ \bibnamefont {Hirsch}},\
  }\bibfield  {title} {\bibinfo {title} {Quantum vs classical dynamics in a
  spin-boson system: manifestations of spectral correlations and scarring},\
  }\href {http://iopscience.iop.org/10.1088/1367-2630/ab8ef8} {\bibfield
  {journal} {\bibinfo  {journal} {New Journal of Physics}\ } (\bibinfo {year}
  {2020})}\BibitemShut {NoStop}%
\bibitem [{\citenamefont {{Moudgalya}}\ \emph
  {et~al.}(2020{\natexlab{c}})\citenamefont {{Moudgalya}}, \citenamefont
  {{Regnault}},\ and\ \citenamefont {{Bernevig}}}]{Moudgalya2020eta}%
  \BibitemOpen
  \bibfield  {author} {\bibinfo {author} {\bibfnamefont {S.}~\bibnamefont
  {{Moudgalya}}}, \bibinfo {author} {\bibfnamefont {N.}~\bibnamefont
  {{Regnault}}},\ and\ \bibinfo {author} {\bibfnamefont {B.~A.}\ \bibnamefont
  {{Bernevig}}},\ }\bibfield  {title} {\bibinfo {title}
  {{{\ensuremath{\eta}}-pairing in {H}ubbard models: From spectrum generating
  algebras to quantum many-body scars}},\ }\href
  {https://doi.org/10.1103/PhysRevB.102.085140} {\bibfield  {journal} {\bibinfo
   {journal} {\prb}\ }\textbf {\bibinfo {volume} {102}},\ \bibinfo {eid}
  {085140} (\bibinfo {year} {2020}{\natexlab{c}})},\ \Eprint
  {https://arxiv.org/abs/2004.13727} {arXiv:2004.13727 [cond-mat.str-el]}
  \BibitemShut {NoStop}%
\bibitem [{\citenamefont {{Mark}}\ and\ \citenamefont
  {{Motrunich}}(2020)}]{Mark2020}%
  \BibitemOpen
  \bibfield  {author} {\bibinfo {author} {\bibfnamefont {D.~K.}\ \bibnamefont
  {{Mark}}}\ and\ \bibinfo {author} {\bibfnamefont {O.~I.}\ \bibnamefont
  {{Motrunich}}},\ }\bibfield  {title} {\bibinfo {title} {{{\ensuremath{\eta}}
  -pairing states as true scars in an extended Hubbard model}},\ }\href
  {https://doi.org/10.1103/PhysRevB.102.075132} {\bibfield  {journal} {\bibinfo
   {journal} {\prb}\ }\textbf {\bibinfo {volume} {102}},\ \bibinfo {eid}
  {075132} (\bibinfo {year} {2020})}\BibitemShut {NoStop}%
\bibitem [{\citenamefont {{van Voorden}}\ \emph {et~al.}(2020)\citenamefont
  {{van Voorden}}, \citenamefont {{Min{\'a}{\v{r}}}},\ and\ \citenamefont
  {{Schoutens}}}]{Van2020}%
  \BibitemOpen
  \bibfield  {author} {\bibinfo {author} {\bibfnamefont {B.}~\bibnamefont {{van
  Voorden}}}, \bibinfo {author} {\bibfnamefont {J.}~\bibnamefont
  {{Min{\'a}{\v{r}}}}},\ and\ \bibinfo {author} {\bibfnamefont
  {K.}~\bibnamefont {{Schoutens}}},\ }\bibfield  {title} {\bibinfo {title}
  {{Quantum many-body scars in transverse field Ising ladders and beyond}},\
  }\href {https://doi.org/10.1103/PhysRevB.101.220305} {\bibfield  {journal}
  {\bibinfo  {journal} {\prb}\ }\textbf {\bibinfo {volume} {101}},\ \bibinfo
  {eid} {220305} (\bibinfo {year} {2020})}\BibitemShut {NoStop}%
\bibitem [{\citenamefont {{Mizuta}}\ \emph {et~al.}(2020)\citenamefont
  {{Mizuta}}, \citenamefont {{Takasan}},\ and\ \citenamefont
  {{Kawakami}}}]{Mizuta2020}%
  \BibitemOpen
  \bibfield  {author} {\bibinfo {author} {\bibfnamefont {K.}~\bibnamefont
  {{Mizuta}}}, \bibinfo {author} {\bibfnamefont {K.}~\bibnamefont
  {{Takasan}}},\ and\ \bibinfo {author} {\bibfnamefont {N.}~\bibnamefont
  {{Kawakami}}},\ }\bibfield  {title} {\bibinfo {title} {{Exact Floquet quantum
  many-body scars under Rydberg blockade}},\ }\href
  {https://doi.org/10.1103/PhysRevResearch.2.033284} {\bibfield  {journal}
  {\bibinfo  {journal} {Phys. Rev. Research}\ }\textbf {\bibinfo {volume}
  {2}},\ \bibinfo {eid} {033284} (\bibinfo {year} {2020})}\BibitemShut
  {NoStop}%
\bibitem [{\citenamefont {{Hart}}\ \emph {et~al.}(2020)\citenamefont {{Hart}},
  \citenamefont {{De Tomasi}},\ and\ \citenamefont {{Castelnovo}}}]{Hart2020}%
  \BibitemOpen
  \bibfield  {author} {\bibinfo {author} {\bibfnamefont {O.}~\bibnamefont
  {{Hart}}}, \bibinfo {author} {\bibfnamefont {G.}~\bibnamefont {{De
  Tomasi}}},\ and\ \bibinfo {author} {\bibfnamefont {C.}~\bibnamefont
  {{Castelnovo}}},\ }\bibfield  {title} {\bibinfo {title} {{From compact
  localized states to many-body scars in the random quantum comb}},\ }\href
  {https://doi.org/10.1103/PhysRevResearch.2.043267} {\bibfield  {journal}
  {\bibinfo  {journal} {Phys. Rev. Research}\ }\textbf {\bibinfo {volume}
  {2}},\ \bibinfo {eid} {043267} (\bibinfo {year} {2020})}\BibitemShut
  {NoStop}%
\bibitem [{\citenamefont {Kao}\ \emph {et~al.}(2020)\citenamefont {Kao},
  \citenamefont {Li}, \citenamefont {Lin}, \citenamefont {Gopalakrishnan},\
  and\ \citenamefont {Lev}}]{Kao2020}%
  \BibitemOpen
  \bibfield  {author} {\bibinfo {author} {\bibfnamefont {W.}~\bibnamefont
  {Kao}}, \bibinfo {author} {\bibfnamefont {K.-Y.}\ \bibnamefont {Li}},
  \bibinfo {author} {\bibfnamefont {K.-Y.}\ \bibnamefont {Lin}}, \bibinfo
  {author} {\bibfnamefont {S.}~\bibnamefont {Gopalakrishnan}},\ and\ \bibinfo
  {author} {\bibfnamefont {B.~L.}\ \bibnamefont {Lev}},\ }\href@noop {}
  {\bibinfo {title} {Creating quantum many-body scars through topological
  pumping of a 1{D} dipolar gas}} (\bibinfo {year} {2020}),\ \Eprint
  {https://arxiv.org/abs/2002.10475} {arXiv:2002.10475 [cond-mat.quant-gas]}
  \BibitemShut {NoStop}%
\bibitem [{\citenamefont {{Medenjak}}\ \emph {et~al.}(2020)\citenamefont
  {{Medenjak}}, \citenamefont {{Bu{\v{c}}a}},\ and\ \citenamefont
  {{Jaksch}}}]{Buca2019_2}%
  \BibitemOpen
  \bibfield  {author} {\bibinfo {author} {\bibfnamefont {M.}~\bibnamefont
  {{Medenjak}}}, \bibinfo {author} {\bibfnamefont {B.}~\bibnamefont
  {{Bu{\v{c}}a}}},\ and\ \bibinfo {author} {\bibfnamefont {D.}~\bibnamefont
  {{Jaksch}}},\ }\bibfield  {title} {\bibinfo {title} {{Isolated Heisenberg
  magnet as a quantum time crystal}},\ }\href
  {https://doi.org/10.1103/PhysRevB.102.041117} {\bibfield  {journal} {\bibinfo
   {journal} {\prb}\ }\textbf {\bibinfo {volume} {102}},\ \bibinfo {eid}
  {041117} (\bibinfo {year} {2020})}\BibitemShut {NoStop}%
\bibitem [{\citenamefont {Lesanovsky}\ and\ \citenamefont
  {Katsura}(2012)}]{Lesanovsky2012}%
  \BibitemOpen
  \bibfield  {author} {\bibinfo {author} {\bibfnamefont {I.}~\bibnamefont
  {Lesanovsky}}\ and\ \bibinfo {author} {\bibfnamefont {H.}~\bibnamefont
  {Katsura}},\ }\bibfield  {title} {\bibinfo {title} {{Interacting Fibonacci
  anyons in a Rydberg gas}},\ }\href
  {https://doi.org/10.1103/PhysRevA.86.041601} {\bibfield  {journal} {\bibinfo
  {journal} {Phys. Rev. A}\ }\textbf {\bibinfo {volume} {86}},\ \bibinfo
  {pages} {041601(R)} (\bibinfo {year} {2012})}\BibitemShut {NoStop}%
\bibitem [{\citenamefont {Fendley}\ \emph {et~al.}(2004)\citenamefont
  {Fendley}, \citenamefont {Sengupta},\ and\ \citenamefont
  {Sachdev}}]{FendleySachdev}%
  \BibitemOpen
  \bibfield  {author} {\bibinfo {author} {\bibfnamefont {P.}~\bibnamefont
  {Fendley}}, \bibinfo {author} {\bibfnamefont {K.}~\bibnamefont {Sengupta}},\
  and\ \bibinfo {author} {\bibfnamefont {S.}~\bibnamefont {Sachdev}},\
  }\bibfield  {title} {\bibinfo {title} {Competing density-wave orders in a
  one-dimensional hard-boson model},\ }\href
  {https://doi.org/10.1103/PhysRevB.69.075106} {\bibfield  {journal} {\bibinfo
  {journal} {Phys. Rev. B}\ }\textbf {\bibinfo {volume} {69}},\ \bibinfo
  {pages} {075106} (\bibinfo {year} {2004})}\BibitemShut {NoStop}%
\bibitem [{\citenamefont {Kramer}\ and\ \citenamefont
  {Saraceno}(1981)}]{Kramer:1981tp}%
  \BibitemOpen
  \bibfield  {author} {\bibinfo {author} {\bibfnamefont {P.}~\bibnamefont
  {Kramer}}\ and\ \bibinfo {author} {\bibfnamefont {M.}~\bibnamefont
  {Saraceno}},\ }\href@noop {} {\emph {\bibinfo {title} {Geometry of the
  Time-Dependent Variational Principle in Quantum Mechanics}}},\ \bibinfo
  {series} {Lecture Notes in Physics}, Vol.\ \bibinfo {volume} {140}\ (\bibinfo
   {publisher} {Springer-Verlag},\ \bibinfo {address} {Berlin Heidelberg},\
  \bibinfo {year} {1981})\BibitemShut {NoStop}%
\bibitem [{\citenamefont {Haegeman}\ \emph {et~al.}(2011)\citenamefont
  {Haegeman}, \citenamefont {Cirac}, \citenamefont {Osborne}, \citenamefont
  {Pi\ifmmode~\check{z}\else \v{z}\fi{}orn}, \citenamefont {Verschelde},\ and\
  \citenamefont {Verstraete}}]{TDVP_QL}%
  \BibitemOpen
  \bibfield  {author} {\bibinfo {author} {\bibfnamefont {J.}~\bibnamefont
  {Haegeman}}, \bibinfo {author} {\bibfnamefont {J.~I.}\ \bibnamefont {Cirac}},
  \bibinfo {author} {\bibfnamefont {T.~J.}\ \bibnamefont {Osborne}}, \bibinfo
  {author} {\bibfnamefont {I.}~\bibnamefont {Pi\ifmmode~\check{z}\else
  \v{z}\fi{}orn}}, \bibinfo {author} {\bibfnamefont {H.}~\bibnamefont
  {Verschelde}},\ and\ \bibinfo {author} {\bibfnamefont {F.}~\bibnamefont
  {Verstraete}},\ }\bibfield  {title} {\bibinfo {title} {Time-dependent
  variational principle for quantum lattices},\ }\href
  {https://doi.org/10.1103/PhysRevLett.107.070601} {\bibfield  {journal}
  {\bibinfo  {journal} {Phys. Rev. Lett.}\ }\textbf {\bibinfo {volume} {107}},\
  \bibinfo {pages} {070601} (\bibinfo {year} {2011})}\BibitemShut {NoStop}%
\bibitem [{\citenamefont {Leviatan}\ \emph {et~al.}(2017)\citenamefont
  {Leviatan}, \citenamefont {Pollmann}, \citenamefont {Bardarson},
  \citenamefont {Huse},\ and\ \citenamefont {Altman}}]{Altman17}%
  \BibitemOpen
  \bibfield  {author} {\bibinfo {author} {\bibfnamefont {E.}~\bibnamefont
  {Leviatan}}, \bibinfo {author} {\bibfnamefont {F.}~\bibnamefont {Pollmann}},
  \bibinfo {author} {\bibfnamefont {J.~H.}\ \bibnamefont {Bardarson}}, \bibinfo
  {author} {\bibfnamefont {D.~A.}\ \bibnamefont {Huse}},\ and\ \bibinfo
  {author} {\bibfnamefont {E.}~\bibnamefont {Altman}},\ }\href@noop {}
  {\bibinfo {title} {Quantum thermalization dynamics with matrix-product
  states}} (\bibinfo {year} {2017}),\ \Eprint
  {https://arxiv.org/abs/1702.08894} {arXiv:1702.08894 [cond-mat.stat-mech]}
  \BibitemShut {NoStop}%
\bibitem [{\citenamefont {Hallam}\ \emph {et~al.}(2019)\citenamefont {Hallam},
  \citenamefont {Morley},\ and\ \citenamefont {Green}}]{Hallam2018ta}%
  \BibitemOpen
  \bibfield  {author} {\bibinfo {author} {\bibfnamefont {A.}~\bibnamefont
  {Hallam}}, \bibinfo {author} {\bibfnamefont {J.~G.}\ \bibnamefont {Morley}},\
  and\ \bibinfo {author} {\bibfnamefont {A.~G.}\ \bibnamefont {Green}},\
  }\bibfield  {title} {\bibinfo {title} {{The Lyapunov spectra of quantum
  thermalisation}},\ }\href {https://doi.org/10.1038/s41467-019-10336-4}
  {\bibfield  {journal} {\bibinfo  {journal} {Nat. Commun.}\ }\textbf {\bibinfo
  {volume} {10}},\ \bibinfo {pages} {2708} (\bibinfo {year}
  {2019})}\BibitemShut {NoStop}%
\bibitem [{\citenamefont {Turner}\ \emph
  {et~al.}(2018{\natexlab{b}})\citenamefont {Turner}, \citenamefont
  {Michailidis}, \citenamefont {Abanin}, \citenamefont {Serbyn},\ and\
  \citenamefont {Papi{\'c}}}]{Turner:2018in}%
  \BibitemOpen
  \bibfield  {author} {\bibinfo {author} {\bibfnamefont {C.~J.}\ \bibnamefont
  {Turner}}, \bibinfo {author} {\bibfnamefont {A.~A.}\ \bibnamefont
  {Michailidis}}, \bibinfo {author} {\bibfnamefont {D.~A.}\ \bibnamefont
  {Abanin}}, \bibinfo {author} {\bibfnamefont {M.}~\bibnamefont {Serbyn}},\
  and\ \bibinfo {author} {\bibfnamefont {Z.}~\bibnamefont {Papi{\'c}}},\
  }\bibfield  {title} {\bibinfo {title} {{Quantum scarred eigenstates in a
  Rydberg atom chain: Entanglement, breakdown of thermalization, and stability
  to perturbations}},\ }\href {https://doi.org/10.1103/PhysRevB.98.155134}
  {\bibfield  {journal} {\bibinfo  {journal} {Phys. Rev. B}\ }\textbf {\bibinfo
  {volume} {98}},\ \bibinfo {pages} {155134} (\bibinfo {year}
  {2018}{\natexlab{b}})}\BibitemShut {NoStop}%
\bibitem [{\citenamefont {Werman}(2020)}]{Werman:2020vd}%
  \BibitemOpen
  \bibfield  {author} {\bibinfo {author} {\bibfnamefont {Y.}~\bibnamefont
  {Werman}},\ }\href@noop {} {\bibinfo {title} {Quantum chaos in a {Rydberg}
  atom system}} (\bibinfo {year} {2020}),\ \Eprint
  {https://arxiv.org/abs/2001.06110} {arXiv:2001.06110 [quant-ph]} \BibitemShut
  {NoStop}%
\bibitem [{\citenamefont {{Khemani}}\ and\ \citenamefont
  {{Nandkishore}}(2019)}]{Khemani2019}%
  \BibitemOpen
  \bibfield  {author} {\bibinfo {author} {\bibfnamefont {V.}~\bibnamefont
  {{Khemani}}}\ and\ \bibinfo {author} {\bibfnamefont {R.}~\bibnamefont
  {{Nandkishore}}},\ }\bibfield  {title} {\bibinfo {title} {{Local constraints
  can globally shatter Hilbert space: a new route to quantum information
  protection}},\ }\href@noop {} {\bibfield  {journal} {\bibinfo  {journal}
  {arXiv e-prints}\ ,\ \bibinfo {eid} {arXiv:1904.04815}} (\bibinfo {year}
  {2019})},\ \Eprint {https://arxiv.org/abs/1904.04815} {arXiv:1904.04815
  [cond-mat.stat-mech]} \BibitemShut {NoStop}%
\bibitem [{\citenamefont {Sala}\ \emph {et~al.}(2020)\citenamefont {Sala},
  \citenamefont {Rakovszky}, \citenamefont {Verresen}, \citenamefont {Knap},\
  and\ \citenamefont {Pollmann}}]{Sala2019}%
  \BibitemOpen
  \bibfield  {author} {\bibinfo {author} {\bibfnamefont {P.}~\bibnamefont
  {Sala}}, \bibinfo {author} {\bibfnamefont {T.}~\bibnamefont {Rakovszky}},
  \bibinfo {author} {\bibfnamefont {R.}~\bibnamefont {Verresen}}, \bibinfo
  {author} {\bibfnamefont {M.}~\bibnamefont {Knap}},\ and\ \bibinfo {author}
  {\bibfnamefont {F.}~\bibnamefont {Pollmann}},\ }\bibfield  {title} {\bibinfo
  {title} {Ergodicity-breaking arising from {Hilbert} space fragmentation in
  dipole-conserving {Hamiltonians}},\ }\href
  {https://doi.org/10.1103/PhysRevX.10.011047} {\bibfield  {journal} {\bibinfo
  {journal} {Phys. Rev. X}\ }\textbf {\bibinfo {volume} {10}},\ \bibinfo
  {pages} {011047} (\bibinfo {year} {2020})}\BibitemShut {NoStop}%
\bibitem [{\citenamefont {{Khemani}}\ \emph {et~al.}(2020)\citenamefont
  {{Khemani}}, \citenamefont {{Hermele}},\ and\ \citenamefont
  {{Nandkishore}}}]{Khemani2019_2}%
  \BibitemOpen
  \bibfield  {author} {\bibinfo {author} {\bibfnamefont {V.}~\bibnamefont
  {{Khemani}}}, \bibinfo {author} {\bibfnamefont {M.}~\bibnamefont
  {{Hermele}}},\ and\ \bibinfo {author} {\bibfnamefont {R.}~\bibnamefont
  {{Nandkishore}}},\ }\bibfield  {title} {\bibinfo {title} {{Localization from
  Hilbert space shattering: From theory to physical realizations}},\ }\href
  {https://doi.org/10.1103/PhysRevB.101.174204} {\bibfield  {journal} {\bibinfo
   {journal} {\prb}\ }\textbf {\bibinfo {volume} {101}},\ \bibinfo {eid}
  {174204} (\bibinfo {year} {2020})}\BibitemShut {NoStop}%
\bibitem [{\citenamefont {Roy}\ and\ \citenamefont
  {Lazarides}(2020)}]{Roy2020}%
  \BibitemOpen
  \bibfield  {author} {\bibinfo {author} {\bibfnamefont {S.}~\bibnamefont
  {Roy}}\ and\ \bibinfo {author} {\bibfnamefont {A.}~\bibnamefont
  {Lazarides}},\ }\bibfield  {title} {\bibinfo {title} {Strong ergodicity
  breaking due to local constraints in a quantum system},\ }\href
  {https://doi.org/10.1103/PhysRevResearch.2.023159} {\bibfield  {journal}
  {\bibinfo  {journal} {Phys. Rev. Research}\ }\textbf {\bibinfo {volume}
  {2}},\ \bibinfo {pages} {023159} (\bibinfo {year} {2020})}\BibitemShut
  {NoStop}%
\bibitem [{\citenamefont {Yang}\ \emph {et~al.}(2020)\citenamefont {Yang},
  \citenamefont {Liu}, \citenamefont {Gorshkov},\ and\ \citenamefont
  {Iadecola}}]{Yang2020}%
  \BibitemOpen
  \bibfield  {author} {\bibinfo {author} {\bibfnamefont {Z.-C.}\ \bibnamefont
  {Yang}}, \bibinfo {author} {\bibfnamefont {F.}~\bibnamefont {Liu}}, \bibinfo
  {author} {\bibfnamefont {A.~V.}\ \bibnamefont {Gorshkov}},\ and\ \bibinfo
  {author} {\bibfnamefont {T.}~\bibnamefont {Iadecola}},\ }\bibfield  {title}
  {\bibinfo {title} {Hilbert-space fragmentation from strict confinement},\
  }\href {https://doi.org/10.1103/PhysRevLett.124.207602} {\bibfield  {journal}
  {\bibinfo  {journal} {Phys. Rev. Lett.}\ }\textbf {\bibinfo {volume} {124}},\
  \bibinfo {pages} {207602} (\bibinfo {year} {2020})}\BibitemShut {NoStop}%
\bibitem [{\citenamefont {Pancotti}\ \emph {et~al.}(2020)\citenamefont
  {Pancotti}, \citenamefont {Giudice}, \citenamefont {Cirac}, \citenamefont
  {Garrahan},\ and\ \citenamefont {Ba{\~n}uls}}]{Pancotti2020}%
  \BibitemOpen
  \bibfield  {author} {\bibinfo {author} {\bibfnamefont {N.}~\bibnamefont
  {Pancotti}}, \bibinfo {author} {\bibfnamefont {G.}~\bibnamefont {Giudice}},
  \bibinfo {author} {\bibfnamefont {J.~I.}\ \bibnamefont {Cirac}}, \bibinfo
  {author} {\bibfnamefont {J.~P.}\ \bibnamefont {Garrahan}},\ and\ \bibinfo
  {author} {\bibfnamefont {M.~C.}\ \bibnamefont {Ba{\~n}uls}},\ }\bibfield
  {title} {\bibinfo {title} {Quantum {East} model: Localization, nonthermal
  eigenstates, and slow dynamics},\ }\href
  {https://doi.org/10.1103/PhysRevX.10.021051} {\bibfield  {journal} {\bibinfo
  {journal} {Phys. Rev. X}\ }\textbf {\bibinfo {volume} {10}},\ \bibinfo
  {pages} {021051} (\bibinfo {year} {2020})}\BibitemShut {NoStop}%
\bibitem [{\citenamefont {{Lychkovskiy}}(2020)}]{Lychkovskiy2020}%
  \BibitemOpen
  \bibfield  {author} {\bibinfo {author} {\bibfnamefont {O.}~\bibnamefont
  {{Lychkovskiy}}},\ }\bibfield  {title} {\bibinfo {title} {{A Remark on the
  Notion of Independence of Quantum Integrals of Motion in the Thermodynamic
  Limit}},\ }\href {https://doi.org/10.1007/s10955-019-02482-2} {\bibfield
  {journal} {\bibinfo  {journal} {J. Stat. Phys.}\ }\textbf {\bibinfo {volume}
  {178}},\ \bibinfo {pages} {1028} (\bibinfo {year} {2020})}\BibitemShut
  {NoStop}%
\bibitem [{\citenamefont {Deutsch}(1991)}]{DeutschETH}%
  \BibitemOpen
  \bibfield  {author} {\bibinfo {author} {\bibfnamefont {J.~M.}\ \bibnamefont
  {Deutsch}},\ }\bibfield  {title} {\bibinfo {title} {{Quantum statistical
  mechanics in a closed system}},\ }\href
  {https://doi.org/10.1103/PhysRevA.43.2046} {\bibfield  {journal} {\bibinfo
  {journal} {Phys. Rev. A}\ }\textbf {\bibinfo {volume} {43}},\ \bibinfo
  {pages} {2046} (\bibinfo {year} {1991})}\BibitemShut {NoStop}%
\bibitem [{\citenamefont {Srednicki}(1994)}]{SrednickiETH}%
  \BibitemOpen
  \bibfield  {author} {\bibinfo {author} {\bibfnamefont {M.}~\bibnamefont
  {Srednicki}},\ }\bibfield  {title} {\bibinfo {title} {{Chaos and quantum
  thermalization}},\ }\href {https://doi.org/10.1103/PhysRevE.50.888}
  {\bibfield  {journal} {\bibinfo  {journal} {Phys. Rev. E}\ }\textbf {\bibinfo
  {volume} {50}},\ \bibinfo {pages} {888} (\bibinfo {year} {1994})}\BibitemShut
  {NoStop}%
\bibitem [{\citenamefont {D'Alessio}\ \emph {et~al.}(2016)\citenamefont
  {D'Alessio}, \citenamefont {Kafri}, \citenamefont {Polkovnikov},\ and\
  \citenamefont {Rigol}}]{dAlessio2016}%
  \BibitemOpen
  \bibfield  {author} {\bibinfo {author} {\bibfnamefont {L.}~\bibnamefont
  {D'Alessio}}, \bibinfo {author} {\bibfnamefont {Y.}~\bibnamefont {Kafri}},
  \bibinfo {author} {\bibfnamefont {A.}~\bibnamefont {Polkovnikov}},\ and\
  \bibinfo {author} {\bibfnamefont {M.}~\bibnamefont {Rigol}},\ }\bibfield
  {title} {\bibinfo {title} {From quantum chaos and eigenstate thermalization
  to statistical mechanics and thermodynamics},\ }\href
  {https://doi.org/10.1080/00018732.2016.1198134} {\bibfield  {journal}
  {\bibinfo  {journal} {Adv. Phys.}\ }\textbf {\bibinfo {volume} {65}},\
  \bibinfo {pages} {239} (\bibinfo {year} {2016})}\BibitemShut {NoStop}%
\bibitem [{\citenamefont {Gogolin}\ and\ \citenamefont
  {Eisert}(2016)}]{Gogolin2016}%
  \BibitemOpen
  \bibfield  {author} {\bibinfo {author} {\bibfnamefont {C.}~\bibnamefont
  {Gogolin}}\ and\ \bibinfo {author} {\bibfnamefont {J.}~\bibnamefont
  {Eisert}},\ }\bibfield  {title} {\bibinfo {title} {Equilibration,
  thermalisation, and the emergence of statistical mechanics in closed quantum
  systems},\ }\href {https://doi.org/10.1088/0034-4885/79/5/056001} {\bibfield
  {journal} {\bibinfo  {journal} {Rep. Prog. Phys.}\ }\textbf {\bibinfo
  {volume} {79}},\ \bibinfo {pages} {056001} (\bibinfo {year}
  {2016})}\BibitemShut {NoStop}%
\bibitem [{\citenamefont {Basko}\ \emph {et~al.}(2006)\citenamefont {Basko},
  \citenamefont {Aleiner},\ and\ \citenamefont {Altshuler}}]{basko2006metal}%
  \BibitemOpen
  \bibfield  {author} {\bibinfo {author} {\bibfnamefont {D.~M.}\ \bibnamefont
  {Basko}}, \bibinfo {author} {\bibfnamefont {I.~L.}\ \bibnamefont {Aleiner}},\
  and\ \bibinfo {author} {\bibfnamefont {B.~L.}\ \bibnamefont {Altshuler}},\
  }\bibfield  {title} {\bibinfo {title} {Metal--insulator transition in a
  weakly interacting many-electron system with localized single-particle
  states},\ }\href {https://doi.org/10.1016/j.aop.2005.11.014} {\bibfield
  {journal} {\bibinfo  {journal} {Ann. Phys. (N. Y.)}\ }\textbf {\bibinfo
  {volume} {321}},\ \bibinfo {pages} {1126} (\bibinfo {year}
  {2006})}\BibitemShut {NoStop}%
\bibitem [{\citenamefont {Serbyn}\ \emph {et~al.}(2013)\citenamefont {Serbyn},
  \citenamefont {Papi{\'c}},\ and\ \citenamefont {Abanin}}]{serbyn2013local}%
  \BibitemOpen
  \bibfield  {author} {\bibinfo {author} {\bibfnamefont {M.}~\bibnamefont
  {Serbyn}}, \bibinfo {author} {\bibfnamefont {Z.}~\bibnamefont {Papi{\'c}}},\
  and\ \bibinfo {author} {\bibfnamefont {D.~A.}\ \bibnamefont {Abanin}},\
  }\bibfield  {title} {\bibinfo {title} {Local conservation laws and the
  structure of the many-body localized states},\ }\href
  {https://doi.org/10.1103/PhysRevLett.111.127201} {\bibfield  {journal}
  {\bibinfo  {journal} {Phys. Rev. Lett.}\ }\textbf {\bibinfo {volume} {111}},\
  \bibinfo {pages} {127201} (\bibinfo {year} {2013})}\BibitemShut {NoStop}%
\bibitem [{\citenamefont {Huse}\ \emph {et~al.}(2014)\citenamefont {Huse},
  \citenamefont {Nandkishore},\ and\ \citenamefont
  {Oganesyan}}]{HuseNandkishoreOganesyan}%
  \BibitemOpen
  \bibfield  {author} {\bibinfo {author} {\bibfnamefont {D.~A.}\ \bibnamefont
  {Huse}}, \bibinfo {author} {\bibfnamefont {R.}~\bibnamefont {Nandkishore}},\
  and\ \bibinfo {author} {\bibfnamefont {V.}~\bibnamefont {Oganesyan}},\
  }\bibfield  {title} {\bibinfo {title} {Phenomenology of fully
  many-body-localized systems},\ }\href
  {https://doi.org/10.1103/PhysRevB.90.174202} {\bibfield  {journal} {\bibinfo
  {journal} {Phys. Rev. B}\ }\textbf {\bibinfo {volume} {90}},\ \bibinfo
  {pages} {174202} (\bibinfo {year} {2014})}\BibitemShut {NoStop}%
\bibitem [{\citenamefont {Sutherland}(2004)}]{sutherland2004beautiful}%
  \BibitemOpen
  \bibfield  {author} {\bibinfo {author} {\bibfnamefont {B.}~\bibnamefont
  {Sutherland}},\ }\href@noop {} {\emph {\bibinfo {title} {Beautiful models: 70
  years of exactly solved quantum many-body problems}}}\ (\bibinfo  {publisher}
  {World Scientific Publishing Company},\ \bibinfo {year} {2004})\BibitemShut
  {NoStop}%
\bibitem [{\citenamefont {Shiraishi}\ and\ \citenamefont
  {Mori}(2017)}]{Shiraishi:2017ek}%
  \BibitemOpen
  \bibfield  {author} {\bibinfo {author} {\bibfnamefont {N.}~\bibnamefont
  {Shiraishi}}\ and\ \bibinfo {author} {\bibfnamefont {T.}~\bibnamefont
  {Mori}},\ }\bibfield  {title} {\bibinfo {title} {Systematic construction of
  counterexamples to the eigenstate thermalization hypothesis},\ }\href
  {https://doi.org/10.1103/PhysRevLett.119.030601} {\bibfield  {journal}
  {\bibinfo  {journal} {Phys. Rev. Lett.}\ }\textbf {\bibinfo {volume} {119}},\
  \bibinfo {pages} {030601} (\bibinfo {year} {2017})}\BibitemShut {NoStop}%
\bibitem [{\citenamefont {Shiraishi}(2019)}]{Shiraishi2019}%
  \BibitemOpen
  \bibfield  {author} {\bibinfo {author} {\bibfnamefont {N.}~\bibnamefont
  {Shiraishi}},\ }\bibfield  {title} {\bibinfo {title} {Connection between
  quantum-many-body scars and the
  {Affleck}{\textendash}{Kennedy}{\textendash}{Lieb}{\textendash}{Tasaki} model
  from the viewpoint of embedded {Hamiltonians}},\ }\href
  {https://doi.org/10.1088/1742-5468/ab342e} {\bibfield  {journal} {\bibinfo
  {journal} {J. Stat. Mech.: Theory Exp}\ }\textbf {\bibinfo {volume} {2019}},\
  \bibinfo {pages} {083103} (\bibinfo {year} {2019})}\BibitemShut {NoStop}%
\bibitem [{\citenamefont {Brandao}\ \emph {et~al.}(2016)\citenamefont
  {Brandao}, \citenamefont {Christandl}, \citenamefont {Harrow},\ and\
  \citenamefont {Walter}}]{Brandao2016}%
  \BibitemOpen
  \bibfield  {author} {\bibinfo {author} {\bibfnamefont {F.~G.}\ \bibnamefont
  {Brandao}}, \bibinfo {author} {\bibfnamefont {M.}~\bibnamefont {Christandl}},
  \bibinfo {author} {\bibfnamefont {A.~W.}\ \bibnamefont {Harrow}},\ and\
  \bibinfo {author} {\bibfnamefont {M.}~\bibnamefont {Walter}},\ }\bibfield
  {title} {\bibinfo {title} {The mathematics of entanglement},\ }\href@noop {}
  {\bibfield  {journal} {\bibinfo  {journal} {arXiv preprint arXiv:1604.01790}\
  } (\bibinfo {year} {2016})}\BibitemShut {NoStop}%
\bibitem [{\citenamefont {Sciolla}\ and\ \citenamefont
  {Biroli}(2011)}]{Sciolla2011}%
  \BibitemOpen
  \bibfield  {author} {\bibinfo {author} {\bibfnamefont {B.}~\bibnamefont
  {Sciolla}}\ and\ \bibinfo {author} {\bibfnamefont {G.}~\bibnamefont
  {Biroli}},\ }\bibfield  {title} {\bibinfo {title} {Dynamical transitions and
  quantum quenches in mean-field models},\ }\href
  {https://doi.org/10.1088/1742-5468/2011/11/p11003} {\bibfield  {journal}
  {\bibinfo  {journal} {J. Stat. Mech.: Theory Exp}\ }\textbf {\bibinfo
  {volume} {2011}},\ \bibinfo {pages} {P11003} (\bibinfo {year}
  {2011})}\BibitemShut {NoStop}%
\bibitem [{\citenamefont {Mori}(2017)}]{Mori2017}%
  \BibitemOpen
  \bibfield  {author} {\bibinfo {author} {\bibfnamefont {T.}~\bibnamefont
  {Mori}},\ }\bibfield  {title} {\bibinfo {title} {Classical ergodicity and
  quantum eigenstate thermalization: Analysis in fully connected ising
  ferromagnets},\ }\href {https://doi.org/10.1103/PhysRevE.96.012134}
  {\bibfield  {journal} {\bibinfo  {journal} {Phys. Rev. E}\ }\textbf {\bibinfo
  {volume} {96}},\ \bibinfo {pages} {012134} (\bibinfo {year}
  {2017})}\BibitemShut {NoStop}%
\bibitem [{\citenamefont {Gazeau}(2009)}]{Gazeau:2009vb}%
  \BibitemOpen
  \bibfield  {author} {\bibinfo {author} {\bibfnamefont {J.-P.}\ \bibnamefont
  {Gazeau}},\ }\href@noop {} {\emph {\bibinfo {title} {{Coherent States in
  Quantum Physics}}}}\ (\bibinfo  {publisher} {Wiley-VCH},\ \bibinfo {year}
  {2009})\BibitemShut {NoStop}%
\bibitem [{\citenamefont {Green}\ \emph {et~al.}(2016)\citenamefont {Green},
  \citenamefont {Hooley}, \citenamefont {Keeling},\ and\ \citenamefont
  {Simon}}]{Green:2016uw}%
  \BibitemOpen
  \bibfield  {author} {\bibinfo {author} {\bibfnamefont {A.~G.}\ \bibnamefont
  {Green}}, \bibinfo {author} {\bibfnamefont {C.~A.}\ \bibnamefont {Hooley}},
  \bibinfo {author} {\bibfnamefont {J.}~\bibnamefont {Keeling}},\ and\ \bibinfo
  {author} {\bibfnamefont {S.~H.}\ \bibnamefont {Simon}},\ }\href@noop {}
  {\bibinfo {title} {Feynman path integrals over entangled states}} (\bibinfo
  {year} {2016}),\ \Eprint {https://arxiv.org/abs/1607.01778} {arXiv:1607.01778
  [cond-mat.str-el]} \BibitemShut {NoStop}%
\bibitem [{\citenamefont {Christensen}\ \emph {et~al.}(2016)\citenamefont
  {Christensen} \emph {et~al.}}]{Christensen2016}%
  \BibitemOpen
  \bibfield  {author} {\bibinfo {author} {\bibfnamefont {O.}~\bibnamefont
  {Christensen}} \emph {et~al.},\ }\href@noop {} {\emph {\bibinfo {title} {An
  introduction to frames and Riesz bases}}}\ (\bibinfo  {publisher}
  {Springer},\ \bibinfo {year} {2016})\BibitemShut {NoStop}%
\bibitem [{\citenamefont {Bengtsson}\ and\ \citenamefont
  {{\.{Z}}yczkowski}(2006)}]{Bengtsson:2006iz}%
  \BibitemOpen
  \bibfield  {author} {\bibinfo {author} {\bibfnamefont {I.}~\bibnamefont
  {Bengtsson}}\ and\ \bibinfo {author} {\bibfnamefont {K.}~\bibnamefont
  {{\.{Z}}yczkowski}},\ }\href@noop {} {\emph {\bibinfo {title} {{Geometry of
  Quantum States}}}}\ (\bibinfo {address} {Cambridge},\ \bibinfo {year}
  {2006})\BibitemShut {NoStop}%
\bibitem [{\citenamefont {Gutzwiller}(2013)}]{GutzwillerBook}%
  \BibitemOpen
  \bibfield  {author} {\bibinfo {author} {\bibfnamefont {M.~C.}\ \bibnamefont
  {Gutzwiller}},\ }\href@noop {} {\emph {\bibinfo {title} {Chaos in classical
  and quantum mechanics}}},\ Vol.~\bibinfo {volume} {1}\ (\bibinfo  {publisher}
  {Springer Science \& Business Media},\ \bibinfo {year} {2013})\BibitemShut
  {NoStop}%
\bibitem [{\citenamefont {{Lin}}\ \emph {et~al.}(2020)\citenamefont {{Lin}},
  \citenamefont {{Calvera}},\ and\ \citenamefont {{Hsieh}}}]{Lin2020}%
  \BibitemOpen
  \bibfield  {author} {\bibinfo {author} {\bibfnamefont {C.-J.}\ \bibnamefont
  {{Lin}}}, \bibinfo {author} {\bibfnamefont {V.}~\bibnamefont {{Calvera}}},\
  and\ \bibinfo {author} {\bibfnamefont {T.~H.}\ \bibnamefont {{Hsieh}}},\
  }\bibfield  {title} {\bibinfo {title} {{Quantum many-body scar states in
  two-dimensional Rydberg atom arrays}},\ }\href
  {https://doi.org/10.1103/PhysRevB.101.220304} {\bibfield  {journal} {\bibinfo
   {journal} {\prb}\ }\textbf {\bibinfo {volume} {101}},\ \bibinfo {eid}
  {220304} (\bibinfo {year} {2020})}\BibitemShut {NoStop}%
\bibitem [{\citenamefont {{Michailidis}}\ \emph {et~al.}(2020)\citenamefont
  {{Michailidis}}, \citenamefont {{Turner}}, \citenamefont {{Papi{\'c}}},
  \citenamefont {{Abanin}},\ and\ \citenamefont {{Serbyn}}}]{Michailidis2020}%
  \BibitemOpen
  \bibfield  {author} {\bibinfo {author} {\bibfnamefont {A.~A.}\ \bibnamefont
  {{Michailidis}}}, \bibinfo {author} {\bibfnamefont {C.~J.}\ \bibnamefont
  {{Turner}}}, \bibinfo {author} {\bibfnamefont {Z.}~\bibnamefont
  {{Papi{\'c}}}}, \bibinfo {author} {\bibfnamefont {D.~A.}\ \bibnamefont
  {{Abanin}}},\ and\ \bibinfo {author} {\bibfnamefont {M.}~\bibnamefont
  {{Serbyn}}},\ }\bibfield  {title} {\bibinfo {title} {{Stabilizing
  two-dimensional quantum scars by deformation and synchronization}},\ }\href
  {https://doi.org/10.1103/PhysRevResearch.2.022065} {\bibfield  {journal}
  {\bibinfo  {journal} {Phys. Rev. Research}\ }\textbf {\bibinfo {volume}
  {2}},\ \bibinfo {eid} {022065} (\bibinfo {year} {2020})}\BibitemShut
  {NoStop}%
\bibitem [{\citenamefont {Surace}\ \emph {et~al.}(2020)\citenamefont {Surace},
  \citenamefont {Mazza}, \citenamefont {Giudici}, \citenamefont {Lerose},
  \citenamefont {Gambassi},\ and\ \citenamefont {Dalmonte}}]{Surace2020}%
  \BibitemOpen
  \bibfield  {author} {\bibinfo {author} {\bibfnamefont {F.~M.}\ \bibnamefont
  {Surace}}, \bibinfo {author} {\bibfnamefont {P.~P.}\ \bibnamefont {Mazza}},
  \bibinfo {author} {\bibfnamefont {G.}~\bibnamefont {Giudici}}, \bibinfo
  {author} {\bibfnamefont {A.}~\bibnamefont {Lerose}}, \bibinfo {author}
  {\bibfnamefont {A.}~\bibnamefont {Gambassi}},\ and\ \bibinfo {author}
  {\bibfnamefont {M.}~\bibnamefont {Dalmonte}},\ }\bibfield  {title} {\bibinfo
  {title} {Lattice gauge theories and string dynamics in rydberg atom quantum
  simulators},\ }\href {https://doi.org/10.1103/PhysRevX.10.021041} {\bibfield
  {journal} {\bibinfo  {journal} {Phys. Rev. X}\ }\textbf {\bibinfo {volume}
  {10}},\ \bibinfo {pages} {021041} (\bibinfo {year} {2020})}\BibitemShut
  {NoStop}%
\bibitem [{\citenamefont {Moessner}\ and\ \citenamefont
  {Raman}(2011)}]{Moessner2011}%
  \BibitemOpen
  \bibfield  {author} {\bibinfo {author} {\bibfnamefont {R.}~\bibnamefont
  {Moessner}}\ and\ \bibinfo {author} {\bibfnamefont {K.~S.}\ \bibnamefont
  {Raman}},\ }\bibfield  {title} {\bibinfo {title} {Quantum dimer models},\
  }in\ \href@noop {} {\emph {\bibinfo {booktitle} {Introduction to Frustrated
  Magnetism}}}\ (\bibinfo  {publisher} {Springer},\ \bibinfo {year} {2011})\
  pp.\ \bibinfo {pages} {437--479}\BibitemShut {NoStop}%
\bibitem [{\citenamefont {Rokhsar}\ and\ \citenamefont
  {Kivelson}(1988)}]{Rokhsar1988}%
  \BibitemOpen
  \bibfield  {author} {\bibinfo {author} {\bibfnamefont {D.~S.}\ \bibnamefont
  {Rokhsar}}\ and\ \bibinfo {author} {\bibfnamefont {S.~A.}\ \bibnamefont
  {Kivelson}},\ }\bibfield  {title} {\bibinfo {title} {Superconductivity and
  the quantum hard-core dimer gas},\ }\href
  {https://doi.org/10.1103/PhysRevLett.61.2376} {\bibfield  {journal} {\bibinfo
   {journal} {Phys. Rev. Lett.}\ }\textbf {\bibinfo {volume} {61}},\ \bibinfo
  {pages} {2376} (\bibinfo {year} {1988})}\BibitemShut {NoStop}%
\bibitem [{\citenamefont {Abramowitz}\ and\ \citenamefont
  {Stegun}(1948)}]{Abramovich}%
  \BibitemOpen
  \bibfield  {author} {\bibinfo {author} {\bibfnamefont {M.}~\bibnamefont
  {Abramowitz}}\ and\ \bibinfo {author} {\bibfnamefont {I.~A.}\ \bibnamefont
  {Stegun}},\ }\href@noop {} {\emph {\bibinfo {title} {Handbook of mathematical
  functions with formulas, graphs, and mathematical tables}}},\ Vol.~\bibinfo
  {volume} {55}\ (\bibinfo  {publisher} {US Government printing office},\
  \bibinfo {year} {1948})\BibitemShut {NoStop}%
\bibitem [{\citenamefont {Ellis}(1984)}]{Ellis:1984iy}%
  \BibitemOpen
  \bibfield  {author} {\bibinfo {author} {\bibfnamefont {R.~S.}\ \bibnamefont
  {Ellis}},\ }\bibfield  {title} {\bibinfo {title} {{Large Deviations for a
  General Class of Random Vectors}},\ }\href@noop {} {\bibfield  {journal}
  {\bibinfo  {journal} {Ann. Probab.}\ }\textbf {\bibinfo {volume} {12}},\
  \bibinfo {pages} {1} (\bibinfo {year} {1984})}\BibitemShut {NoStop}%
\bibitem [{\citenamefont {Kadison}\ and\ \citenamefont
  {Ringrose}(1983)}]{Kadison:1983to}%
  \BibitemOpen
  \bibfield  {author} {\bibinfo {author} {\bibfnamefont {R.~V.}\ \bibnamefont
  {Kadison}}\ and\ \bibinfo {author} {\bibfnamefont {J.~R.}\ \bibnamefont
  {Ringrose}},\ }\href@noop {} {\emph {\bibinfo {title} {{Fundamentals of the
  Theory of Operator Algebras}}}},\ \bibinfo {series} {Elementary Theory},
  Vol.~\bibinfo {volume} {1}\ (\bibinfo  {publisher} {Academic Press},\
  \bibinfo {year} {1983})\BibitemShut {NoStop}%
\end{thebibliography}%

\clearpage{}

\end{document}